\documentclass[preprint2,longabstract]{aastex}

\slugcomment{}

\shorttitle{Hickson Compact Groups}
\shortauthors{Johnson et al.}

\begin{document}


\title{The Infrared Properties of Hickson Compact Groups}


\author{Kelsey E. Johnson \altaffilmark{1}}
\affil{Dept. of Astronomy, University of Virginia, P.O. Box 400325,
    Charlottesville, VA 22904}
\email{kej7a@virginia.edu}

\author{John E. Hibbard}
\affil{National Radio Astronomy Observatory, 520 Edgemont Road, 
Charlottesville, VA 22903}

\author{Sarah C. Gallagher}
\affil{Dept. of Physics and Astronomy, University of California, 430
Portola Plaza, Box 951547, Los Angeles CA, 90095-1547}

\author{Jane C. Charlton}
\affil{Dept. of Astronomy and Astrophysics, Pennsylvania State 
University, University Park, PA 16802}

\author{Ann E. Hornschemeier}
\affil{Laboratory for X-Ray Astrophysics, NASA Goddard Space Flight Center, 
Code 662.0, Greenbelt, MD 20771}

\author{Thomas H. Jarrett}
\affil{Spitzer Science Center, California Institute of Technology, Pasadena, CA 91125}

\and

\author{Amy E. Reines}
\affil{Dept. of Astronomy, University of Virginia, P.O. Box 400325,
    Charlottesville, VA 22904}

\altaffiltext{1}{Adjunct at National Radio Astronomy Observatory, 520 
Edgemont Road, Charlottesville, VA 22903}

\pagebreak

\begin{abstract}

Compact groups of galaxies provide a unique environment to study the
mechanisms by which star formation occurs amid continuous
gravitational encounters.  We present 2MASS (JHK), {\it Spitzer} IRAC
(3.5-8$\mu$m) and MIPS (24$\mu$m) observations of a sample of twelve
Hickson Compact Groups (HCGs 2, 7, 16, 19, 22, 31, 42, 48, 59, 61, 62,
and 90) that includes a total of 45 galaxies.  The infrared colors of
the galaxies in this sample span a range of parameter space, and some
trends are apparent in the data.  The near-infrared colors of the
sample galaxies are largely consistent with being dominated by
slightly reddened normal stellar populations. There is also some
evidence for a K-band excess in a few cases, which likely indicates
the presence of hot dust at or near the sublimation temperature
associated with AGN or star formation activity.  Galaxies that have
the most significant PAH and/or hot dust emission (as inferred from
excess 8$\mu$m flux) also tend to have larger amounts of extinction
and/or K-band excess and stronger 24$\mu$m emission, all of which
suggest ongoing star formation activity.  We separate the twelve HCGs
in our sample into three types based on the ratio of the group HI mass
to dynamical mass.  We find evidence that galaxies in the most
gas-rich groups tend to be the most actively star forming.  Galaxies
in the most gas-poor groups tend to be tightly clustered around a
narrow range in colors consistent with the integrated light from a
normal stellar population.  We interpret these trends as indicating
that galaxies in gas-rich groups experience star formation and/or
nuclear actively until their neutral gas consumed, stripped, or
ionized.  The galaxies in this sample exhibit a ``gap'' between
gas-rich and gas-poor groups in infrared color space that is sparsely
populated and not seen in the {\it Spitzer} First Look Survey sample.
This gap may suggest a rapid evolution of galaxy properties in
response to dynamical effects.  These results suggest that the global
properties of the groups and the local properties of the galaxies are
connected.

\end{abstract}


\section{Introduction}
Compact groups of galaxies provide a rich environment in which to
study the effect of galaxy interactions and mergers on galaxy
formation and evolution.  Compact groups are among the densest
concentrations of galaxies known, comparable to the centers of rich
clusters \citep{hickson82}.  However, they also have relatively low
velocity dispersions, increasing the duration of gravitational
interactions between the galaxies in these groups \citep{hickson92}.
In fact, a large fraction of galaxies within Hickson Compact Groups
(HCGs) have morphological peculiarities of some kind \citep{mendes94}
or abnormal rotation curves \citep{rubin91}.  Because of their galaxy
densities and velocity dispersions, compact groups may provide local
analogs to hierarchical galaxy formation in the earlier universe.
Moreover, these groups provide laboratories to study a variety of
physical processes, including the formation of active galactic nuclei
(AGN), super star clusters, and dwarf galaxies.

It is currently unclear how the frequent tidal encounters that occur
in compact groups induce or inhibit activity and transformation of the
member galaxies.  Studies of elliptical galaxies in HCGs indicate that
interactions have modified their isophotes and/or that some
ellipticals in these groups are bluer than normal \citep{zepf91}.
Surveys of subsets of HCGs have yielded apparently contradictory
results relating the role of interactions and star formation.  For
example, based on a sample of 95 galaxies within 31 HCGs, H$\alpha$
emission seems to be anti-correlated with the velocity dispersion of
HCGs \citep{severgnini01}, suggesting that higher levels of star
formation activity are associated with slower interactions between
group members.  However, based on a sample of 30 galaxies in 20 HCGs,
\citet{nishiura00} do not find any significant correlation between the
dynamical properties of a group and the presence of disturbed
morphology or nuclear activity of group members.

The compact group environment may lead to exhaustion or stripping of
molecular gas, or perhaps concentration of molecular gas into the
nuclei of the member galaxies.  Individual galaxies in HCGs as a class
may be relatively deficient in both atomic and molecular gas compared
to isolated spirals \citep{yun97, huchtmeier97, verdes-montenegro98,
verdes-montenegro01}, although \citet{leon98} find the molecular gas
content in HCGs to be similar to that in pairs and starburst samples.
Leon et al. (1998) also find enhanced H$_2$ for the most compact
groups, which suggests that the galaxy interactions drive the gas
inwards.  Groups that are more deficient in HI also have a higher
detection rate in X-ray observations \citep{verdes-montenegro01,
ponman96}, possibly indicating that ionization via shocks, heating, or
the intergalactic radiation field contributes to the lack of neutral
gas observed in these systems.  An example of strong large-scale
shocks due to encounters between galaxies and the IGM is seen in
Stephan's Quintet (HCG92) \citep{appleton06}.  There is also evidence
for the formation of tidal dwarf galaxies in HCGs
\citep[e.g.][]{hunsberger98}, supporting the idea that significant
amounts of matter may be stripped from member galaxies during their
interactions.

Based on HI observations of 72 HCGs, Verdes-Montenegro et al. (2001)
propose an evolutionary scenario that is somewhat analogous to the
famous ``Toomre sequence'' for merging galaxy pairs.  First a loose
group gradually contracts into a compact group \citep{barton98}.  This
stage is followed by a ``transformation'' period in which galaxy
morphologies tend to evolve to earlier types, during which HI gas is
extracted and dispersed, and X-ray gas is created.  Throughout this
scenario, the gas is actively redistributed from the relatively
quiescent interstellar medium (ISM), to fuel star formation, power AGN
activity, and contribute to the intercluster medium (ICM) --- giving
rise to a variety of observable processes.

Many of the processes that could be taking place within compact groups
will have a signature in the infrared emission of the galaxies --- AGN
activity, star formation, and the presence of warm/cold dust should
all be readily observable in the thermal infrared.  Based on IRAS
observations, \citet{allam95} and \citet{verdes-montenegro98} find
that compact group galaxies exhibit a normal level of infrared
emission compared to control samples of isolated galaxies, which
suggests that the tidal activity in these groups does not generally
enhance star formation.  Thermal infrared observations with the {\it
  Infrared Space Observatory} (ISO) were carried out on HCG~31 and
HCG~92 \citep{xu99, sulentic01, o'halloran02, xu03}.  These
observations indicate that several modes of star formation related to
the compact group environment are simultaneously present in these
groups (including intense star formation at collision interfaces, star
formation in the tidal debris, and star formation due to galaxy
interactions with the ICM), which demonstrates that in at least some
HCGs the interaction of group members has affected star formation.
These observations also potentially provide insight into the nature of
gas deficiency in HCGs; the far-IR 60$\mu$m and 100$\mu$m observations
of HCG~92 indicate diffuse emission from dust outside of the galaxies
\citep[e.g.][]{xu99}, plausibly powered by stripped or tidally formed
stars.

These case studies have been valuable for gaining insight into
specific groups.  However, compact groups present very complex
environments, and studies of individual groups cannot probe the range
of processes and environments that may contribute to an ``evolutionary
sequence''.  As part of a multi-wavelength effort to study compact
groups, and the variety of physical processes that take place within
them, we have obtained {\it Spitzer Space Telescope} observations of a
sample of twelve nearby ($<4500$~km~s$^{-1}$) compact groups chosen
to span a range in evolutionary properties.  In addition to the {\it
Spitzer} data, we also present the 2MASS \citep[The Two Micron All Sky
Survey,][]{skrutskie06} fluxes of the galaxies in our sample.  The
goal of this initial paper is to present an overview of the main infrared
properties of the sample.

\section{Observations \label{Observations}}

\begin{deluxetable}{ccccccccc}
\tabletypesize{\scriptsize}
\tablecaption{Sample of Hickson Compact Groups used in this study \label{sample}}
\tablewidth{0pt}
\tablehead{

& \colhead{$<v_\odot>$} & \colhead{R.A.} & \colhead{Dec.} & 
\multicolumn{3}{c}{Galaxy Membership} & & \colhead{Group}\\
\colhead{ID} & \colhead{(km/s)} & \colhead{(J2000)} & \colhead{(J2000)} & 
\colhead{E/SO} & \colhead{Sp} & \colhead{Other} & 
\colhead{$log(M_{{\rm HI}})/log(M_{{\rm dyn}})$ \tablenotemark{a}} 
& \colhead{type \tablenotemark{b}}
}
\startdata
HCG~02 & 4309 & 00:31:30.0 & +08:25:52 & 0 & 2 & 1 & 0.96$\pm$0.07 & I \\
HCG~07 & 4233 & 00:39:23.9 & +00:52:41 & 0 & 4 & 0 & 0.85$\pm$0.06 & II \\
HCG~16 & 3957 & 02:09:31.3 & -10:09:31 & 0 & 2 & 2 & 0.90$\pm$0.05 & I \\
HCG~19 & 4245 & 02:42:45.1 & -12:24:43 & 1 & 1 & 1 & 0.81$\pm$0.10 & II \\
HCG~22 & 2686 & 03:03:31.3 & -15:40:32 & 1 & 2 & 0 & 0.84$\pm$0.05 & II \\
HCG~31 & 4094 & 05:01:38.3 & -04:15:25 & 0 & 2 & 5 & 1.01$\pm$0.07 & I \\
HCG~42 & 3976 & 10:00:21.8 & -19:38:57 & 3 & 1 & 0 & 0.75$\pm$0.06 & III \\
HCG~48 & 3162 & 10:37:45.6 & -27:04:50 & 2 & 2(0) 
\tablenotemark{c} & 0 & 0.71$\pm$0.06 & III \\
HCG~59 & 4058 & 11:48:26.6 & +12:42:40 & 1 & 2 & 1 & 0.81$\pm$0.05 & II \\
HCG~61 & 3907 & 12:12:24.9 & +29:11:21 & 2 & 1 & 0 & 0.89$\pm$0.08 & I \\
HCG~62 & 4122 & 12:53:08.1 & -09:13:27 & 4 & 0 & 0 & 0.70$\pm$0.07 & III \\
HCG~90 & 2644 & 22:02:06.0 & -31:55:48 & 2 & 1 & 1 & 0.74$\pm$0.06 & III \\
\enddata

\tablenotetext{a}{Dynamical masses calculated from velocity
dispersions and median galaxy separations in
\citet{hickson92,ribeiro98,zimer03}.  HI masses are from
\citet{verdes-montenegro01}.}

\tablenotetext{b}{Group type based on $log(M_{{\rm
HI}})/log(M_{{\rm dyn}})/log(M_{{\rm HI}})$ as discussed in \S~\ref{stages}.}  

\tablenotetext{c}{Galaxies HCG~48b and HCG~48c have slightly
discordant velocities and may not be proper group members.}
\end{deluxetable}

We selected a sample of twelve HCGs from the Hickson Compact Group
catalog \citep{hickson92} based on distance ($<4500$~km~s$^{-1}$),
angular extent ($< 8'$), and membership (at least three concordant
members).  The resulting sample is shown in Table~\ref{sample}.  We
include several compact groups in each of the three stages of the
proposed evolutionary sequence \citep{verdes-montenegro01}:
pre-interaction, shocked intergroup medium, and smooth intergroup
medium (see \S~\ref{stages} for explanation of evolutionary states).
Nevertheless, within each of these broad categories, there is a large
variety of group properties.

{\it Spitzer} observations of the twelve HCGs were obtained with the
IRAC \citep{fazio} and MIPS \citep{rieke} cameras as part of a GO-1
program.  The IRAC observations were taken in high dynamic range mode,
with one 12~sec exposure and three 30~sec exposures for each dithered
position.  The total number of exposures taken for each group was
either 24 (HCGs 19, 31, 59, 61, and 62) or 48 (HCGs 2, 7, 16, 22, 42,
48, 90) for each set of cameras (3.6$\mu$m and 5.8$\mu$m together,
then 4.5$\mu$m and 8.0$\mu$m together), depending on the spatial
extent of the group.  After the basic calibrated data (BCD) were
obtained, they were run through both artifact mitigation and stray
light correction routines as part of the {\it Spitzer} MOPEX package
\citep[version 093005,][]{makovoz06}.  The 30~sec exposures were then
combined onto a finer grid scale using the {\it Spitzer} mosaic
software, in order to better sample the point spread functions (PSFs).
The resulting grids have a pixel scale of $0.6\arcsec \times
0.6\arcsec$.  The 12~sec exposures were used only in the case of
saturation in the 30~sec exposures.  MIPS 24$\mu$m images were
obtained with three cycles of 3~sec duration with the number of
exposures ranging from two to four.  The post-BCD observations were
downloaded and mosaiced onto the same grid with a pixel scale of
$2.5\arcsec \times 2.5\arcsec$.  The sensitivity limits vary slightly for each
of the compact groups: for 3.6$\mu$m, 4.5$\mu$m, 5.8$\mu$m, 8.0$\mu$m,
and 24$\mu$m respectively, typical 3-$\sigma$ limits are $\sim
3\mu$Jy, 5$\mu$Jy, 30$\mu$Jy, 40$\mu$Jy, and 250$\mu$Jy and 3-$\sigma$
extended source sensitivity limits are $\sim 0.04$~MJy~ster$^{-1}$,
0.05~MJy~ster$^{-1}$, 0.2~MJy~ster$^{-1}$, 0.2~MJy~ster$^{-1}$, and
0.3~MJy~ster$^{-1}$.

For the purposes of comparing photometry from IRAC and 24$\mu$m
observations, the 2MASS and IRAC images were first convolved to the
24$\mu$m PSF. Photometry was then performed with a specialized IDL
routine developed by A. Reines and R. Indebetouw (see description of
code in Reines et al. in prep) that detects a specified contour level
in a reference image, assigns a background annulus that has the same
shape as the aperture contour (but is expanded out by a specified
factor), and then carries out the photometry using identical apertures
and annuli at each wavelength for a given galaxy.  For the photometry
in this paper, apertures were determined by combining all four IRAC
bands (weighted by $\lambda^{-1}$), and determining a contour level of
1.5-2$\sigma$.  A background annulus was used that had inner and outer
scales 2$\times$ and 2.5$\times$ the scale of the aperture.  Extended
source aperture corrections are necessary due to the significant
contribution from scattered light, particularly in the 5.8$\mu$m and
8.0$\mu$m bands; these were determined following the prescription of
T.~Jarrett using the ``average'' radii of the irregular apertures used
\footnote{Further information can be found on Jarrett's IRAC website:
http://spider.ipac.caltech.edu/staff/jarrett/irac/calibration/ }.
Typical corrections were 7\%, 3\%, 33\%, and 24\% at 3.6$\mu$m,
4.5$\mu$m, 5.8$\mu$m, and 8.0$\mu$m, respectively, which are close to
the values for an ``infinite'' aperture determined by T. Jarrett.
Uncertainties in the flux densities greater than 10~mJy are dominated
by calibration issues and are estimated to be $\sim 10$\%.  Flux
densities of the fainter sources are more strongly subject to
statistical variations that cause the exact contour shape to change,
and we estimate the total uncertainties empirically to be $\sim 20$\%
for flux densities of 5-10~mJy, $\sim 50$\% for flux densities
1-5~mJy, and $\sim 100$\% for flux densities $< 1$~mJy (which are only

\onecolumn

\begin{figure}
\epsscale{1.0}
\plottwo{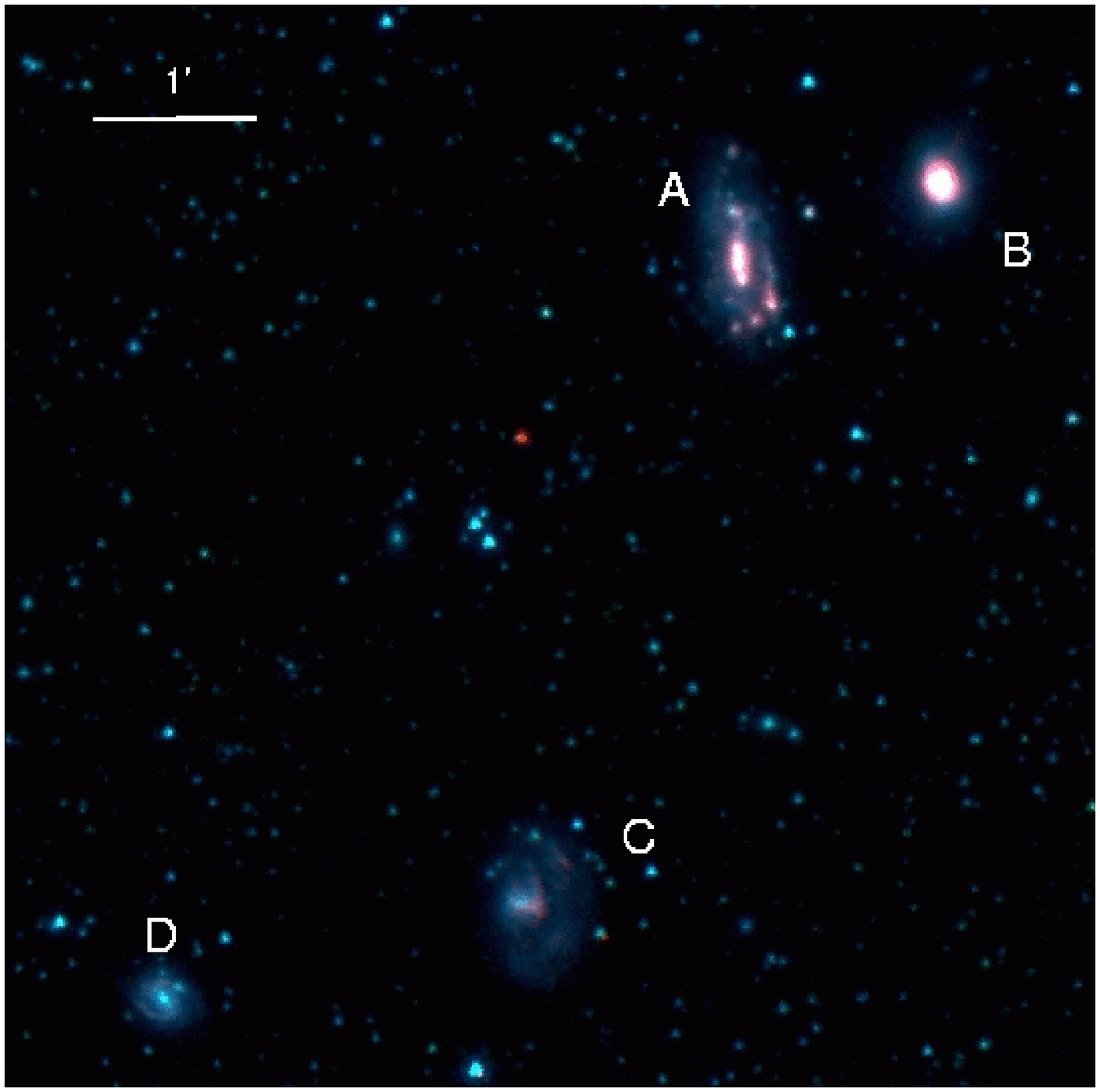}{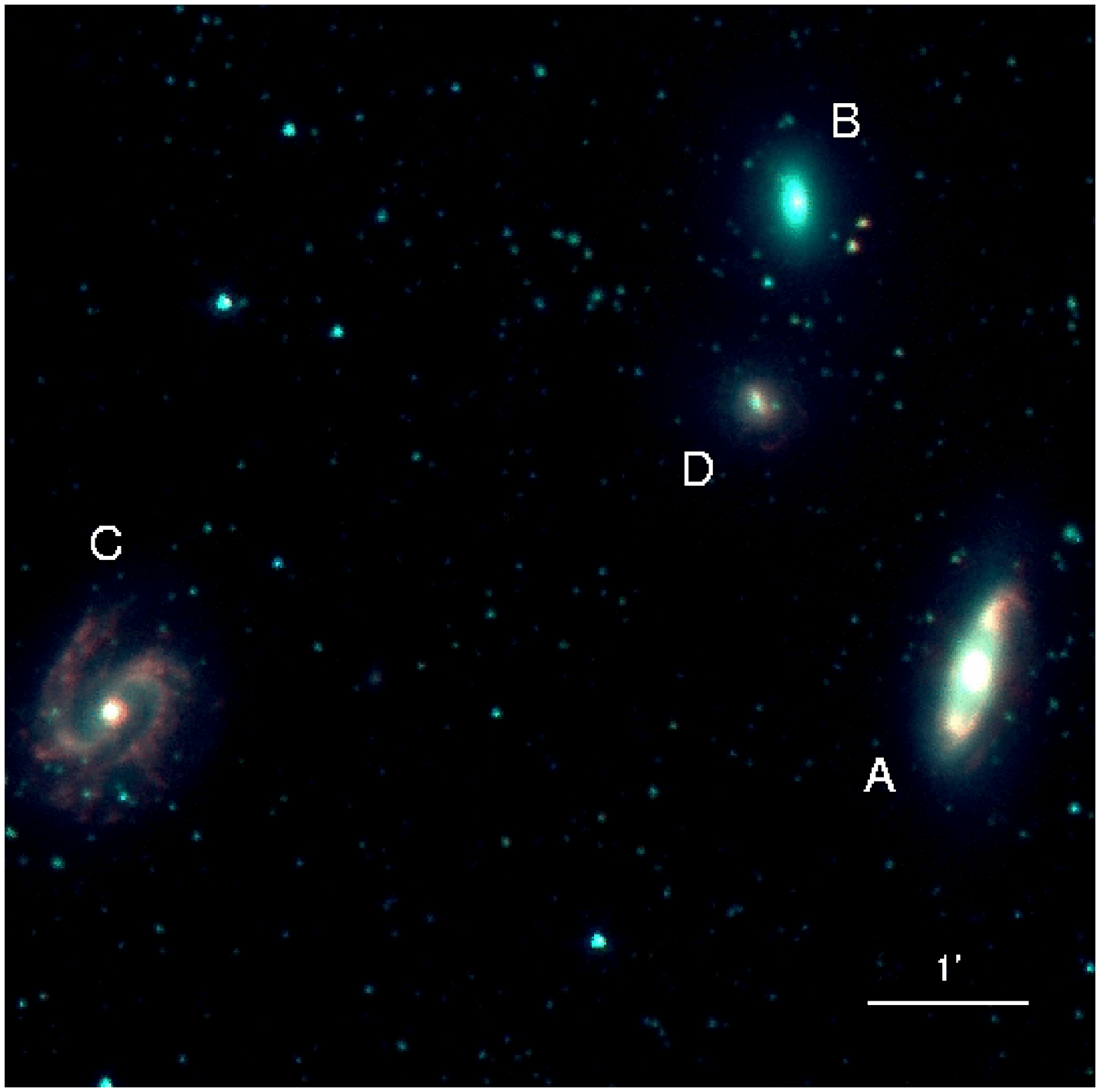}
\caption{Spitzer Space Telescope images of the Hickson Compact Groups
in this Sample.  Blue corresponds to 3.6$\mu$m, green corresponds to
4.5$\mu$m, and red corresponds to 8$\mu$m.  North is up in all
cases. (left) HCG~2. Galaxy D is a background object. (right) HCG~7.
\label{images}}
\end{figure}

\begin{figure}
\epsscale{1.0}
\plottwo{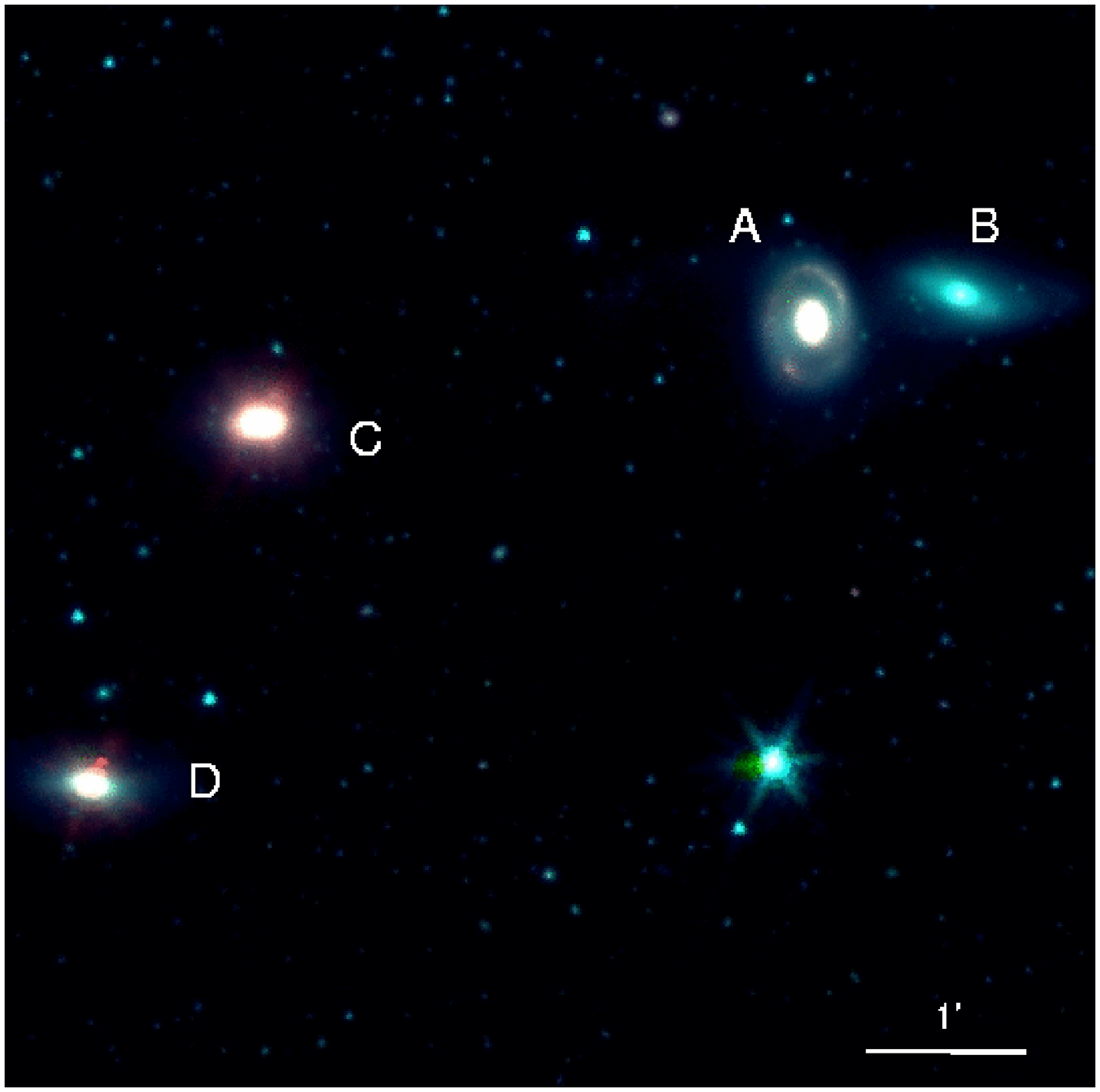}{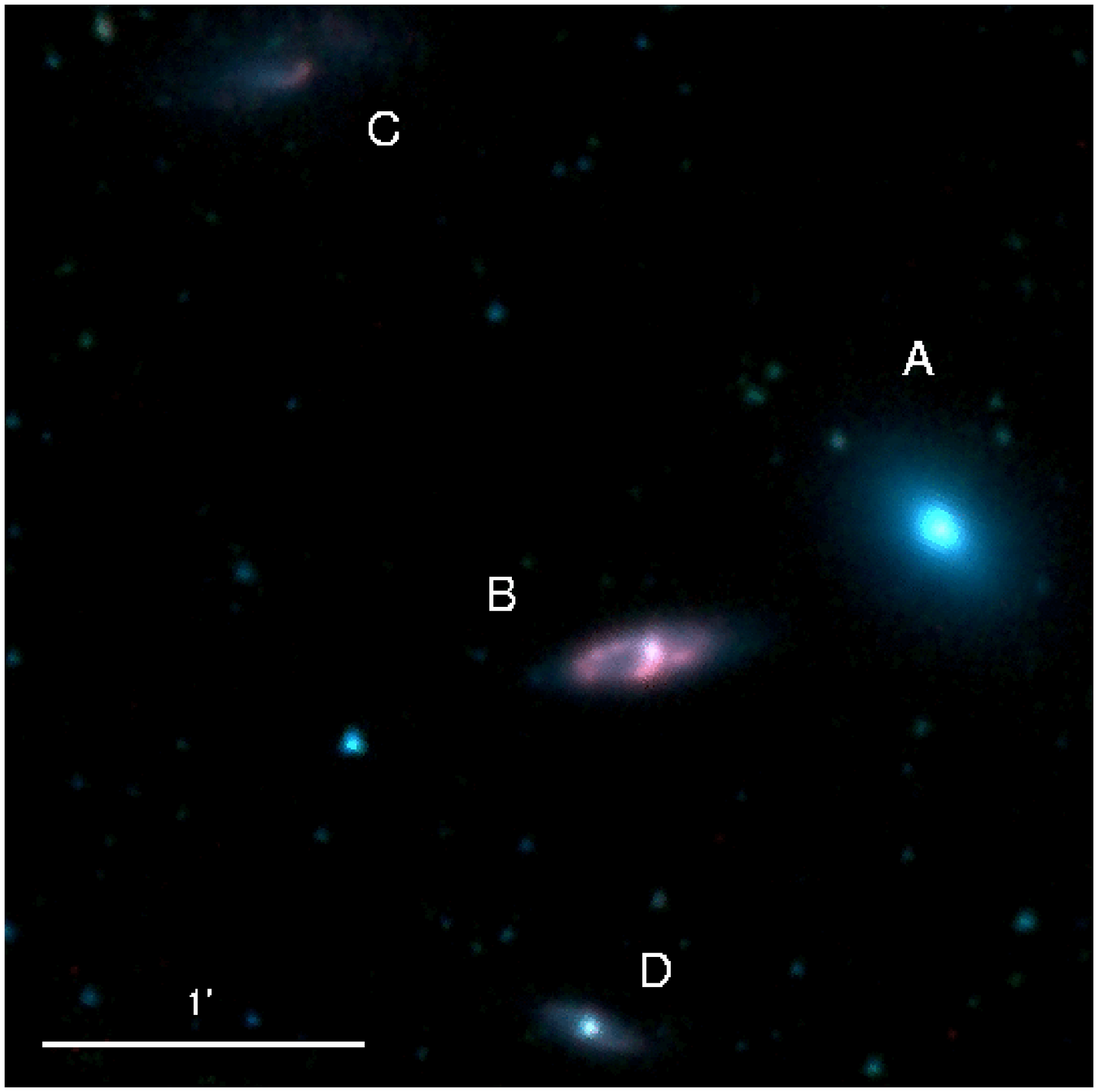}
\caption{Spitzer Space Telescope images with colors as in
Fig.~\ref{images}.  (left) HCG~16. (right) HCG~19.  Galaxy~D is a
background object.}
\end{figure}

\begin{figure}
\epsscale{1.0}
\plottwo{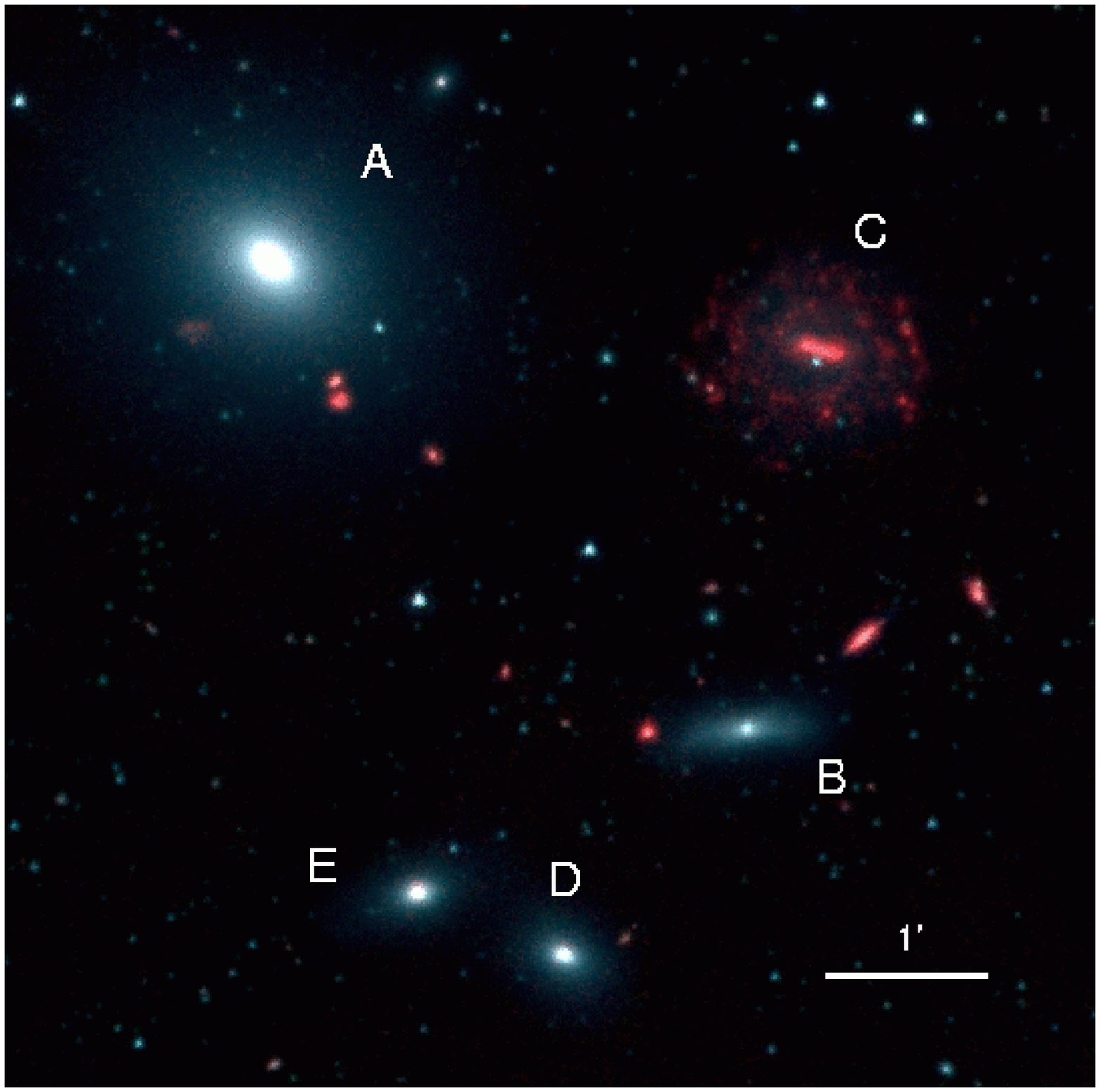}{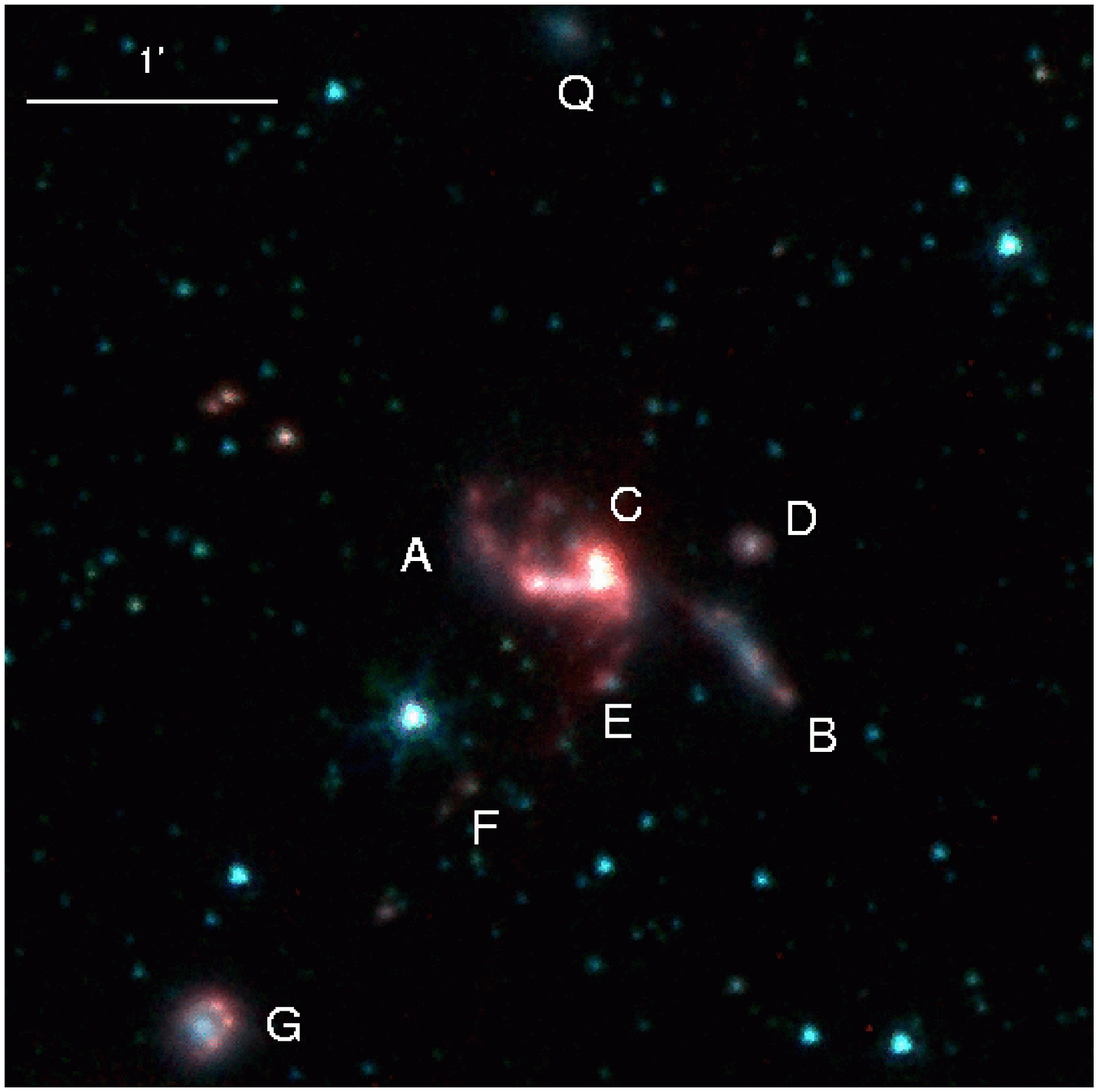}
\caption{Spitzer Space Telescope images with colors as in
Fig.~\ref{images}. (left) HCG~22.  Galaxies~D and E are background
objects. (right) HCG~31.  Galaxy~D is a background object.}
\end{figure}

\begin{figure}
\epsscale{1.0}
\plottwo{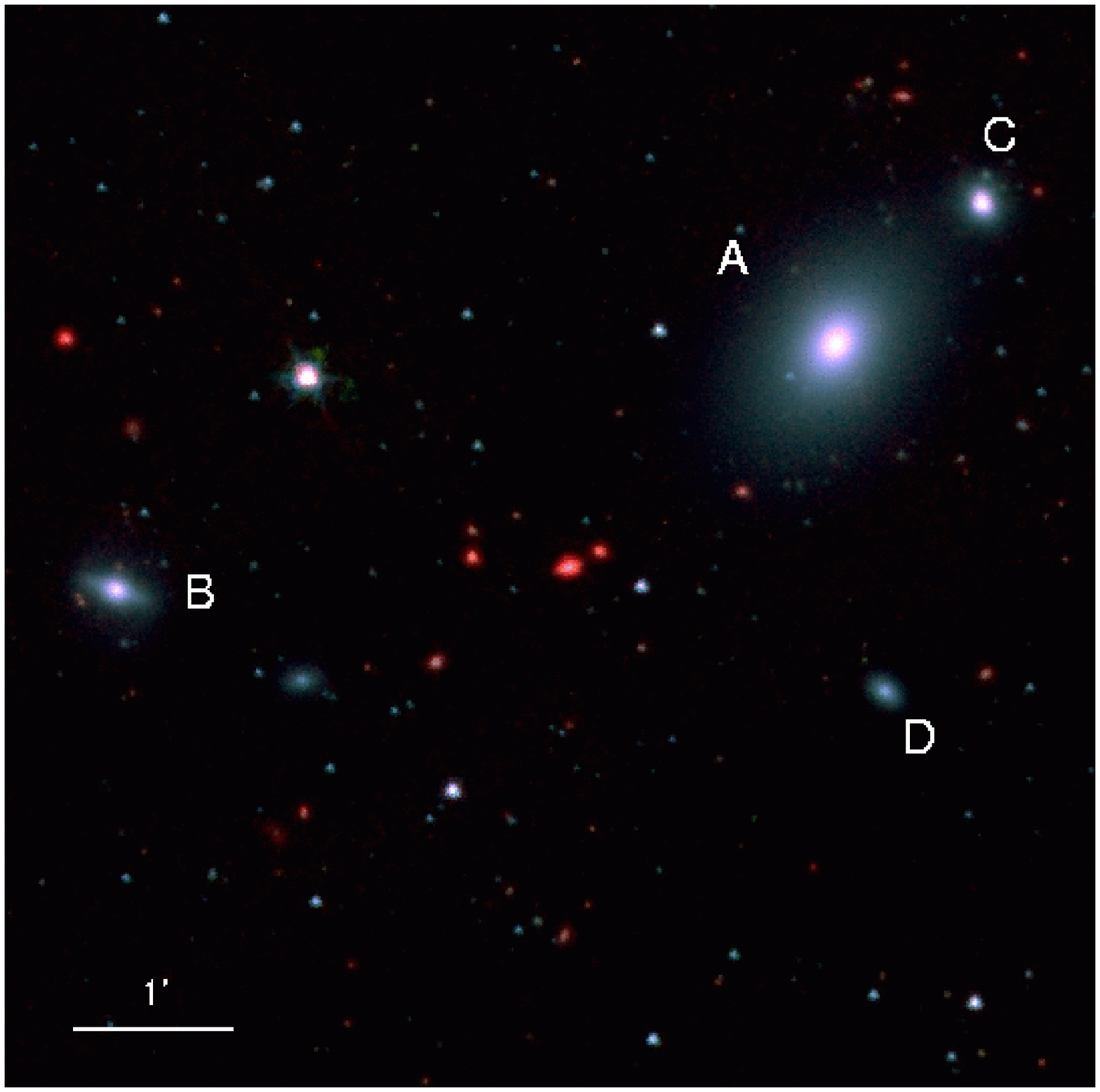}{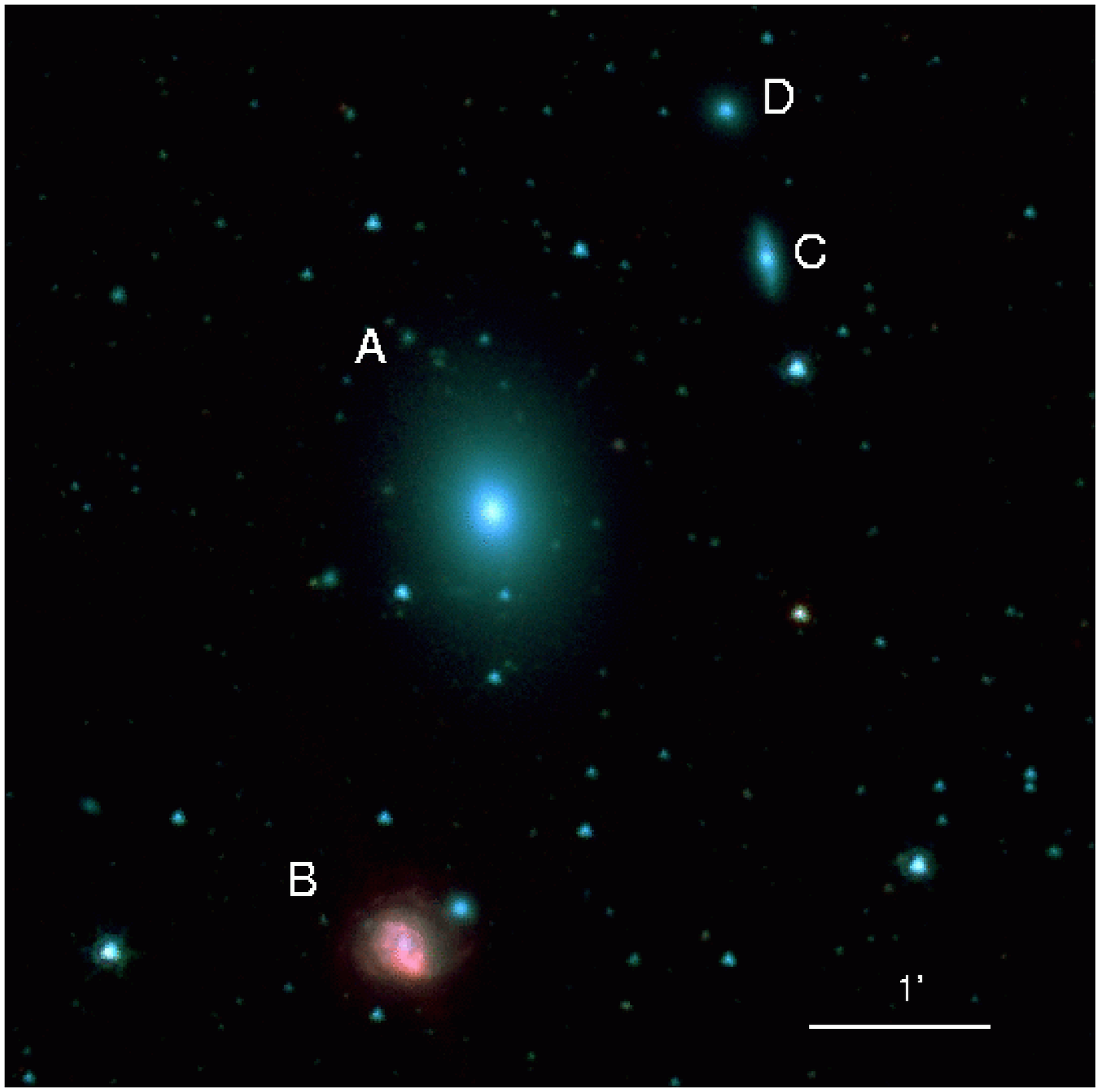}
\caption{Spitzer Space Telescope images with colors as in
Fig.~\ref{images}. (left) HCG~42. (right) HCG~48.  Galaxies~B and D
appear to be in different subgroups of the larger cluster Abell 1060.}
\end{figure}

\begin{figure}
\epsscale{1.0}
\plottwo{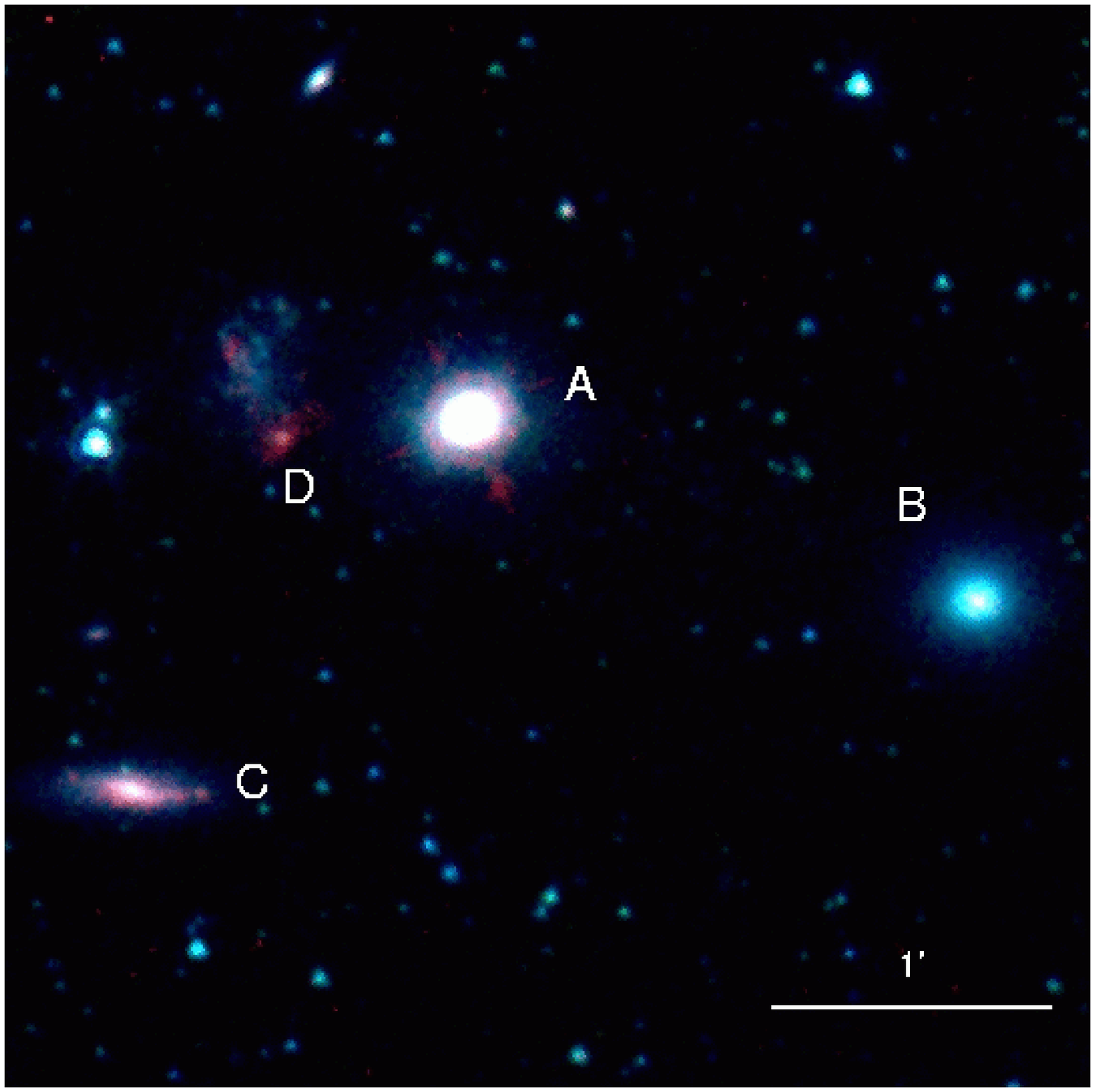}{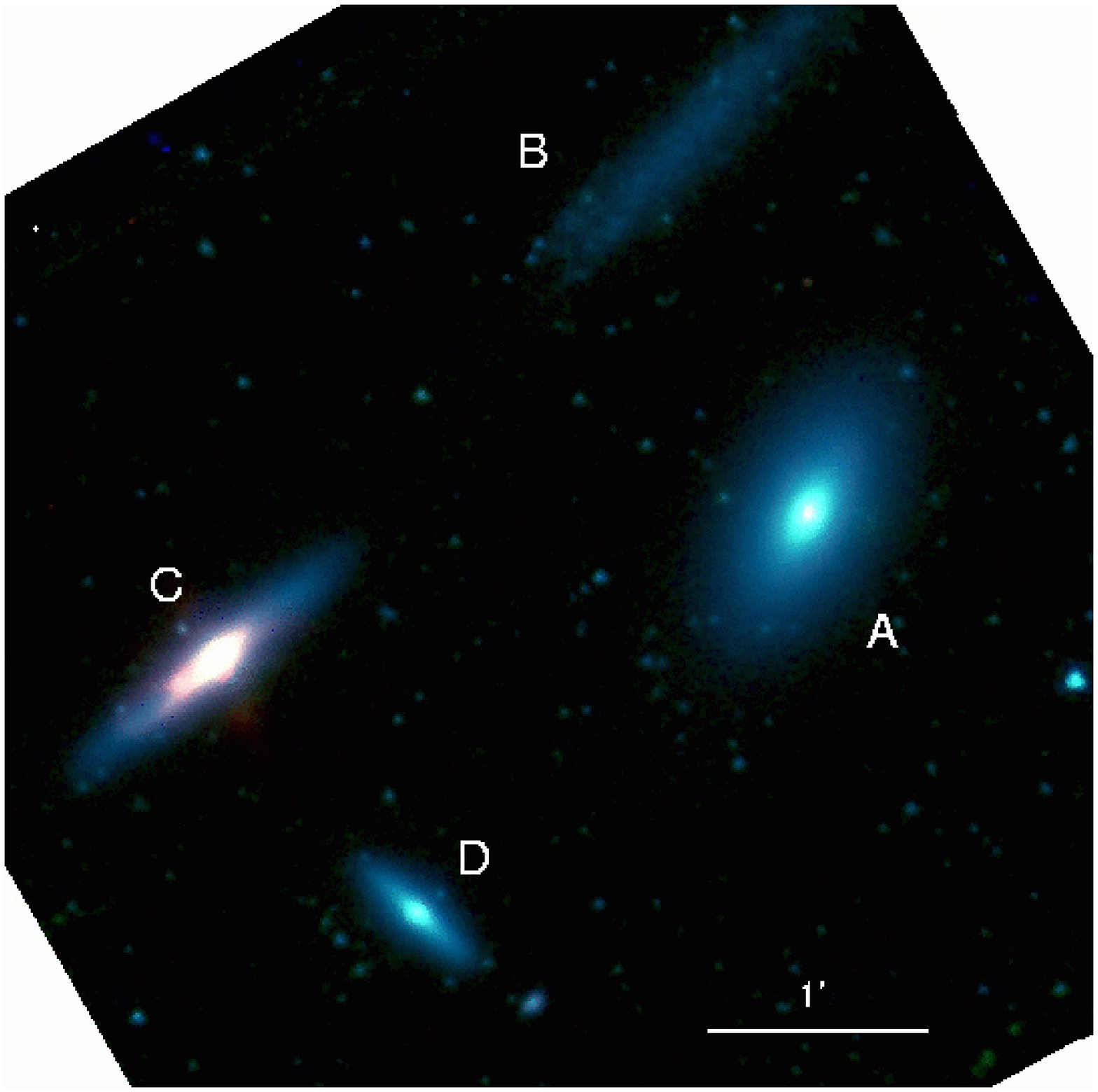}
\caption{Spitzer Space Telescope images with colors as in
Fig.~\ref{images}. (left) HCG~59. (right) HCG~61.
Galaxy~B is a foreground object.}
\end{figure}

\begin{figure}
\epsscale{1.0}
\plottwo{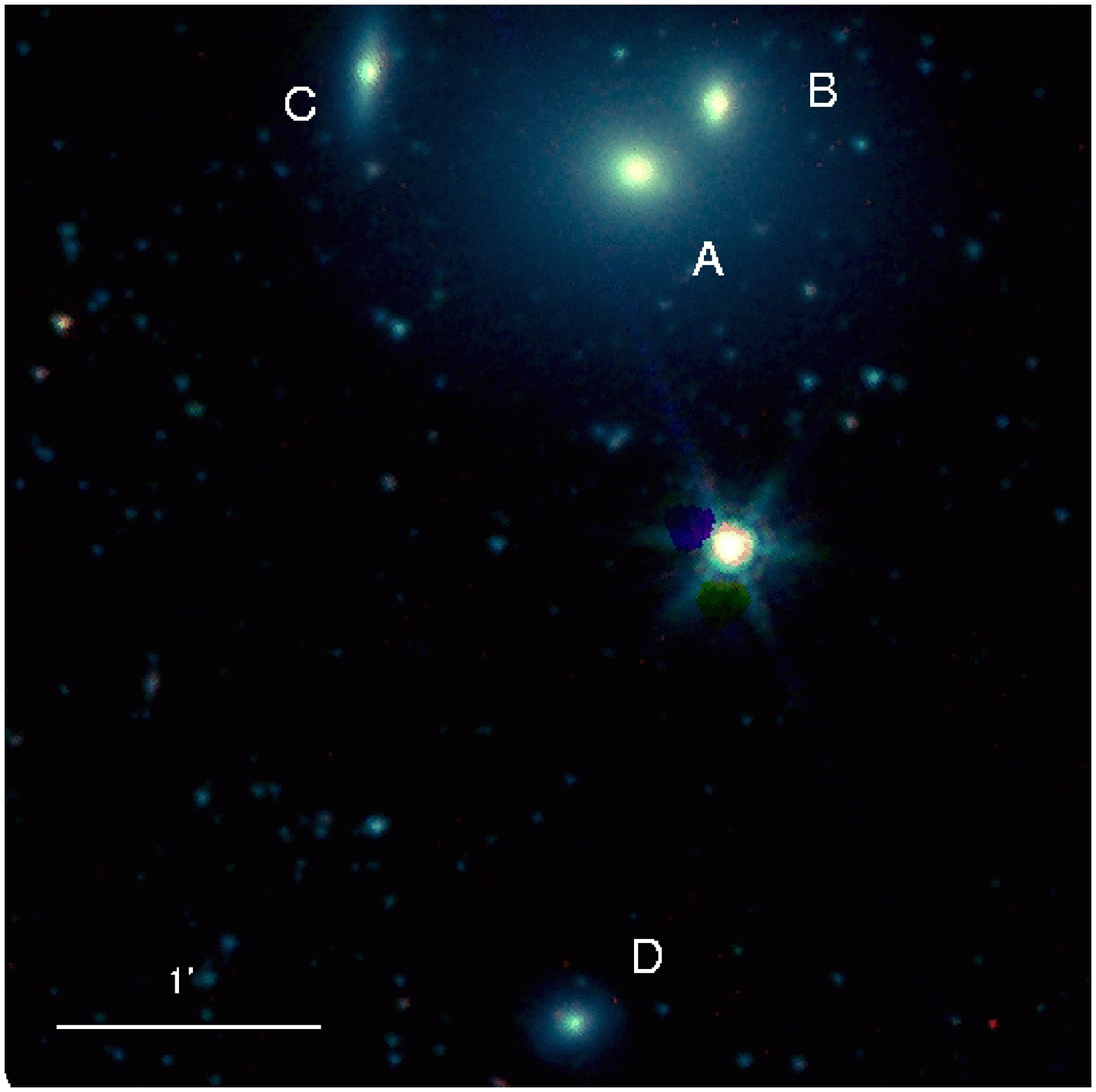}{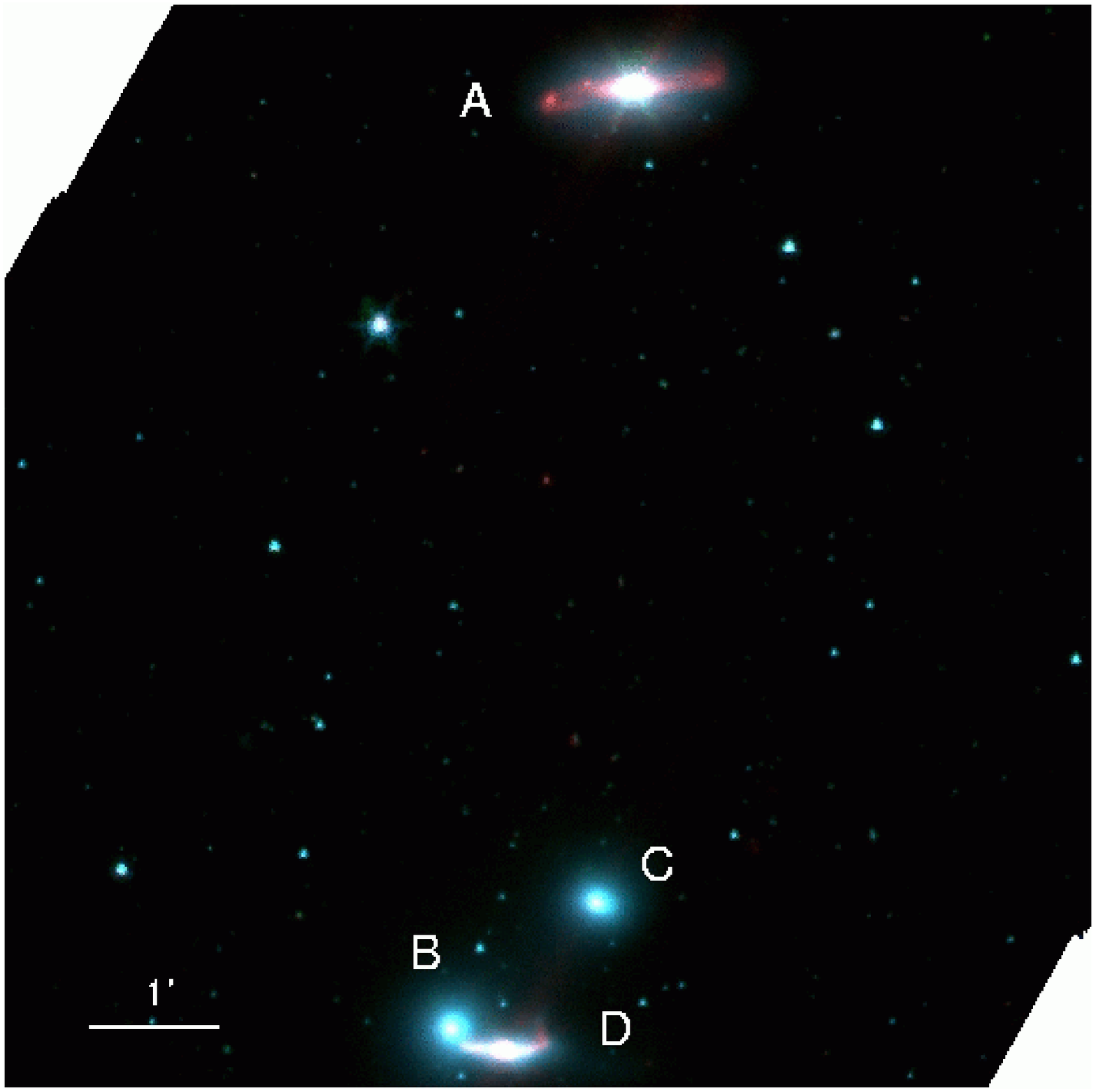}
\caption{Spitzer Space Telescope images with colors as in
Fig.~\ref{images}. (left) HCG~62. (right) HCG~90. \label{images6}}
\end{figure}

\twocolumn

\noindent marginally detected).  The uncertainties for faint objects
are largely driven by surface brightness issues, where small
variations in the local background can dramatically affect the
measured flux densities.

\subsection{Evolutionary Stages of Compact Groups in this Sample \label{stages}}
Compact groups are complex environments with heterogeneous properties
and present a challenge to any kind of morphological or evolutionary
classification; placing compact groups into evolutionary stages is
more complex than for the two-galaxy interactions found in the
analogous Toomre Sequence.  As evidence of the difficulties in
classifying HCGs, Verdes-Montenegro et al. note the lack of one-to-one
correspondence between the presence of tidal features and the atomic
gas content of Hickson Compact Groups.  \citet{ribeiro98} classify
compact groups in three separate families including (1) ``real''
compact groups, (2) core~$+$~halo systems, and (3) parts of loose
groups; this classification only reflects the galaxy populations and
space densities.  \citet{verdes-montenegro01} proposed an evolutionary
sequence based on the content and distribution of HI gas for Hickson
Compact Groups: pre-interaction, shocked intergroup medium, and smooth
intergroup medium.  In this scenario, the evolutionary stage of a
group is postulated to be associated with the HI deficiency of the
group members and the group as a whole.  HI deficiency is defined as
$Def_{{\rm HI}}={\rm log}[M({\rm HI)_{pred}}] - {\rm log}[M({\rm
    HI)_{obs}}]$, where the ``predicted'' HI mass of a galaxy is
determined from the typical HI mass of a field galaxy of similar
Hubble type and luminosity \citep{haynes84}.

We wish to classify Hickson Compact Groups in a quantifiable way while
also mitigating the effects of possible morphological
misclassification of galaxies in determining a group's evolutionary
state.  Here we adopt a variation of the Verdes-Montenegro et
al. method for classification of the Hickson Compact Groups; as a
proxy for evolutionary state, we use ${\rm log}(M_{{\rm HI}})/{\rm
log}(M_{{\rm dyn}})$.  Using dynamical masses calculated from velocity
dispersions and median galaxy separations in \citet{hickson92},
\citet{ribeiro98}, and/or \citet{zimer03} and HI masses from
\citet{verdes-montenegro01} (with the exception of HCG~90, for which
the HI mass was provided by J. Hibbard, private communication), each
group was placed in one of three categories: (I) relatively HI-rich,
${\rm log}(M_{{\rm HI}})/{\rm log}(M_{{\rm dyn}}) \geq 0.9$, (II)
intermediate HI, $0.9 > {\rm log}(M_{{\rm HI}})/{\rm log}(M_{{\rm
dyn}}) \geq 0.8$, and (III) relatively HI poor, ${\rm log}(M_{{\rm
HI}})/{\rm log}(M_{{\rm dyn}}) < 0.8$.  The boundaries of these
categories are somewhat arbitrary and were chosen to divide our sample
of groups roughly in equal thirds.  The resulting classifications are
listed in Table~\ref{sample}.  Only two of the groups in our sample,
HCG~16 and HCG~31, were explicitly classified by Verdes-Montenegro et
al.; both of these groups were identified as ``phase 2'' in that
paper.  In the classification scheme used in this paper HCG~16 and
HCG~31 are both type~I (relatively HI rich).

\begin{figure}[t]
\epsscale{1}
\plotone{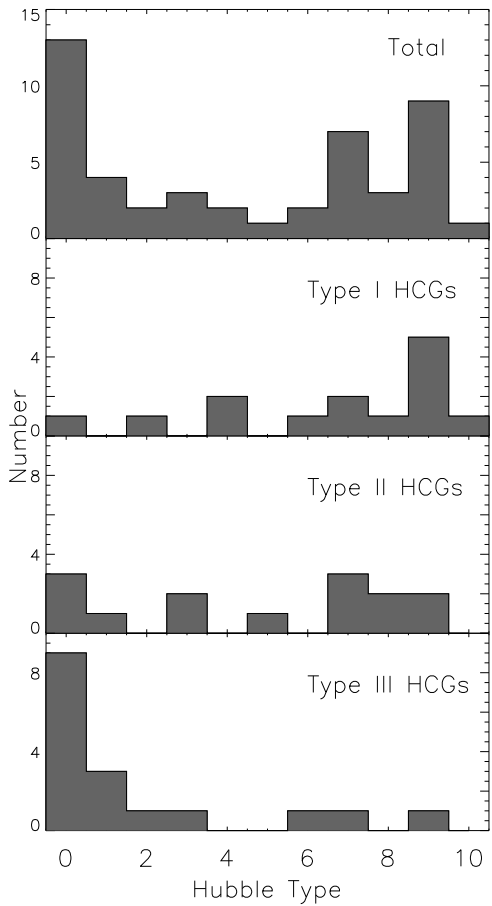}
\caption{Distribution of Hubble type \citep[using the convention
of][]{haynes84} with respect to group type for the galaxies in 
this sample of twelve Hickson Compact Groups.   \label{types}}
\end{figure}

The distributions of the Hubble types \citep[using the convention
of][]{haynes84} of the constituent galaxies with respect to group type
are shown in Figure~\ref{types}.  The histogram resulting from the
total sample of galaxies in the twelve HCGs studied in this program
shows a clear bimodal structure, with peaks in the relative number of
galaxies around Hubble type~0 and Hubble types~6-8.  Given that only
twelve groups are included in this sample, this histogram is
remarkably consistent with the distribution of morphological types
found in all HCGs \citep{hickson88}.  However, it is interesting to 
note that this distribution is distinctly different from general 
samples of galaxies such as the Third Reference Catalog 
\citep[RC3, e.g.][]{vandenbergh02}, which has a uni-modal distribution 
that peaks around type Sbc galaxies.

In Figure~\ref{types} it is
clear that early Hubble types tend to be found in groups of type~III,
while late Hubble types are preferentially found in groups of type~I.
Galaxies in type~II groups are relatively evenly distributed among
Hubble types.  These trends suggest that the degree to which a group
is relatively gas rich or gas poor may simply reflect the Hubble types
of a group's constituent galaxies.  However, the physical scenario may
be more subtle than this, especially given that HCGs as a class appear
to be gas deficient, and the causal relationship between a group's
galaxy content and gas properties is not straight forward.

We note that of the type~I and type~II groups in this sample, only one
of them was detected in the X-ray survey of \citet{ponman96} (HCG~16,
which is known to contain active galaxies), while all four of the
type~III groups were detected (HCG~42, HCG~48, HCG~62, and HCG~90).
The relative contribution to a group's total X-ray flux from the
individual galaxies from the IGM is important to distinguish in this
context.  For example, HCG~42 and HCG~62 have bright emission from the
IGM, while the X-ray emission HCG~90 is dominated by the contribution
from individual galaxies.  HCG~16 also has evidence for a faint IGM in
addition to the contribution from AGN activity \citep{belsole03}.  In
fact, while all of the type~III groups have X-ray detections, none of
the galaxies in type~III groups have strong infrared emission as seen
in their SEDs shown in Figures~\ref{hcg2sed}-\ref{hcg90sed}.  This
result is consistent with what is seen in brightest cluster galaxies
in clusters without strong cooling flows by \citet{egami06}.  Those
clusters frequently have evidence for AGN activity that apparently
injects energy into the intracluster medium and prevents cooling; the
large X-ray bubbles in HCG 62 indicate similar activity
\citep[e.g.][]{morita06}.
In the case of type~III groups in this sample, they could be observed
as ``gas deficient'' in HI simply because their intra-group gas has
largely been heated and ionized.  There have been several interesting
suggestions that a hot cluster ICM plays an important role in the
truncation of star formation activity in clusters (e.g., Poggianti et
al. 2004, Moran et al. 2006).

\section{Results}
\subsection{The Infrared Colors of Galaxies in Hickson Compact Groups}
The infrared colors of galaxies in the Hickson Compact Groups in this
sample span a wide range of parameter space.  Here we discuss trends
and correlations that are apparent in this data set. {\it Spitzer}
IRAC three-color (3.6$\mu$m, 4.5$\mu$m, and 8$\mu$m) images are shown
in Figures~\ref{images}-\ref{images6}. The infrared colors of the
galaxies in this sample are shown in
Figures~\ref{plotIRAC1}-\ref{plotJHK2}.

The near- to mid-IR colors of a source will largely be dependent on
the temperature of the dust, the relative strength of the underlying
stellar blackbody emission, and polycyclic aromatic hydrocarbon (PAH)
emission, which can contribute significantly to three of the four IRAC
bands \citep{draine06}.  The near-IR 2MASS observations at the J, H,
and K-bands will largely be dominated by stellar blackbody emission
with a possible contribution from very hot dust near the sublimation
temperature ($\sim 1600$~K).  Of the four IRAC bands, the only band
that is not subject to a possible strong contribution by PAH emission
is the 4.5$\mu$m band.  The strongest PAH contribution is expected in
the 8.0$\mu$m band.  Emission in the 24$\mu$m band is dominated by
thermal emission from dust.

These three dominant components of emission in the infrared can, to
some extent, be disentangled by examining the spectral energy
distributions of galaxies. A source with emission dominated by hot
dust will result in a K-band ``excess'', and a rising continuum from
3.6$\mu$m to 24$\mu$m.  A source with strong PAH emission will exhibit
a falling spectral energy distribution from 3.6$\mu$m to 4.5$\mu$m and
a rising spectral energy distribution from 4.5$\mu$m to 8.0$\mu$m.  A
source that is dominated by stellar light will have a spectral energy
distribution that falls continuously from the near-IR to 24$\mu$m with
the Rayleigh-Jeans tail of the stellar blackbody emission.

\subsubsection{IRAC colors \label{IRAC_colors}}

\begin{figure}[t]
\epsscale{1}
\plotone{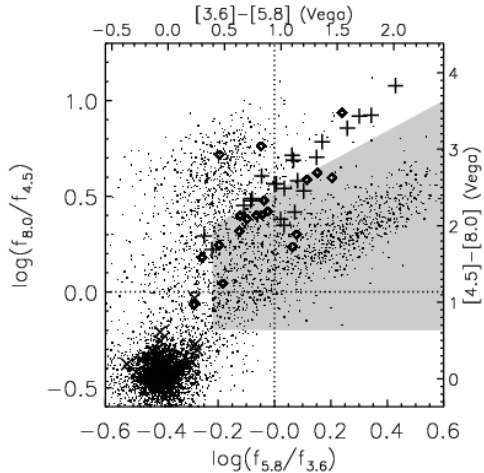}
\caption{IRAC color-color diagram.  The shaded area indicates the
color selection criteria for AGN from \citet{lacy04}.  The ellipse
indicates the region occupied by unreddened main-sequence stars, and
the arrow indicates an $A_V = 10$.  The galaxies in our sample are
indicated as ``$+$'' symbols (if log$[f_{8.0}/f_{4.5}]>0$), and
``$\times$'' symbols (if log$[f_{8.0}/f_{4.5}]<0$); these two symbols
are used again with the same meaning in Figures~\ref{plotMIX1} and
\ref{plotJHK1}.  The KISS sample of star-forming galaxies are shown
for comparison as diamonds \citep{rosenberg06}, and the galaxies from
the FLS are shown as dots \citep{lacy04}.
\label{plotIRAC1}}
\end{figure}

Figure~\ref{plotIRAC1} shows the IRAC colors of the galaxies, which we
have separated into four quadrants that are divided by lines
delineating whether the spectral energy distribution of a source is
rising or falling with increasing wavelength for a given flux density
ratio.  Galaxies with emission dominated by stellar light with little
dust or PAH emission are expected to be located in the lower left
quadrant.  For reference, we show an ellipse that indicates the region
in this color space that would be occupied by unreddened main-sequence
and giant stars along with a reddening vector that corresponds to
$A_V=10$.  

Galaxies with active star formation and/or AGN activity are largely
expected to be located on the top portion of this plot
\citep[e.g.][]{lacy04}; contributions to their SEDs from PAH emission
or hot dust will result in stronger emission at 8$\mu$m than at
4.5$\mu$m.  In order to differentiate the galaxies that fall in the
top or bottom of this plot (reflecting the ratio of flux densities in
the 8.0$\mu$m and 4.5$\mu$m bands) in Figures~\ref{plotMIX1} and
\ref{plotJHK1}, we consistently use $\times$ and $+$ symbols,
respectively, to represent them.  For comparison, we plot the galaxies
from the KISS survey of star-forming galaxies as diamonds
\citep{rosenberg06}, and the {\it Spitzer} First Look Survey (FLS) as
dots \citep{lacy04}.  The color selection criteria for AGN defined by
\citet{lacy04} are indicated by a shaded area; while a significant
fraction of the galaxies in the KISS sample are in this region of
color space, only a handful of the galaxies in this sample fall in
that area, including HCG~7A, HCG~7D, HCG~22C, HCG~31G, HCG~59A, and
HCG~90A.  However, the integrated colors of galaxies can obscure
low-level AGN activity, and several of the galaxies in this sample are
consistent with hosting AGN based on their nuclear colors (Gallagher
et al., in prep).

The galaxies that fall in the lower left quadrant (e.g. dominated by
stellar emission) are tightly clustered near the stellar ellipse.  By
contrast, the galaxies that fall in the upper right quadrant exhibit a
significant spread in their colors, which is consistent with variable
contributions from hot dust and PAH emission, plausibly reflecting
different levels of star formation and/or AGN activity in the
galaxies.  As expected, the KISS sample is almost exclusively located
in the upper portion of this plot.

A major difference between the IRAC colors of galaxies in this sample
in and the FLS sample of \citet{lacy04} is the lack of a vertical
plume in our data (as seen in Fig.~\ref{plotIRAC1}).  Lacy et
al. attribute the galaxies in this plume to occupy this area of color
space because they are at redshifts where the 6.2$\mu$m PAH feature
has shifted out of the 5.8$\mu$m band and into the 8.0$\mu$m band.
The galaxies in the sample of HCGs presented here do not have
redshifts sufficient to cause this effect (by definition, the sample
was selected to have velocities $<4500$~km~s$^{-1}$).

There are a few curious features in the HCG galaxy distribution shown
in Fig.~\ref{plotIRAC1}, which are anomalous when compared to the FLS
galaxy sample of \citet{lacy04}.  First, the HCG galaxies follow a
relatively narrow finger along the upper edge of the AGN-starburst
range of FLS galaxies.  The narrowness of this trend suggests a
reasonably tight relationship between PAH emission and dust emission
in the HCG sample.  Our comparison between the HCG galaxies and the
FLS galaxies indicates an interesting $\sim 0.3$~dex {\it offset}
between this finger and the one in the FLS plot.  The vertical offset
suggests that the HCG galaxies have stronger PAH emission in the
8.0$\mu$m band than typical galaxies in the FLS.  This could be due to
generally elevated star formation and/or dust content.  However,
another very plausible explanation is simply that the bulk of the
8$\mu$m PAH feature has red-shifted out of the 8$\mu$m band for a
significant fraction of the galaxies in the FLS, which would only
require a moderate redshift of $z\sim 0.2$.

We also note that the gap that we see in our colors in
Fig.~\ref{plotIRAC1} is {\it not} present in the FLS sample
\citep[][Figure~1]{lacy04}.  This sample of HCGs has a relative lack
of galaxies that fall in the color space between systems dominated by
stellar light and systems with significant hot dust and/or PAH
contributions.  This ``gap'' is not apparent in the FLS sample, which
could be masked by the shifting of the 6.2$\mu$m PAH feature between
bands, as discussed above.  However, this gap is also not seen in the
galaxies of the SINGS sample, which are not at significant redshifts.
The nature of this gap will be discussed further in the context of
evolutionary stages in \S~\ref{stages_trends}.

\subsubsection{Pan-Infrared colors}

\begin{figure}[t]
\epsscale{1}
\plotone{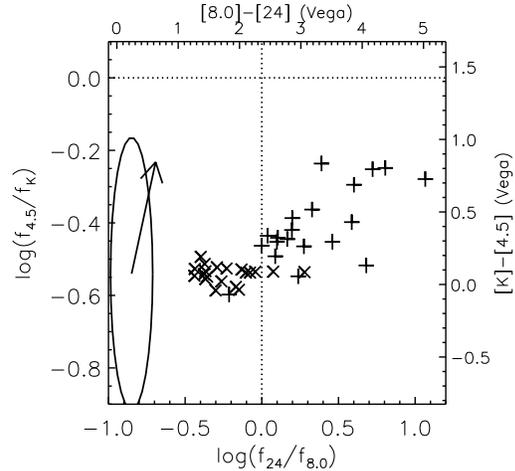}
\caption{Pan-IR color-color diagram.  Galaxies to the right of
the vertical line are likely to have hot dust, and galaxies above the
horizontal line would have an extremely red near-IR color, possibly
indicating AGN activity.  The ellipse indicates the region occupied by
unreddened main-sequence stars, and the arrow indicates an $A_V = 10$.
Galaxies indicated with ``$+$'' symbols are those that have a
rising spectrum from 4.5$\mu$m to 8.0$\mu$m as shown in
Figure~\ref{plotIRAC1}.  
\label{plotMIX1}}
\end{figure}

Figure~\ref{plotMIX1} shows a color combination spanning the K-band to
24$\mu$m.  As in Figure~\ref{plotIRAC1}, we have separated this color
space into four quadrants with lines delineating whether the spectral
energy distribution of a source is rising or falling with increasing
wavelength for a given flux density ratio.  All of the galaxies fall
in the lower two quadrants.  There is a clear trend in this color
space for the sources identified in Figure~\ref{plotIRAC1} as having
significant PAH emission and/or hot dust at 8$\mu$m (indicated with
the $+$ symbols) to also have relatively strong 24$\mu$m emission in
this plot.  There is also a slight trend in the data for galaxies that
have relatively strong thermal dust emission, as manifest in a higher
log$(f_{24}/f_8)$ ratio, to also have relatively high ratios of
log$(f_{4.5}/f_K)$, which provides further evidence for a hot dust
contribution.

\subsubsection{2MASS colors \label{2MASS_colors}}

\begin{figure}[t]
\epsscale{1}
\plotone{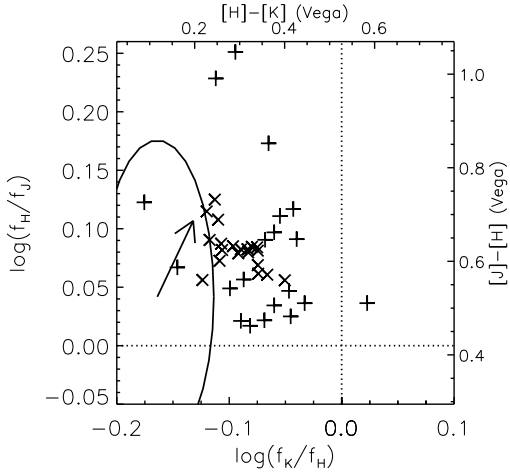}
\caption{2MASS J, H, K color-color diagram.  The ellipse
indicates the region occupied by unreddened main-sequence stars, and
the arrow indicates an $A_V = 1$ (note that Figures~\ref{plotIRAC1} and
\ref{plotMIX1} show an $A_V=10$).  Most of the galaxies are consistent
with having a slightly reddened normal main-sequence population.
Galaxies indicated with ``$+$'' symbols are those that have a
rising spectrum from 4.5$\mu$m to 8.0$\mu$m as shown in
Figure~\ref{plotIRAC1}. 
\label{plotJHK1}}
\end{figure}

Figure~\ref{plotJHK1} shows the near-IR colors of the galaxies in this
sample of HCGs.  As in Figures~\ref{plotIRAC1} and \ref{plotMIX1}, the
ellipse indicates the region of color space occupied by unreddened
main-sequence and giant stars, and a reddening vector for $A_V=1$ is
shown.  The J-H and H-K colors of nearly all of the galaxies in the
sample are consistent with being dominated by normal stellar
populations that are either slightly reddened or alternatively
affected by RGB and AGB star contributions to the integrated near-IR
colors.  The marginally outlying points may reflect an infrared excess
due to a hot dust contribution.  There is a trend in the near-IR
colors for the sources identified in Figure~\ref{plotIRAC1} as having
significant PAH emission at 8$\mu$m (indicated with the $+$ symbols)
to also have larger amounts of reddening and/or hot dust
contributions.  In addition these sources exhibit a much larger
scatter in their near-IR colors than the sources identified in
Figure~\ref{plotIRAC1} as having relatively weak PAH and/or hot dust
emission (indicated with the $\times$ symbols).

\subsection{The Relationship Between Infrared Properties and Evolutionary State \label{stages_trends}}

\begin{figure}[t]
\epsscale{1}
\plotone{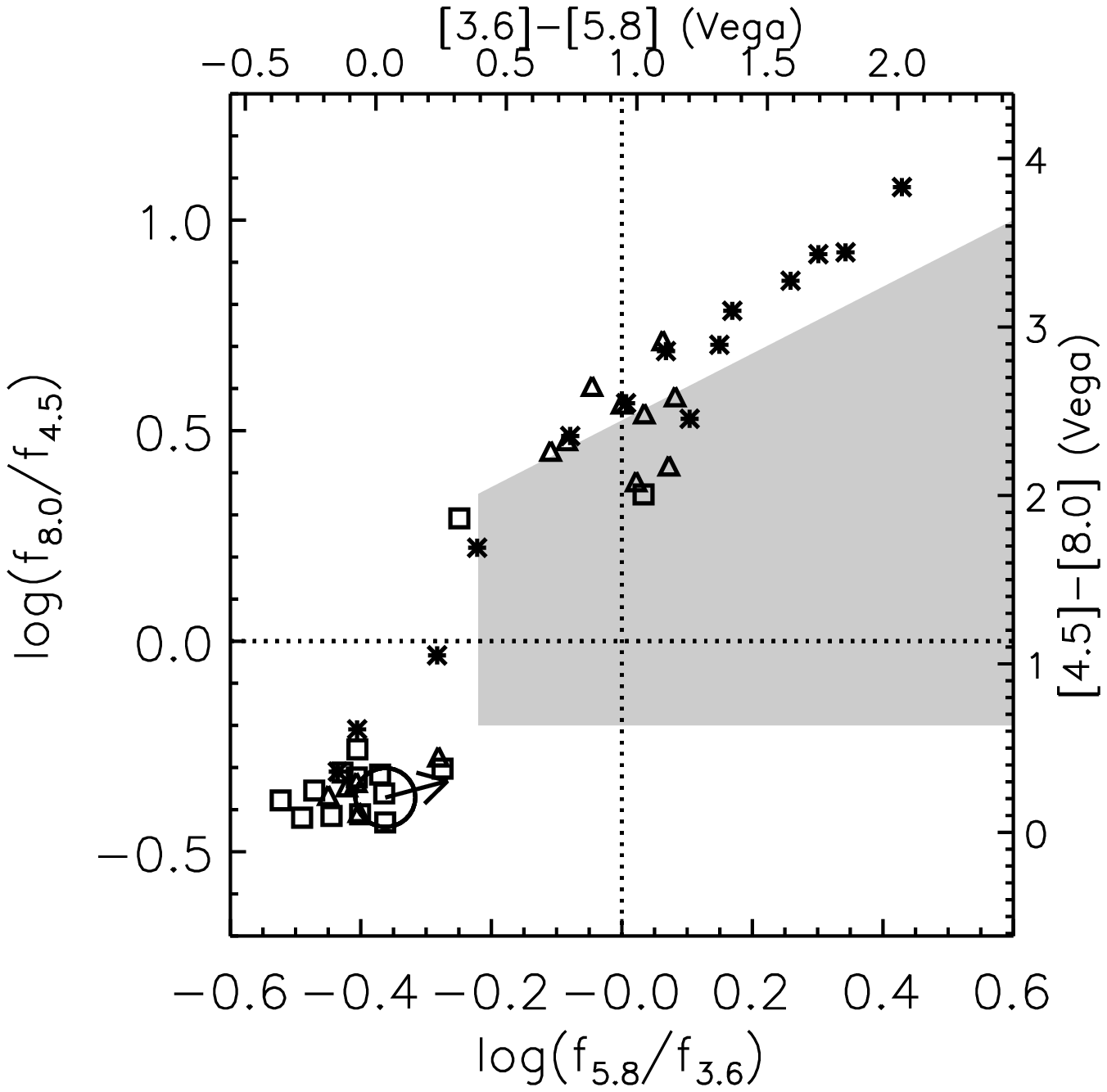}\\
\epsscale{0.9}
\plottwo{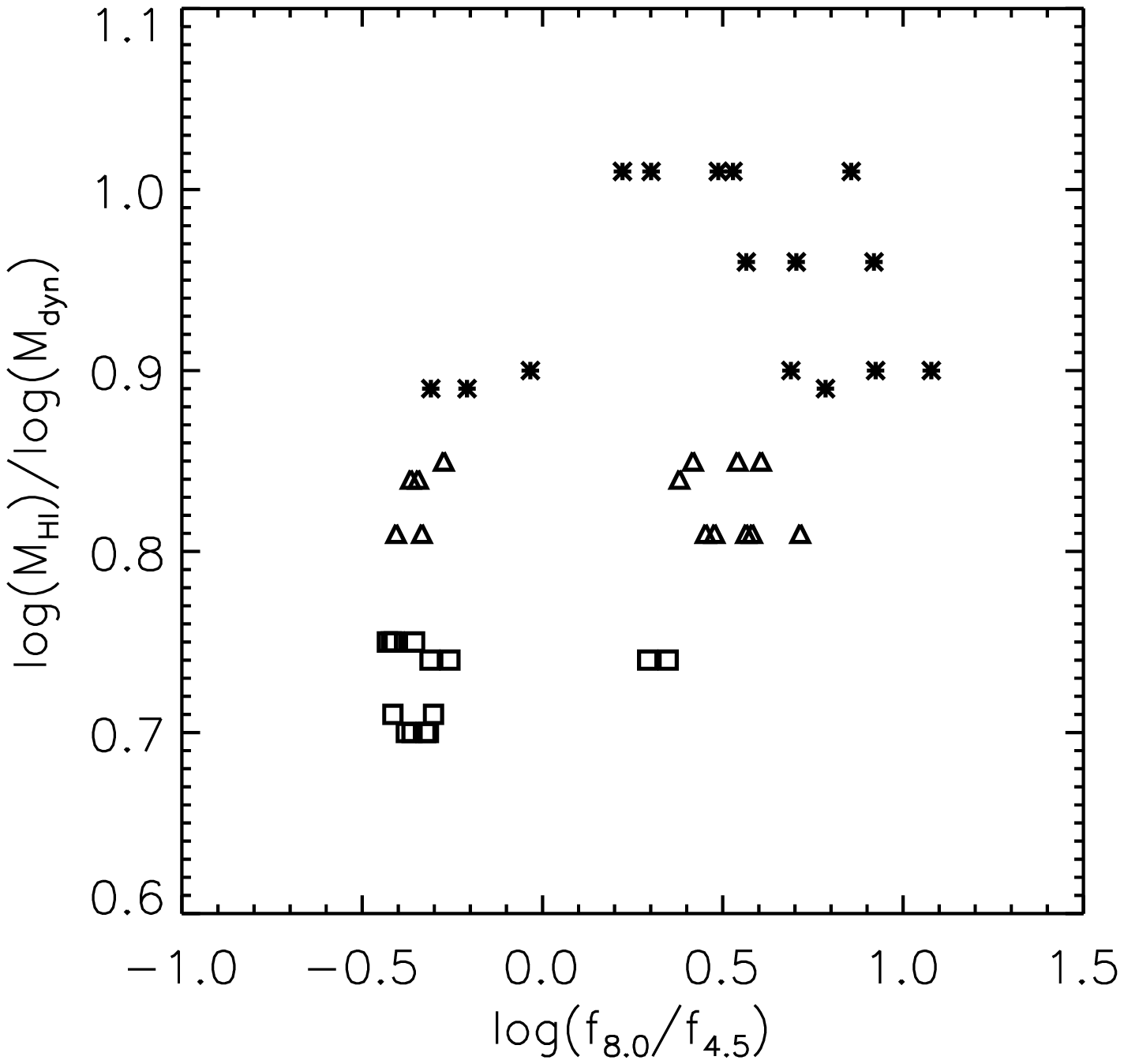}{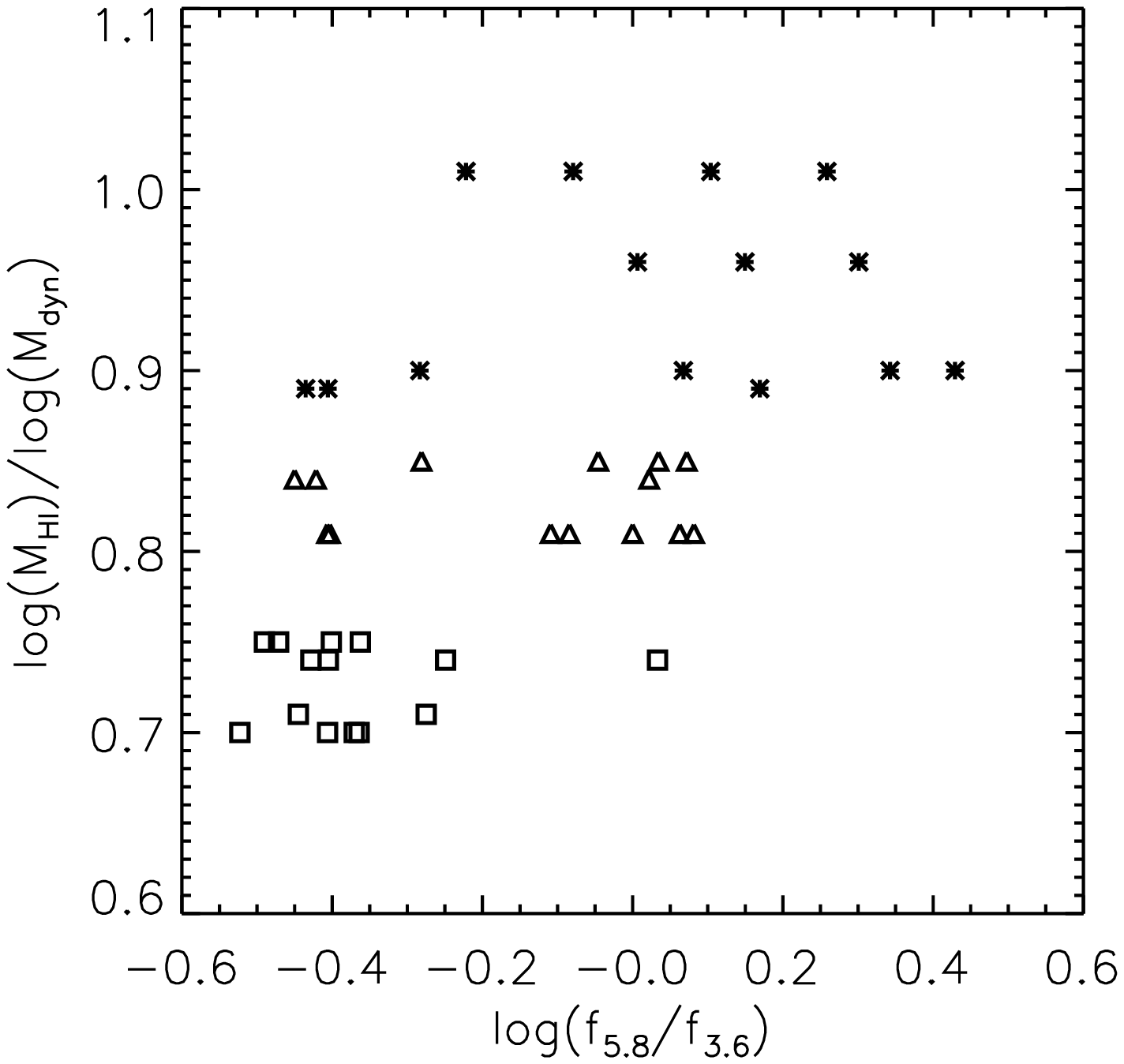}
\caption{(top) IRAC color-color diagram with the same features as
Fig.~\ref{plotIRAC1}, but with the galaxies separated into three
categories: ${\rm log(M_{HI})/log(M_{dyn}})< 0.8$ as squares,
$0.8 \leq {\rm log(M_{HI})/log(M_{dyn}})<0.9$ as triangles, and ${\rm
log(M_{HI})/log(M_{dyn}})\geq 0.9$ as stars.  (bottom) The relationship
between the colors (used for the x-axis and y-axis above) of the
galaxies in this sample and the ${\rm log(M_{HI})/log(M_{dyn}})$ ratio
for the parent groups using the same symbols as above. See discussion
in section~\ref{stages_trends}.
\label{plotIRAC2}}
\end{figure}

\begin{figure}[t]
\epsscale{1}
\plotone{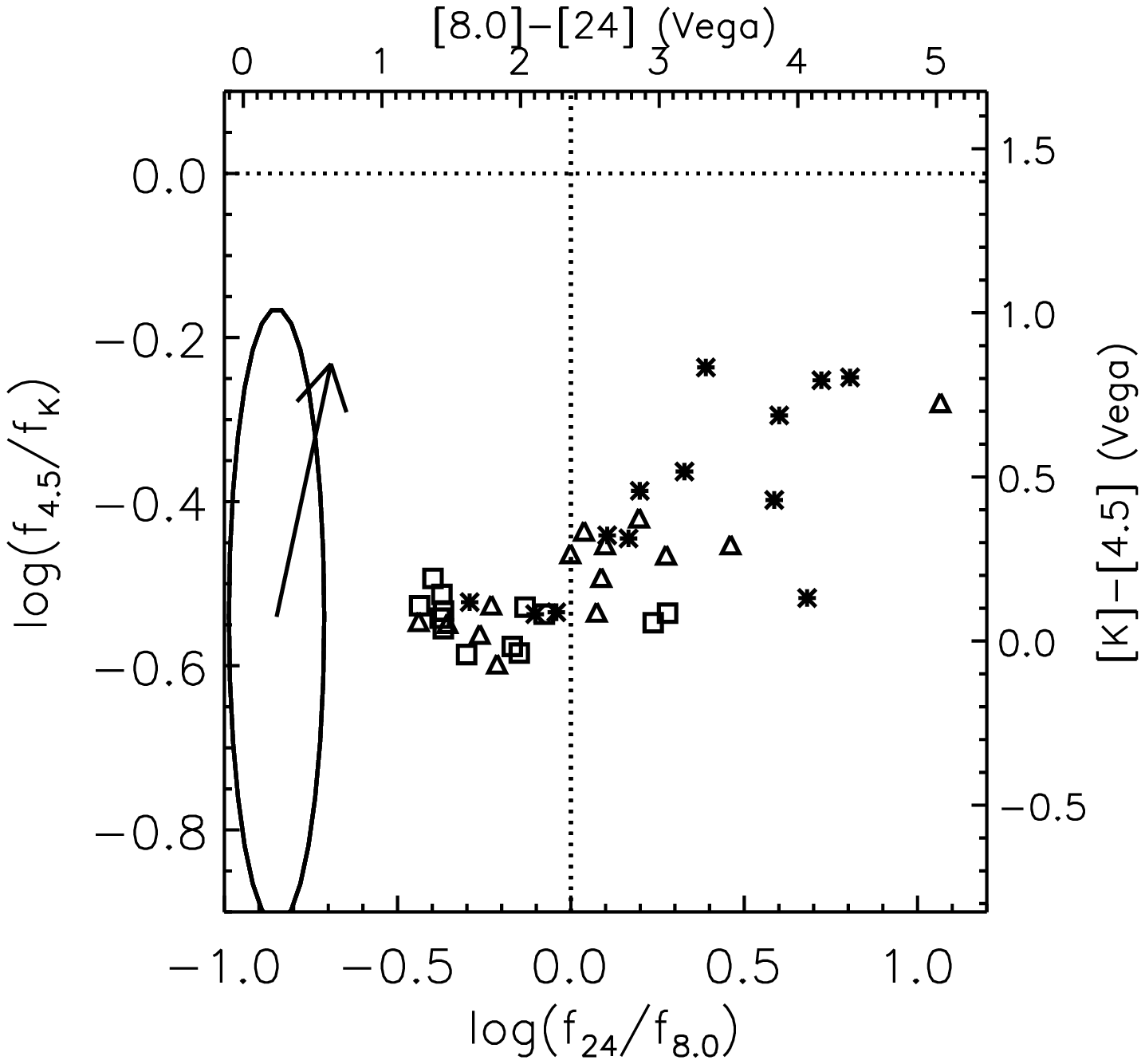}\\
\plottwo{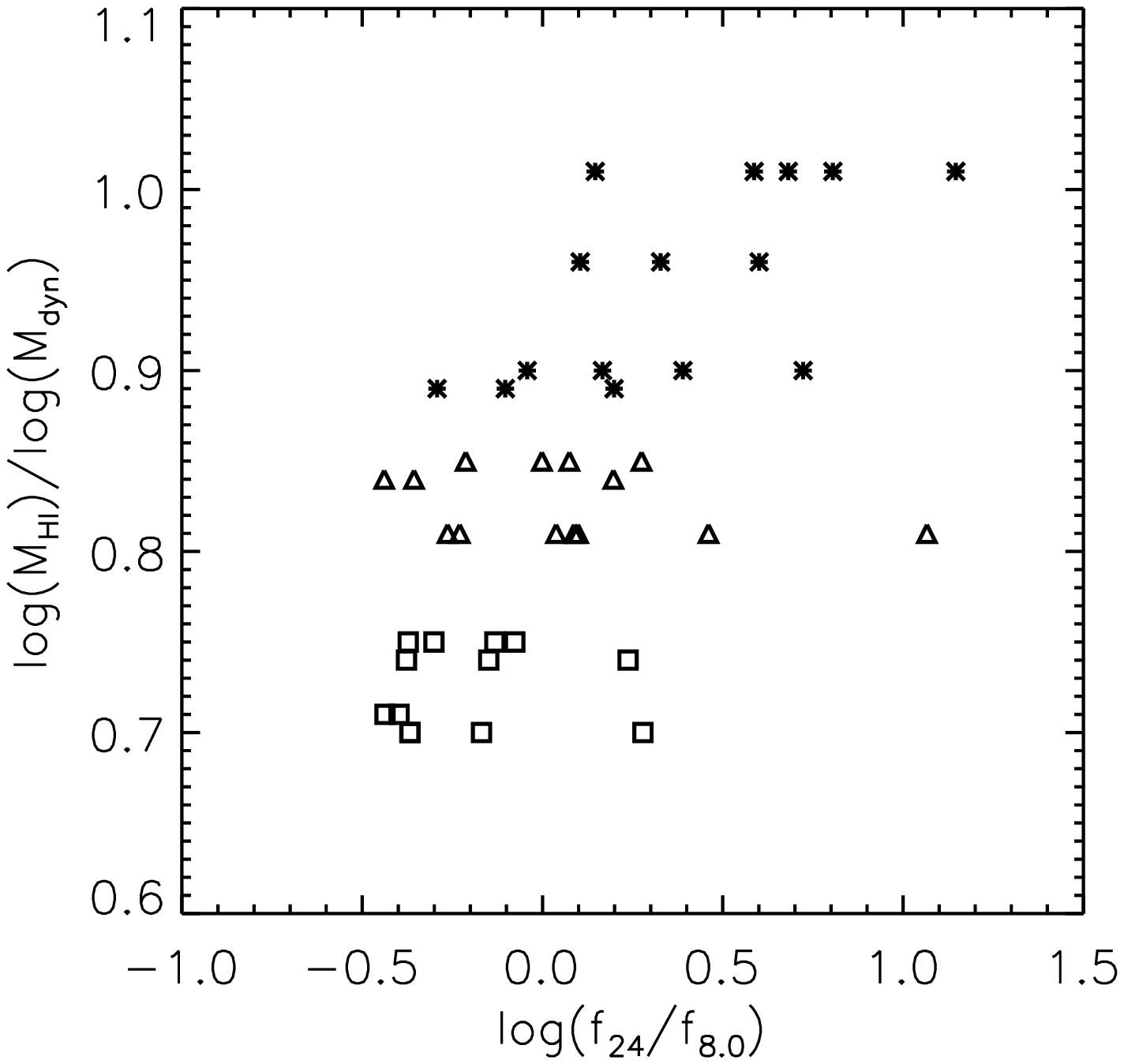}{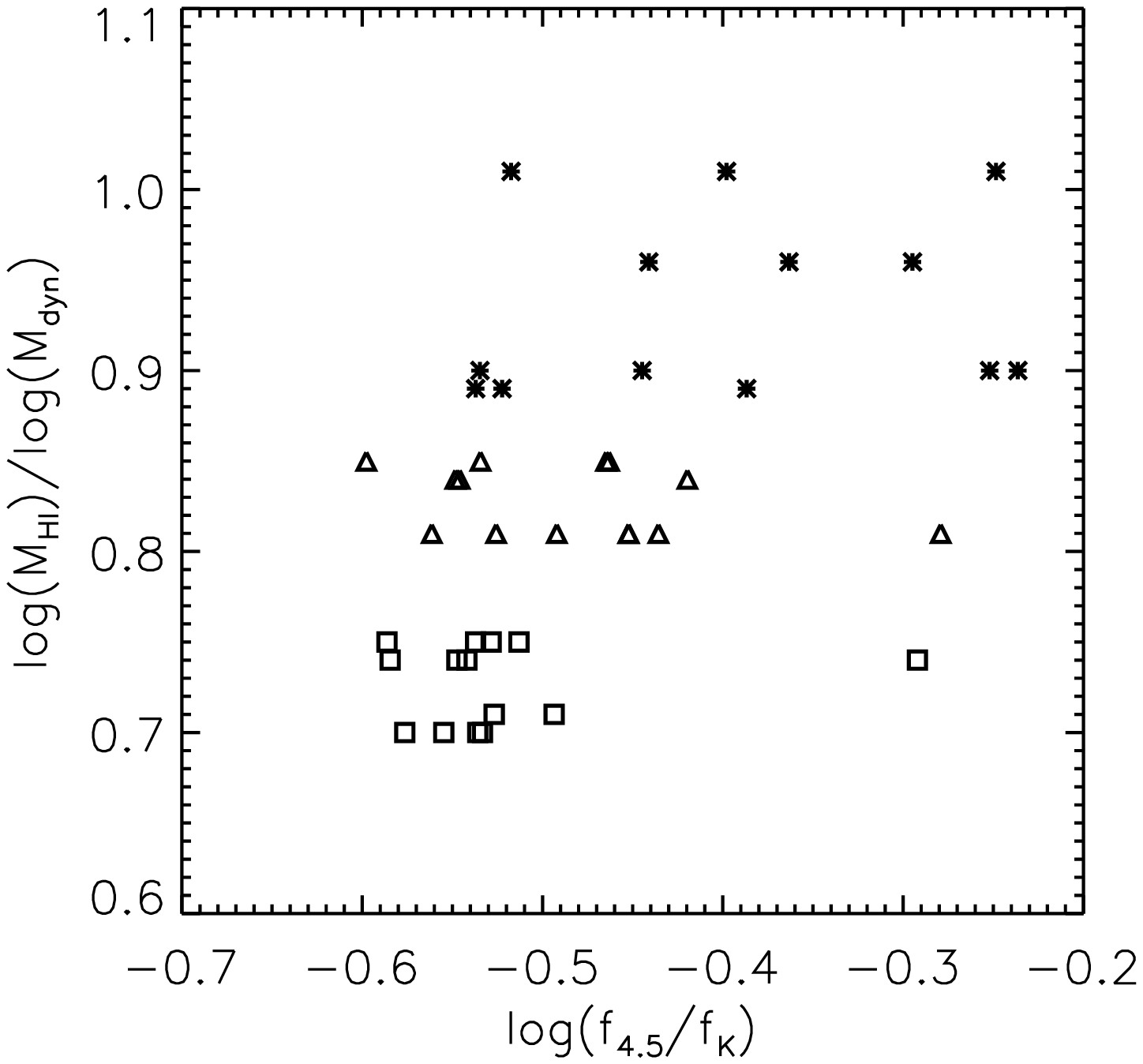}
\caption{(top) Pan-IR color-color diagram with the same features as 
Fig.~\ref{plotMIX1}, but with the galaxies separated
by ${\rm log(M_{HI})/log(M_{dyn}})$ as in Fig.~\ref{plotIRAC2}.
(bottom) The relationship between the colors (used for the x-axis and
y-axis above) of the galaxies in this sample and the ${\rm
log(M_{HI})/log(M_{dyn}})$ ratio for the parent groups using the same
symbols as above. See discussion in section~\ref{stages_trends}.
\label{plotMIX2}}
\end{figure}

\begin{figure}[t]
\epsscale{1}
\plotone{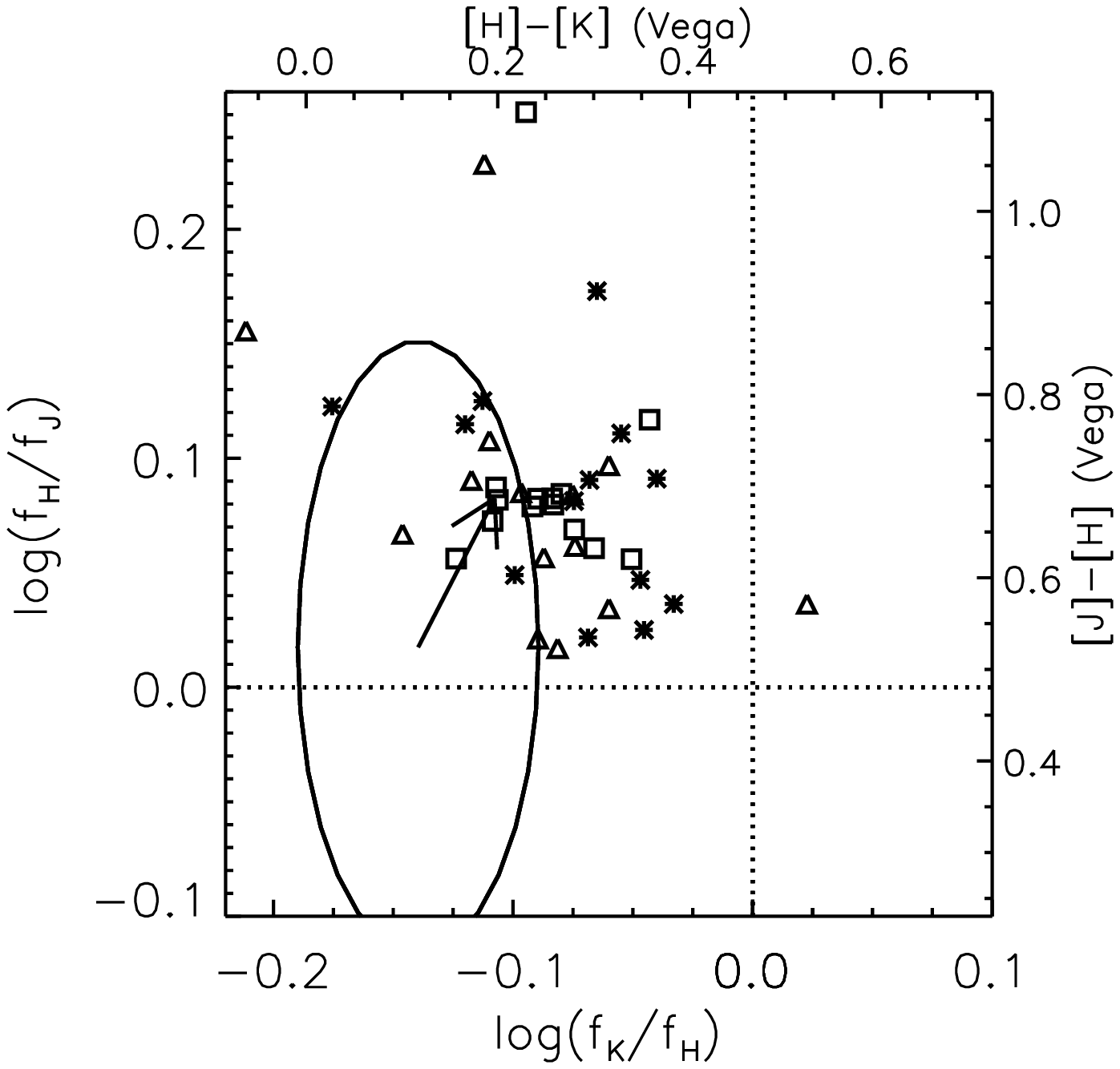}\\
\plottwo{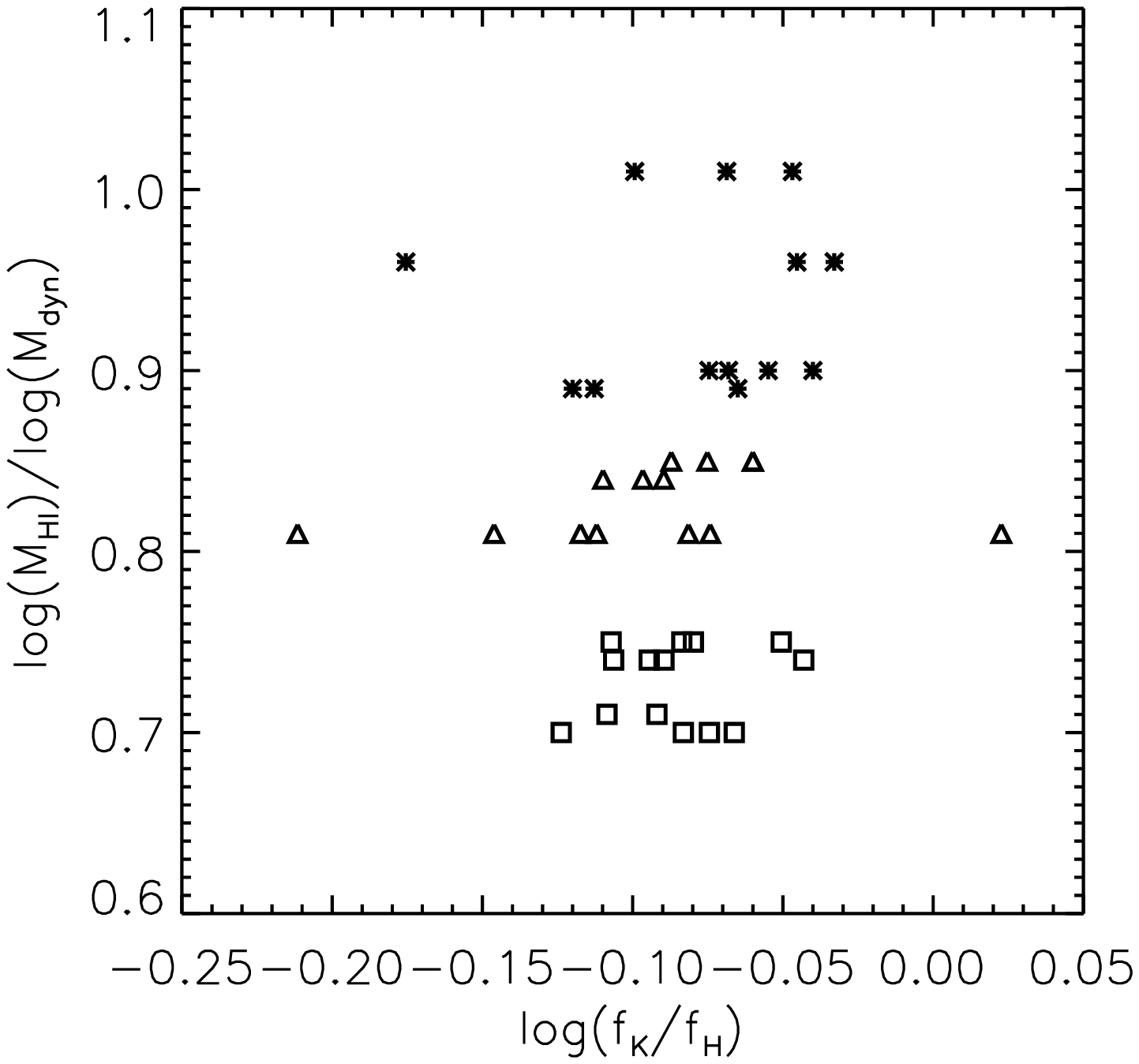}{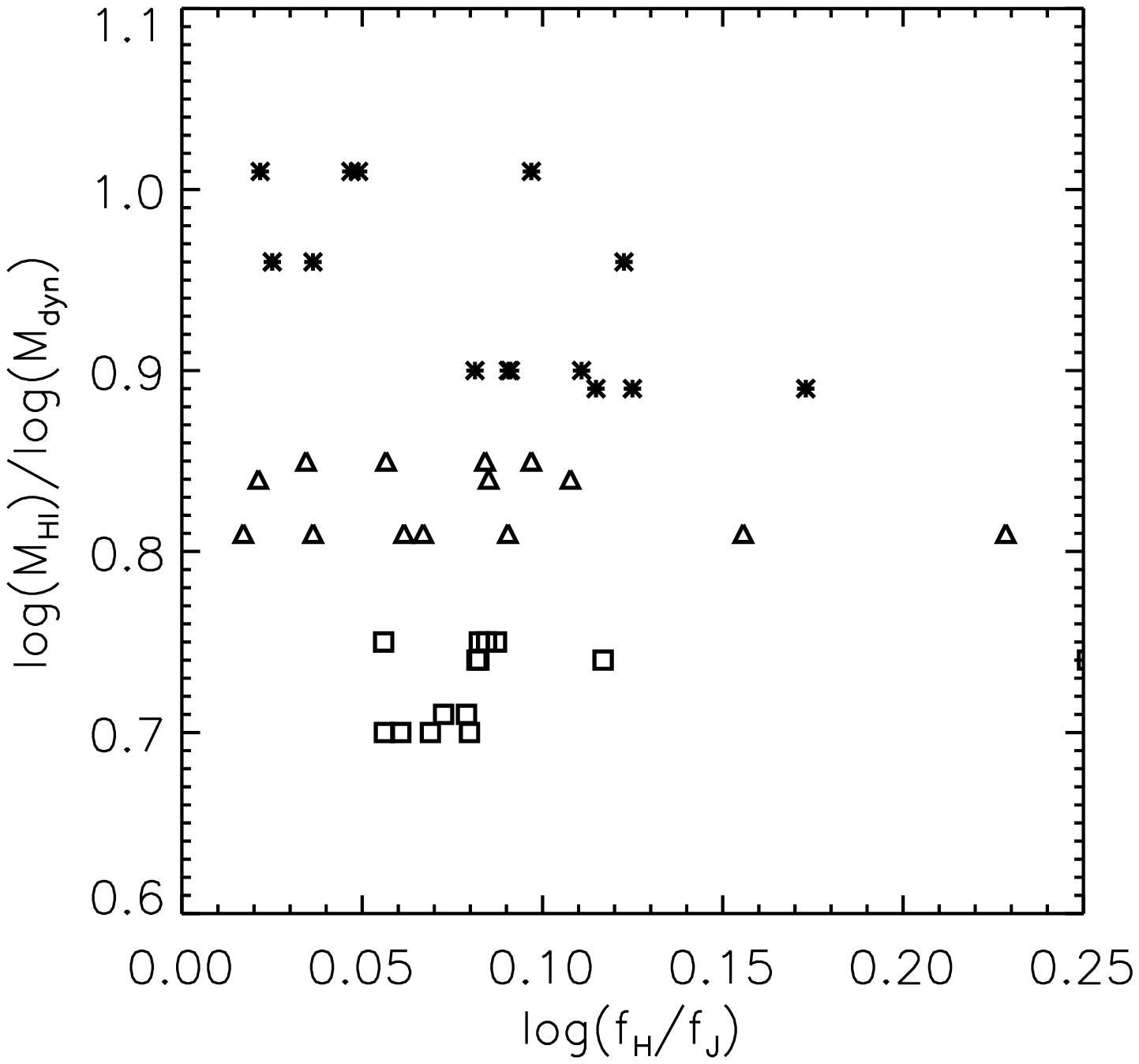}
\caption{(top) 2MASS J, H, K color-color diagram with the same
features as Fig.~\ref{plotJHK1}, but with the galaxies separated by
${\rm log(M_{HI})/log(M_{dyn}})$ as in Fig.~\ref{plotIRAC1}.  (bottom)
The relationship between the colors (used for the x-axis and y-axis
above) of the galaxies in this sample and the ${\rm
log(M_{HI})/log(M_{dyn}})$ ratio for the parent groups using the same
symbols as above. See discussion in sections~\ref{stages_trends}.
\label{plotJHK2}}
\end{figure}

In order to assess whether there are any trends in the infrared
properties of HCG galaxies with evolutionary state, we have divided
our sample into three types as discussed in \S~\ref{stages}: \\
(I) ${\rm log}(M_{{\rm HI}})/{\rm log}(M_{{\rm dyn}}) \geq 0.9$, \\
(II) $0.9 >{\rm log}(M_{{\rm HI}})/{\rm log}(M_{{\rm dyn}}) \geq 0.8$, \\
(III)${\rm log}(M_{{\rm HI}})/{\rm log}(M_{{\rm dyn}}) < 0.8$.  \\
In other
words, type~I groups are relatively the most gas-rich by mass, and
type~III groups are relatively the most gas-poor by mass.  In
Figures~\ref{plotIRAC2}-\ref{plotJHK2} these types are respectively
shown as stars, triangles, and squares.  It is important to note that
these types are based on the properties of the {\it group}, whereas
the data points are {\it individual galaxies} within a group of a
given type.  In other words, we are attempting to probe whether the
local properties of a member galaxy are related to the global
properties of the group in which it resides.

The infrared colors of the constituent galaxies appear to be related
to the type of group in which they reside, and Figure~\ref{plotIRAC2}
reveals a few striking trends.  First, the HI gas-deficient group
galaxies (type~III) are relatively tightly clustered near the nexus of
stellar colors, which is expected if there is little star formation,
reddening, or AGN activity.  The only outliers are known AGN galaxies
in HCG~90 (galaxies A and D).  The galaxies in more gas-rich groups
span a much larger range in colors, with type~I groups extending
toward much redder colors, indicating higher levels of star formation
and AGN activity overall.  It is particularly noteworthy that the
galaxies in this sample appear to follow a sequence in the infrared
color-color space, with a relative gap in the galaxy population found
at moderate colors ($ -0.2 \lesssim {\rm log}(f_{8.0}/f_{4.5})
\lesssim 0.2$, and $-0.3 \lesssim {\rm log}(f_{5.8}/f_{3.6}) \lesssim
-0.1$).

This ``sequence'' and ``gap'' are more significant in comparison to
the FLS galaxies and SINGS galaxies, which do not appear to exhibit
such a gap.  We note that this gap is reminiscent of the so-called
``green-valley'' that is apparent in color-magnitude diagrams of
galaxy samples such as that presented by \citet{hogg04}.  This gap may
also reflect the bimodal structure of the distribution of Hubble types
seen in Fig.~\ref{types}.  If this gap in the HCG galaxies
distribution is dynamical in nature, it suggests that star formation
and AGN activity in HCG galaxies responds to the dynamical processes
taking place on rapid timescales.

Trends similar to those in Fig.~\ref{plotIRAC2} are seen in
Figure~\ref{plotMIX2}, with the median color of galaxies in type~I,
type~II, and type~III groups slightly offset.  The near-IR colors
alone, as shown in Figure~\ref{plotJHK2}, do not reveal any strong
trends with group type.  However, it is interesting to note that the
outlying sources are either from groups of type~I or type~II, or
known AGN galaxies (HCG~90A and HCG~90D).

\begin{figure}[t]
\plotone{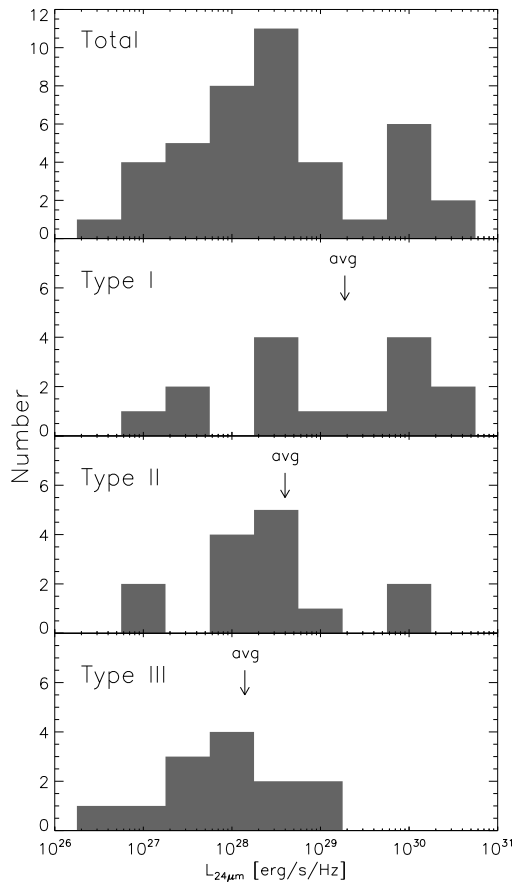}
\caption{Histograms of the 24$\mu$m luminosities for the total sample
  of galaxies (top panel) as well as galaxies belonging to compact
  groups of each type (see \S~\ref{stages_trends} for discussion).
\label{lum_hist}}
\end{figure}

The 24$\mu$m luminosity of the member galaxies also exhibits a slight
trend with group type, as shown in Figure~\ref{lum_hist}.  Galaxies in
gas-rich groups of type~I have the highest average 24$\mu$m
luminosity, while galaxies in gas-poor groups of type~III have the
lowest average 24$\mu$m luminosity.  This trend is in accord with
type~I groups hosting the most actively star-forming galaxies.
\citet{smith07} examined 24$\mu$m emission of the Arp interactive and
field spiral samples of galaxies, but see little statistical evidence
for a difference between those two populations.  This is not
necessarily inconsistent with the trends seen here, as our sample of
compact groups contains many early type galaxies, which the
\citet{smith07} sample was not designed to investigate.

\section{Summary}

We present new {\it Spitzer} observations of a sample of twelve
Hickson Compact Groups comprising 45 galaxies and combine these with
existing 2MASS survey observations.  The goal of this project is to
examine the relationship between the compact group environments and
various physical processes.  The main results of this paper are
summarized below.

The amount of PAH emission, reddening, and 24$\mu$m emission in each
galaxy appear to be related; these observables tend to be similarly
strong or weak for a given galaxy, reflecting the amount of recent
star formation activity.  Galaxies that strongly exhibit these
features, also have much greater scatter in their colors, plausibly
reflecting variable contributions to the integrated light from AGN and
starburst activity; galaxies dominated by stellar light with little
evidence for recent starburst and/or AGN activity have a much smaller
range of infrared colors.  The near-infrared (J,H,K) colors of the
member galaxies are largely consistent with being dominated by
slightly reddened normal stellar populations.

We separate the compact groups into three types based on the ratio of
their dynamical and HI masses, with type~I being the most gas-rich and
type~III being the most gas-poor.  An analysis of morphological types
indicates that late-type galaxies are preferentially found in type~I
groups, and early-type galaxies are preferentially found in type~III
groups.  However, the causal relationship that underlies these trends
is not clear.  The near-IR colors do not show any strong trends with
group type, although the outlying points with a K-band excess tend to
be galaxies in type~I and type~II groups.  Galaxies in type~I groups
tend to be the most actively star forming as manifest in their
thermal-IR colors, with type~II also showing a moderate level of
activity, and type~III groups are relatively quiescent.  This result
could be interpreted as type~I groups actively consuming their gas
(and forming stars), while type~II groups may be generally somewhat
further along in this process.  Galaxies in type~III groups tend to be
relatively tightly clustered around the locus of colors for normal
stellar populations, with the few outliers known to be active
galaxies.  The sensitivity of thermal infrared colors to neutral gas
content suggests that the gas {\it content} of a Hickson compact group
is at least as important as the neutral gas {\it distribution} in
promoting or inhibiting star formation.  Clearly both gas content and
gas distribution are important to a group's evolutionary state,
although it is not yet clear how independent or related these
parameters are.

The group type also appears to be reflected in the 24$\mu$m luminosity
of the member galaxies.  Galaxies in groups of type~I have the highest
average 24$\mu$m luminosity, while galaxies in groups of type~III have
the lowest average 24$\mu$m luminosity.  This trend is in accord with
the most gas-rich groups hosting the most actively star-forming
galaxies.  Of the type~I and type~II groups in this sample, only one
of them has an X-ray detection \citep[HCG~16, which is known to
contain active galaxies as well as faint X-ray emission from the
IGM][]{belsole03}, while all four of the gas-poor groups have been
detected in the X-ray \citep[HCG~42, HCG~48, HCG~62, and
HCG~90][]{ponman96}.  These results are consistent with the hypothesis
that the missing HI gas in the most ``deficient'' groups may have been
converted to X-ray emitting hot gas.

There are striking trends seen in this study between the ${\rm
log}(M_{{\rm HI}})/{\rm log}(M_{{\rm dyn}})$ ratio for an entire
Hickson Compact Group and the infrared colors of the individual member
galaxies, which suggest that the constituent galaxies in Hickson
Compact Groups are related to the type of group in which they reside.
However it is not clear from these observations whether the
``gas-richness'' of a group simply reflects its fraction of late-type
galaxies, or whether the gas-richness of a group affects the
constituent galaxy types.  Clearly the morphology of the gas within a
group must have a role, as pointed out by
\citet{verdes-montenegro01}. 

The galaxy sample here has characteristics that are distinctly
different from the sample of galaxies in the {\it Spitzer} First Look
Survey.  Most notably, Hickson Compact Group galaxies exhibit a
``gap'' in infrared color space that is sparsely populated.  This gap
may suggest a rapid evolution of galaxy properties in response to
dynamical effects in HCGs.

\acknowledgments

We are grateful to Remy Indebetouw for his help developing the IDL
program used in this project.  We also thank Amanda Heiderman for
helpful discussions on compact groups.  The referee, Phil Appleton,
provided insightful feedback that significantly enhanced this paper.
This work was supported in part by the {\it Spitzer Space Telescope}
grant for project \#3596.  K.~E.~J.  gratefully acknowledges partial
support for this research provided by NSF/CAREER grant AST-0548103.
Support for S.~C.~G. was provided by NASA through the {\it Spitzer}
Fellowship Program, under award 1256317.  This publication makes use
of data products from the Two Micron All Sky Survey, which is a joint
project of the University of Massachusetts and the Infrared Processing
and Analysis Center/California Institute of Technology, funded by the
National Aeronautics and Space Administration and the NSF.  This
research has made use of the NASA/IPAC Extragalactic Database (NED)
which is operated by the Jet Propulsion Laboratory, California
Institute of Technology, under contract with the National Aeronautics
and Space Administration.

{\it Facilities:} \facility{Spitzer}, \facility{VLA}.

\begin{deluxetable}{llllllllllllllllll}
\tabletypesize{\small}
\tablecaption{2MASS and Spitzer flux densities of the galaxies in the sample 
\label{fluxes}}
\tablewidth{0pt}
\tablehead{\colhead{HCG}
& & \colhead{J} & \colhead{H} & \colhead{K} & \colhead{3.6$\mu$m} & \colhead{4.5$\mu$m} 
& \colhead{5.8$\mu$m} & \colhead{8.0$\mu$m} & \colhead{24$\mu$m} \\

\colhead{Galaxy} &\colhead{Morph} & \colhead{(mJy)} & 
\colhead{(mJy)} & \colhead{(mJy)} & \colhead{(mJy)} & \colhead{(mJy)} & 
\colhead{(mJy)} & \colhead{(mJy)} &  \colhead{(mJy)}
}
\startdata

2a & SBd & 27.9 &   37.0 &   24.7 &  15.1 &   10.7 &   21.3 &   54.1 &   115 \\

2b & cI & 21.9 &   23.2 &   20.9 &   14.5 &   10.6 &   29.0 &   88.0 &   351 \\

2c & SBc & 12.6 &   13.7 &   12.7 &   6.7 &   4.6 &   6.8 &   16.9 &   21.5  \\

& & & & & & & & & \\

7a & Sb & 124 &  155 &    13. &  70.3 &   46.3 &   76.1 &   161 &    303 \\

7b\tablenotemark{d} & SB0 & 67.4 &   81.8 &   68.8 &   31.9 &   20.1 &   16.7 &   10.7 &   12.7 \\

7c & SBc & 58.9 &   67.1 &   54.9 &   29.0 &   18.9 &   26.1 &   76.3 &   76.0 \\

7d\tablenotemark{d} & SBc & 31.5 &   34.1 &   29.7 &   11.1 &   7.5 &   13.1 &   19.6 &   12.0 \\

& & & & & & & & & \\

16a\tablenotemark{d} & SBa & 151 &    186 &    159 &    85.6 &   57.1 &   100 &    279 &    409 \\

16b\tablenotemark{d} & Sab & 90.4 &   109 &    91.8 &   43.0 &   26.8 &   22.4 &   24.8 &   22.5 \\

16c & Im & 73.7 &   90.9 &   82.9 &   65.9 &   48.1 &   177 &    576 &   1412 \\

16d & Im & 63.3 &   81.7 &   72.0 &   47.7 &   40.3 &   105 &    338 &    1785 \\

& & & & & & & & & \\

19a & E2 & 43.2 &   53.2 &   40.6 &   19.9 &   12.1 &   7.8 &   5.6 &   3.3 \\

19b & Scd & 10.3 &   11.2 &   11.8 &   5.8 &   3.8 &   6.7 &   19.7 &   24.1 \\

19c & Sdm & 5.8 &   8.3 &   5.1 &   2.7 &   1.8 &   2.1 &   5.1 &   6.4 \\

& & & & & & & & & \\

22a & E2  & 305 &    371 &    297 &    142 &    84.5 &   53.8 &   38.2 &   13.9 \\

22b & Sa  & 20.6 &   26.4 &   20.5 &   9.3 &   5.8 &   3.3 &   2.5 &   1.1 \\

22c & SBc & 24.0 &   25.2 &   20.5 &   11.6 &   7.8 &   12.2 &   18.7 &   29.4 \\

& & & & & & & & & \\

31ace\tablenotemark{d} & (pec) & 20.1 &   22.5 &   17.9 &   13.4 &   10.1 &   24.3 &   72.5 &   463 \\

31b\tablenotemark{d} & Sm  & 3.9 &   4.1 &   3.5 &   1.8 &   1.4 &   1.5 &   4.3 &   16.6 \\

31f &(Im) & 0.4 &   0.4 &   $\cdots$ &   0.1 &   0.1 &   0.1 &   0.2 &   2.8 \\

31g & Im  & 7.9 &   8.8 &   7.9 &   3.7 &   2.4 &   4.7 &   8.1 &   38.9 \\

31q &(Im) & 1.2 &   1.5 &   0.9 &   0.5 &   0.3 &   0.3 &   0.5 &   0.7 \\

& & & & & & & & & \\

42a\tablenotemark{d} & E3 & 373 &    451 &    372 &    177 &    108 &    76.8 &   40.1 &   33.6 \\

42b &SB0 & 46.1 &   56.0 &   46.6 &   22.5 &   13.8 &   7.6 &   6.1 &   4.5 \\

42c\tablenotemark{d} &E2  & 58.0 &   70.9 &   55.4 &   27.2 &   17.0 &   10.8 &   6.6 &   2.8 \\

42d &E2  & 8.0 &   9.1 &   8.1 &   3.4 &   2.1 &   1.1 &   0.8 &   0.4 \\

& & & & & & & & & \\

48a \tablenotemark{d} &E2 & 306 &    367 &    297 &    142 &    88.3 &   75.4 &   44.0 &   16.1 \\
48b \tablenotemark{d,e} &Sc  & 42.1 &   47.0 &   37.7 &   19.5 &   14.6 &   23.5 &   69.1 &   81.9 \\

48c \tablenotemark{e}& S0a  & 16.2 &   21.3 &   18.6 &   7.9 &   4.9 &   3.3 &   2.1 &   0.8 \\

48d & E1 & 8.8 &   10.4 &   8.1 &   3.9 &   2.6 &   1.4 &   1.0 &   0.4 \\

& & & & & & & & & \\

59a\tablenotemark{d} &Sa  & 22.5 &   23.4 &   19.4 &   12.1 &   10.2 &   14.6 &   38.9 &   453 \\

59b &E0  & 10.5 &   12.1 &   10.2 &   4.3 &   2.8 &   1.7 &   1.1 &   0.6 \\

59c &Sc  & 3.6 &   4.2 &   3.0 &   1.7 &   1.1 &   1.4 &   3.3 &   3.6 \\

59d\tablenotemark{d} &Im  & 2.6 &   4.4 &   3.4 &   1.8 &   1.2 &   1.8 &   4.4 &   12.7 \\

& & & & & & & & & \\

61a &S0a  & 141 &    188 &    145 &    68.2 &   42.1 &   26.8 &   26.0 &   20.5 \\

61c &Sbc  & 70.5 &   105 &    90.4 &   52.7 &   37.1 &   77.8 &   226 &    357 \\

61d &S0  & 33.7 &   43.9 &   33.3 &   15.8 &   10.0 &   5.8 &   4.9 &   2.5 \\

& & & & & & & & & \\

62a\tablenotemark{d} &E3  & 157 &    184 &    155 &    73.7 &   45.4 &   31.4 &   21.9 &   9.4 \\

62b\tablenotemark{d} &S0  & 64.0 &   76.9 &   63.5 &   29.3 &   17.7 &   11.5 &   8.4 &   3.6 \\

62c\tablenotemark{d} &S0 & 32.6 &   37.1 &   27.9 &   13.0 &   7.4 &   3.9 &   3.1 &   2.1 \\

62d & E2 & 8.0 &   9.2 &   7.9 &   3.7 &   2.3 &   1.6 &   1.0 &   1.9 \\

& & & & & & & & & \\

90a &Sa  & 243 &    318 &    288 &    176 &    147 &    190 &   328 &   $\cdots$ \\

90b\tablenotemark{d} &E0  & 182 &    220 &    179 &    78.8 &   46.6 &   31.0 &   25.8 &   18.3 \\

90c\tablenotemark{d} &E0  & 164. &    198 &    155 &    74.0 &   44.5 &   27.6 &   21.7 &   9.1 \\

90d\tablenotemark{d} & Im & 152. &    271 &    218 &    96.7 &   61.8 &   54.5 &   121 &    209 \\
\enddata

\tablenotetext{a}{Morphological types from \citet{hickson89,mendes94} except where noted by parenthesis.}

\tablenotetext{b}{Near-IR flux densities in the J-, H-, and K-bands were 
directly measured from the 2MASS Atlas images in the same manner as 
described in \S~\ref{Observations} for the {\it Spitzer} images.}

\tablenotetext{c}{Uncertainties are estimated to be $\sim 10\%$ for
flux densities greater than 10~mJy, $\sim 20\%$ for flux densities of
5-10~mJy, and $\sim 50\%$ for flux densities of 1-5~mJy, and $\sim
100\%$ for marginal detections less than 1~mJy.}

\tablenotetext{d}{These galaxies were not cleanly seperated at the
1$\sigma$ contour level, and photometry was carried out using
apertures defined by higher contour levels.}
\tablenotetext{e}{Galaxies HCG~48b and HCG~48c may not be proper group
members \citep[see][]{ribeiro98}.}
\end{deluxetable}

\appendix
\section{NOTES ON INDIVIDUAL HICKSON COMPACT GROUPS}

\subsection{HCG~2 \label{hcg2sed}}
\begin{figure}
\epsscale{1.}
\plottwo{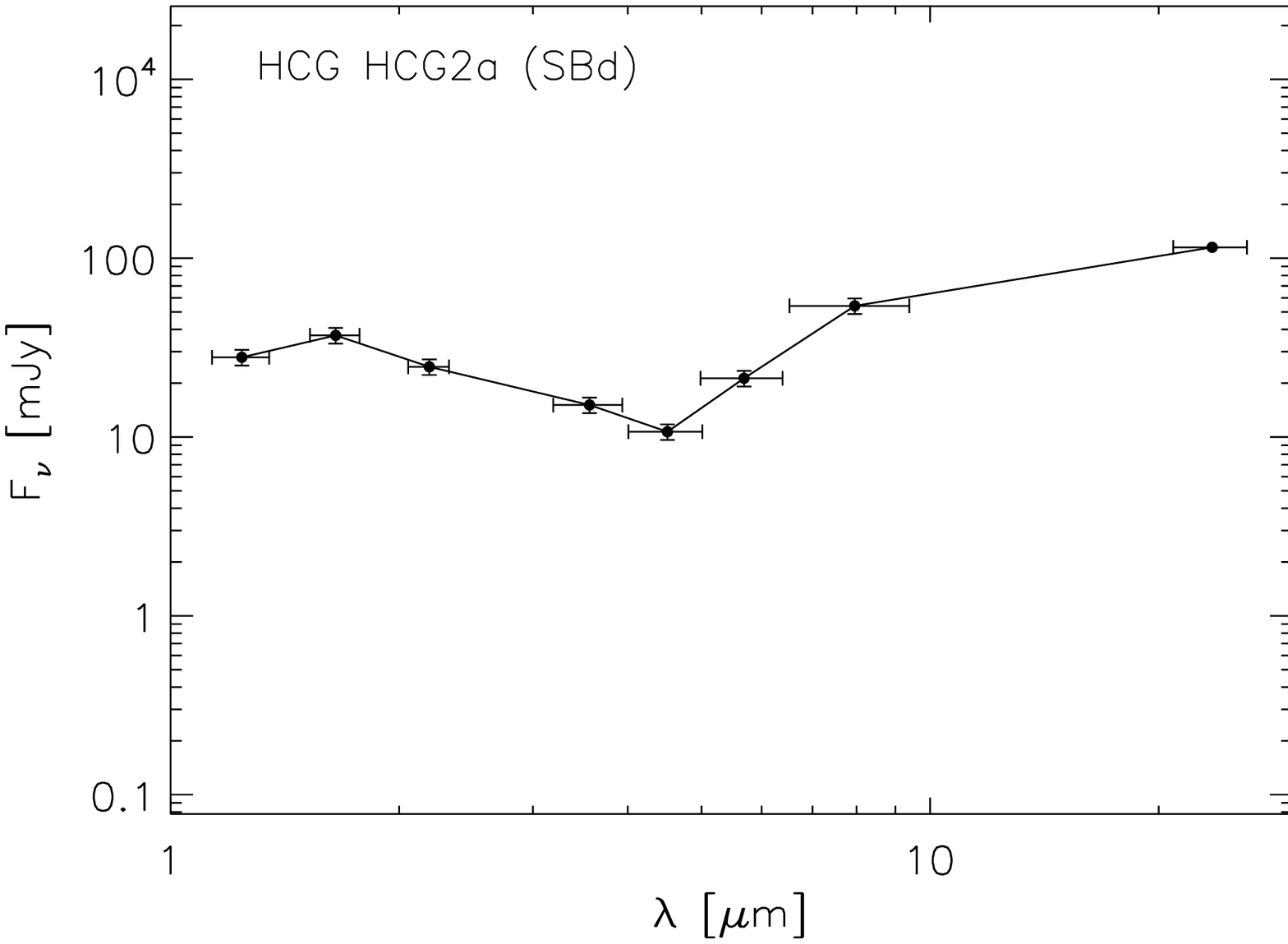}{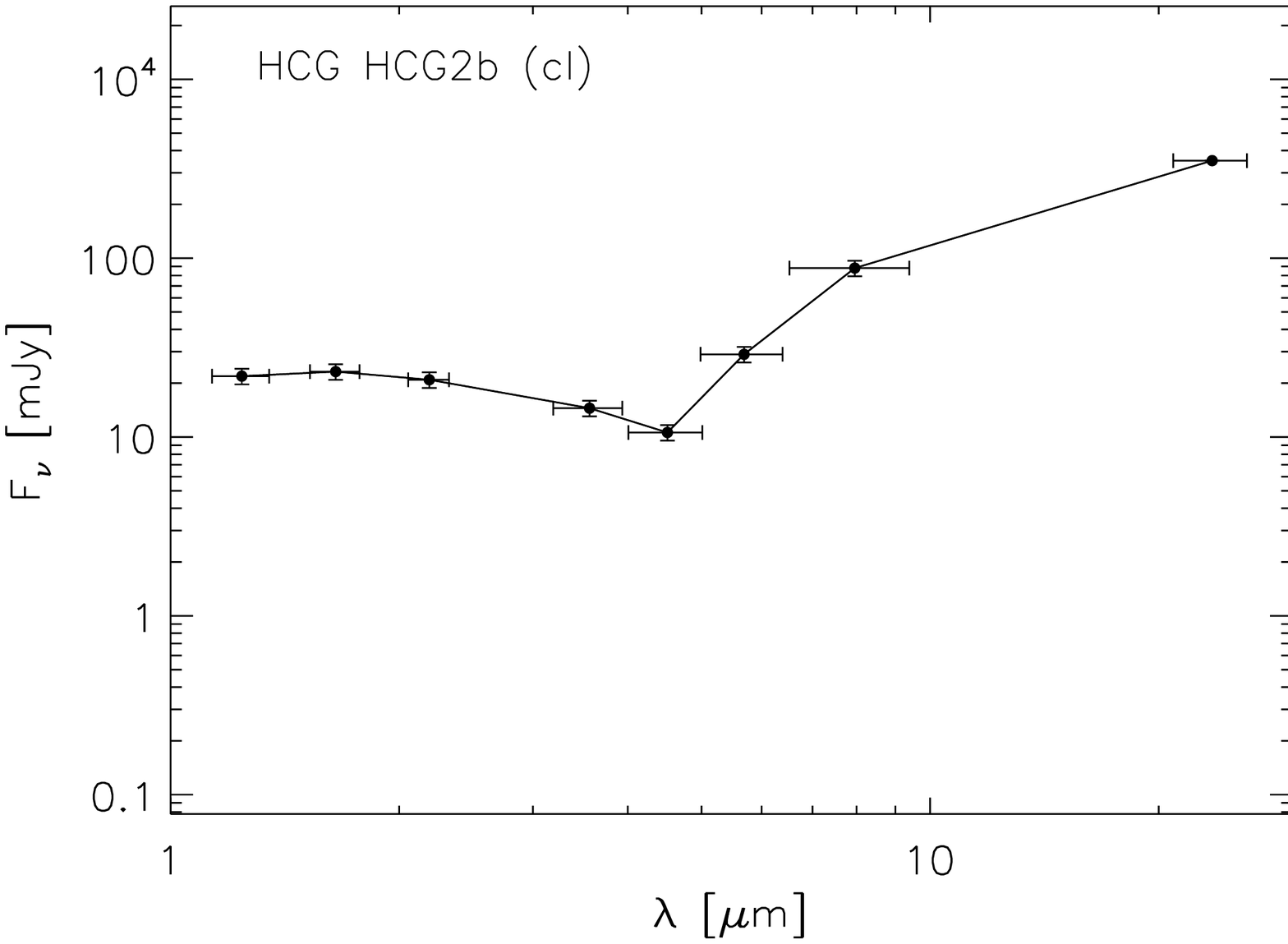}\\
\epsscale{0.46}
\plotone{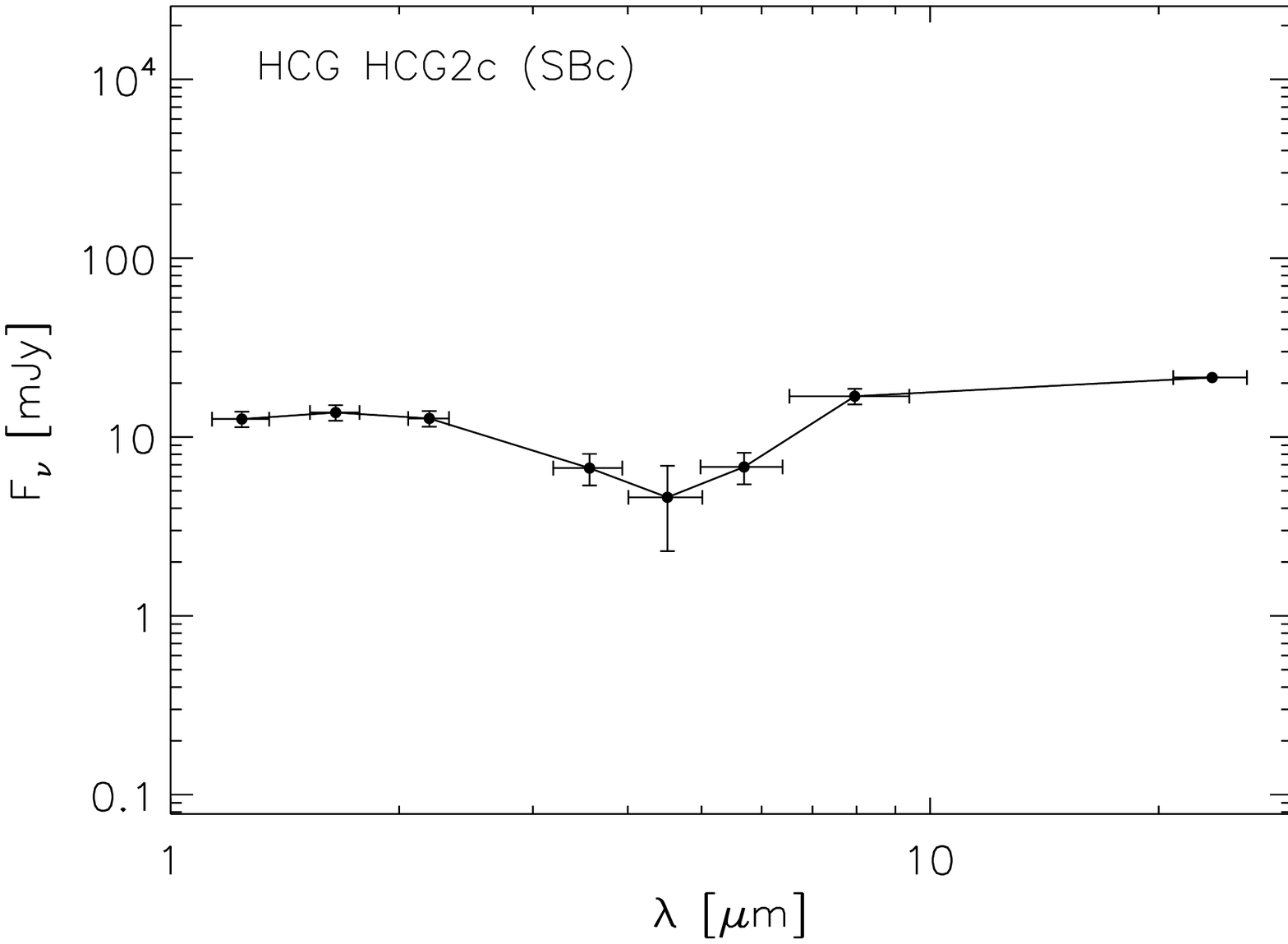}
\caption{Infrared spectral energy distributions for the galaxies
in HCG~2.  Horizontal error bars reflect the filter widths. \label{hcg2}}
\end{figure}

The spectral energy distributions for the galaxies in HCG~2 are shown
in Figure~\ref{hcg2}.  This group is classified as type~I according to
${\rm log}(M_{{\rm HI}})/{\rm log}(M_{{\rm dyn}})$, indicating it is
relatively gas-rich.  Galaxy~A hosts a number of bright star forming
knots as evident by their 4.5$\mu$m, 8.0$\mu$m, and 24$\mu$m emission.
Galaxy~B (MRK~552) is compact, but luminous in the infrared and the dominant
source of 24$\mu$m emission in the group.  Galaxy~C has a hint of
low-level star formation to the west of its nuclear region.  Galaxy~D
is a background source.  All three have similar infrared SED's that are
typical of star-forming galaxies.

\subsection{HCG~7}
\begin{figure}
\epsscale{1.0}
\plottwo{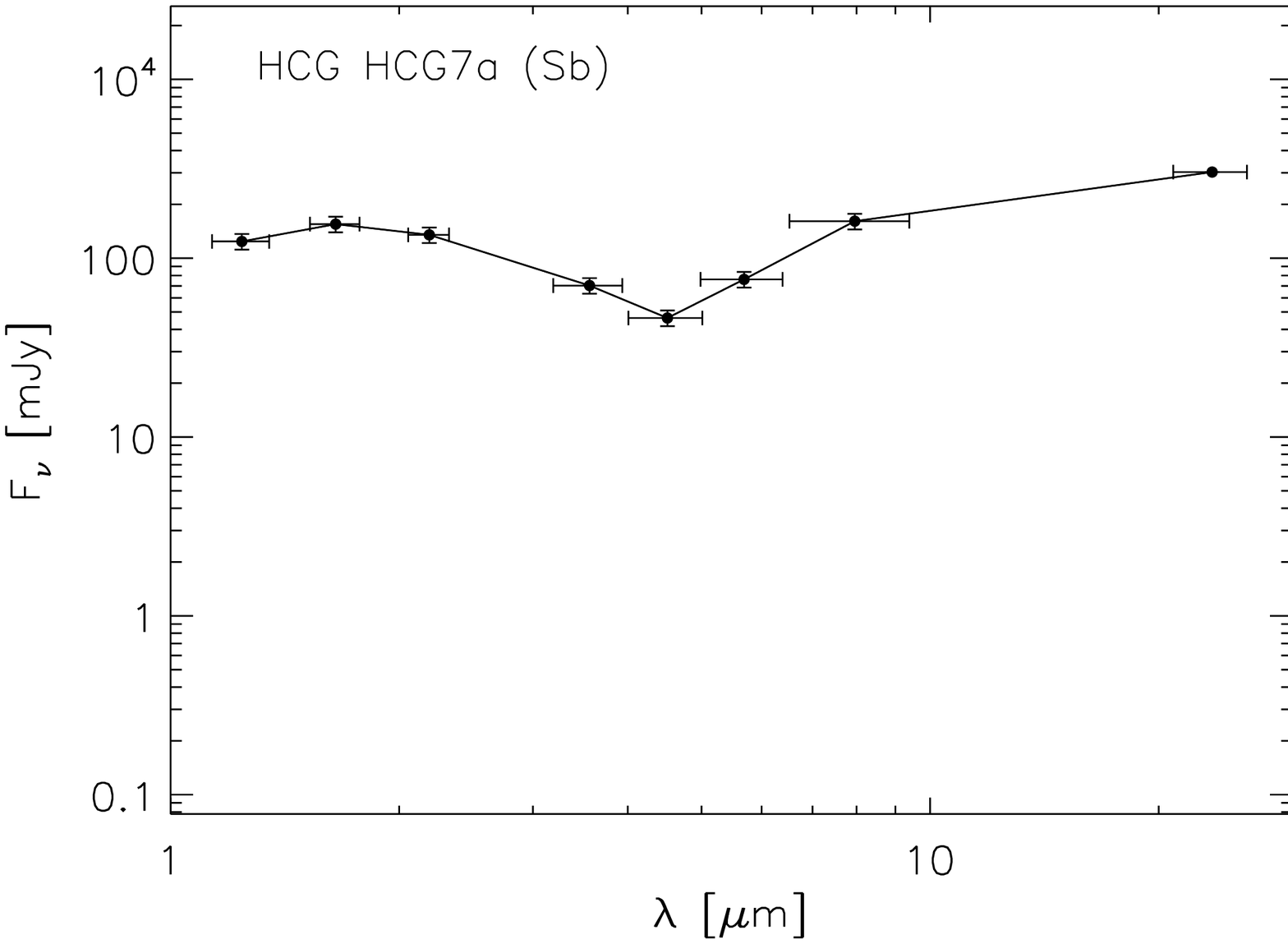}{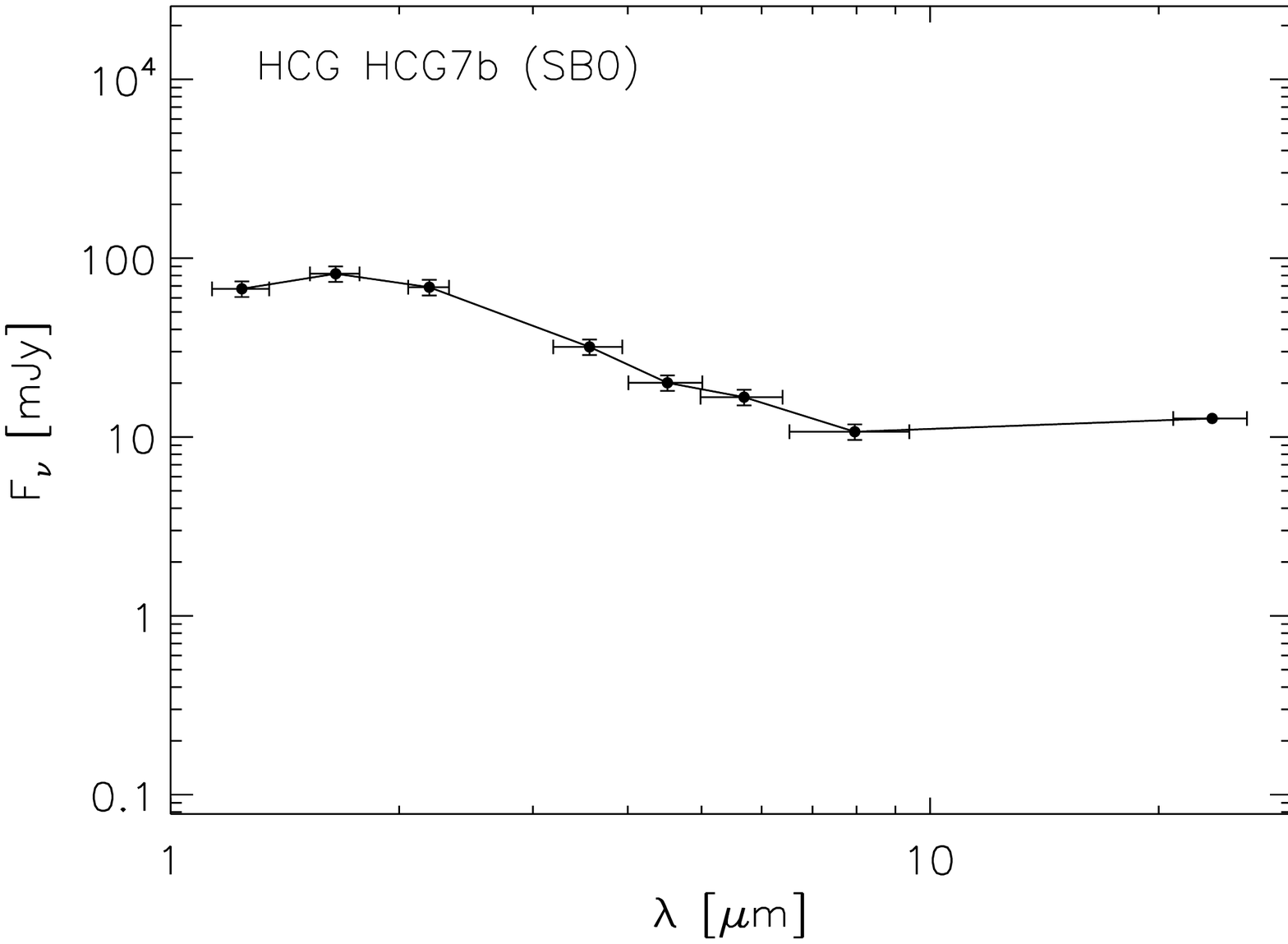}\\
\plottwo{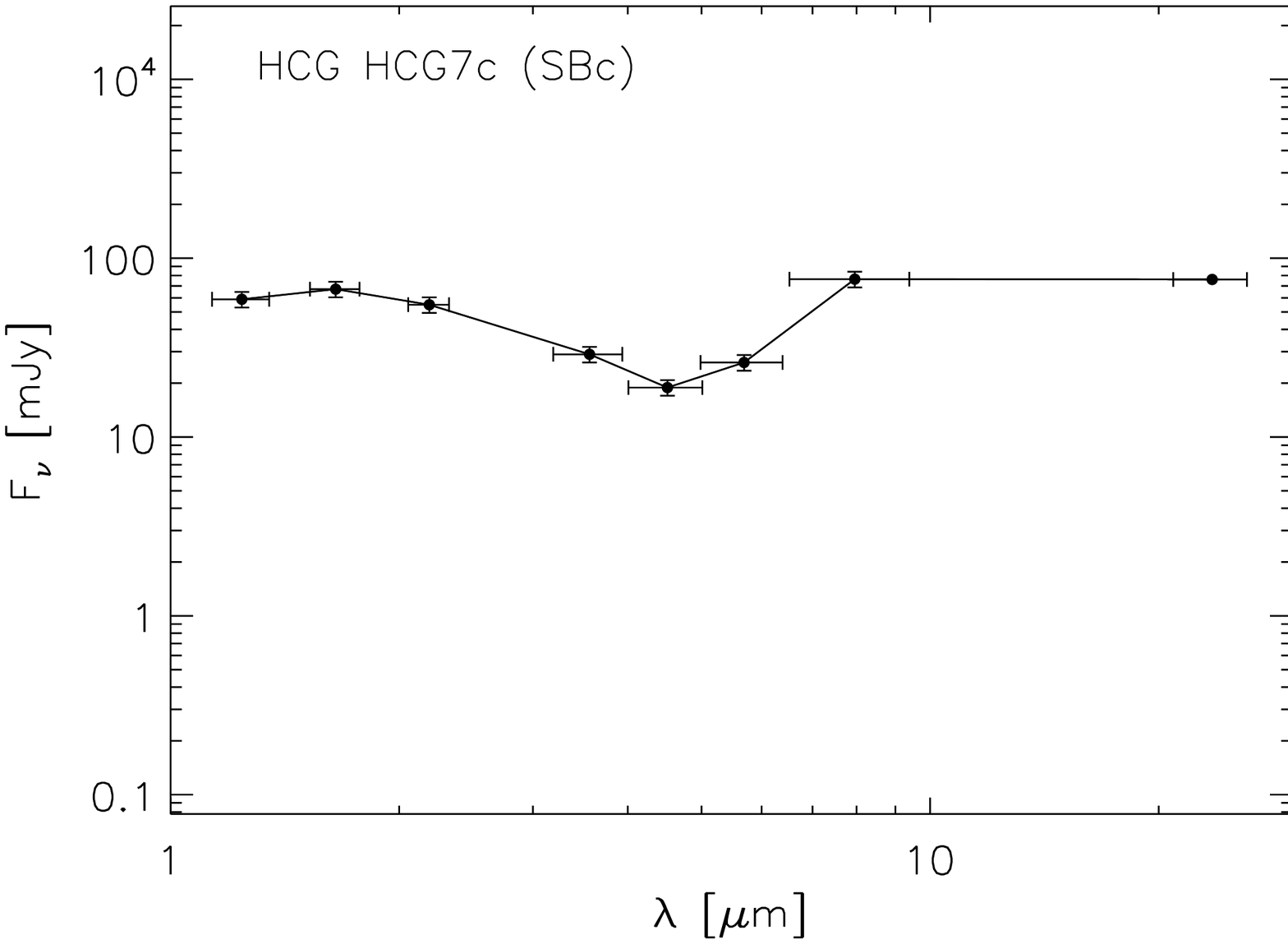}{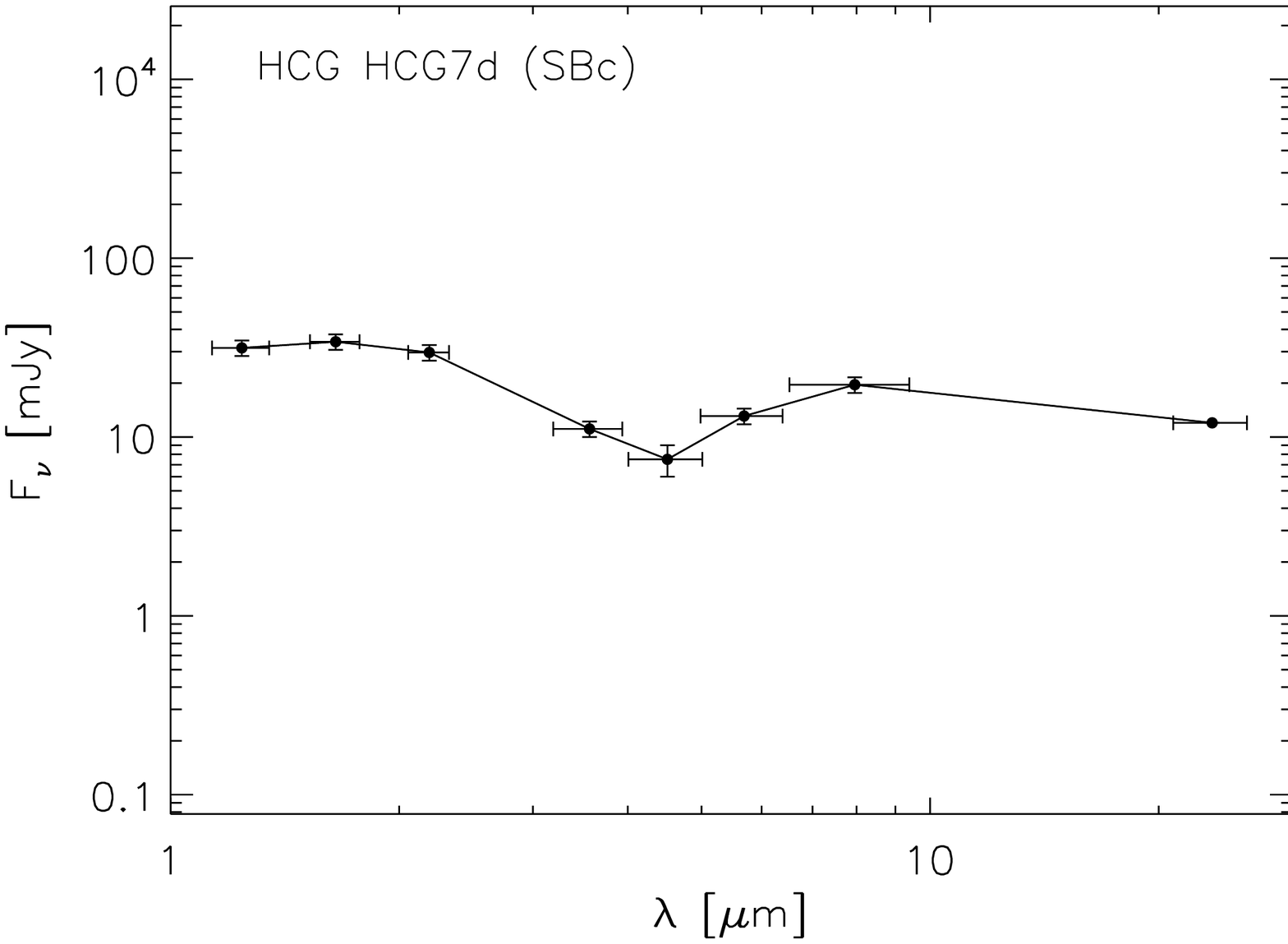}
\caption{Infrared spectral energy distributions for the galaxies
in HCG~7.  Horizontal error bars reflect the filter widths. \label{hcg7}}
\end{figure}

The spectral energy distributions for the galaxies in HCG~7 are shown
in Figure~\ref{hcg7}.  This group is classified as type~II according
to ${\rm log}(M_{{\rm HI}})/{\rm log}(M_{{\rm dyn}})$, indicating a
moderate amount of gas is present in the group.  Galaxy~A (NGC~192)
has a disturbed morphology with substantial infrared emission
throughout.  The nuclear region of galaxy~A is a luminous 24$\mu$m
source.  Galaxy~B (NGC~196) is dominated by stellar light with little
evidence of recent star formation.  Galaxy~C (NGC~201) has elevated
4.5$\mu$m and 8.0$\mu$m emission throughout the spiral arms, and the
nuclear region stands out at 24$\mu$m, possibly indicating low level
AGN activity.  Galaxy~D (NGC~197) is slightly disturbed and appears to
have a modest level of star formation.

\subsection{HCG~16}
\begin{figure}
\epsscale{1.0}
\plottwo{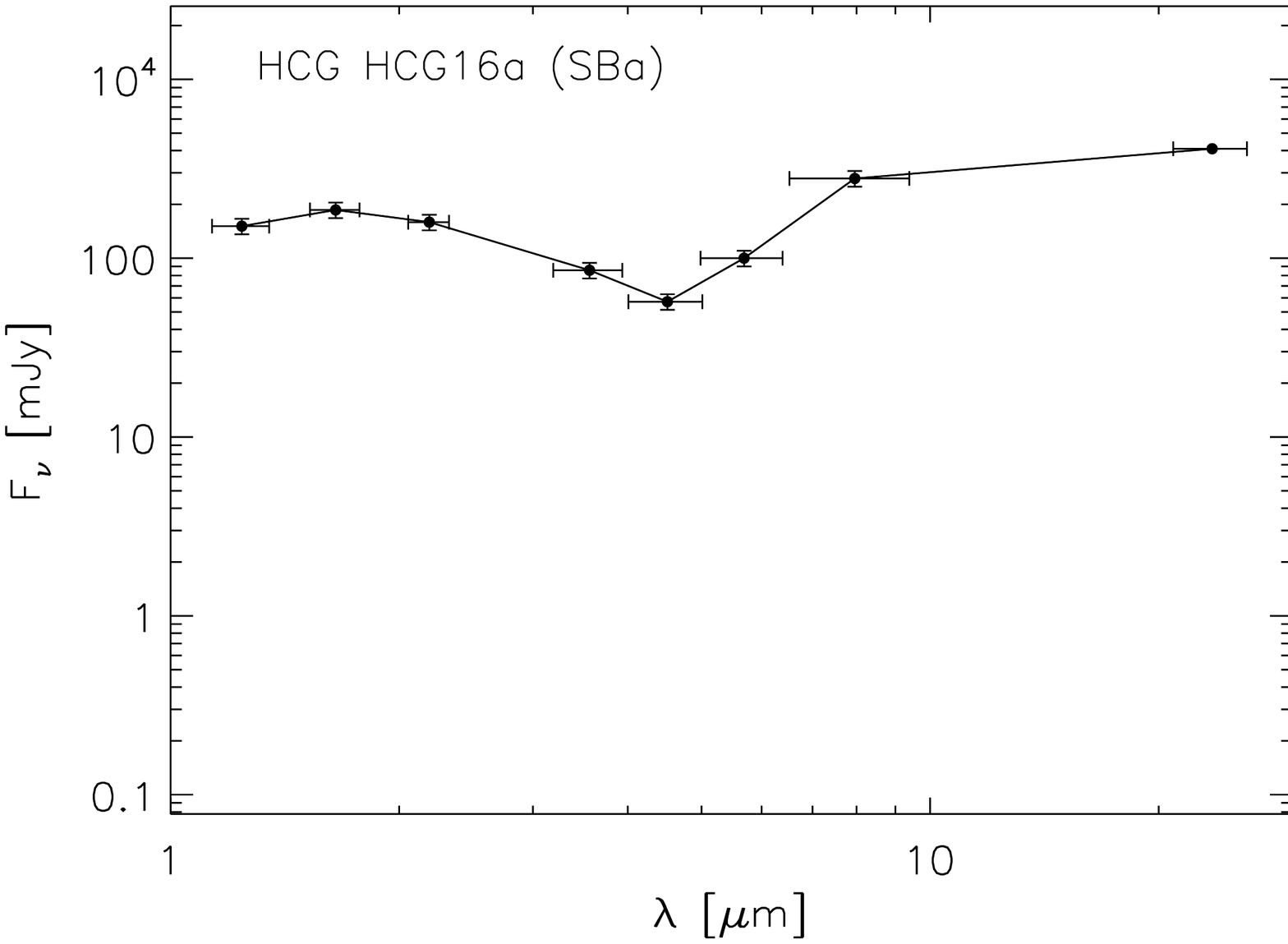}{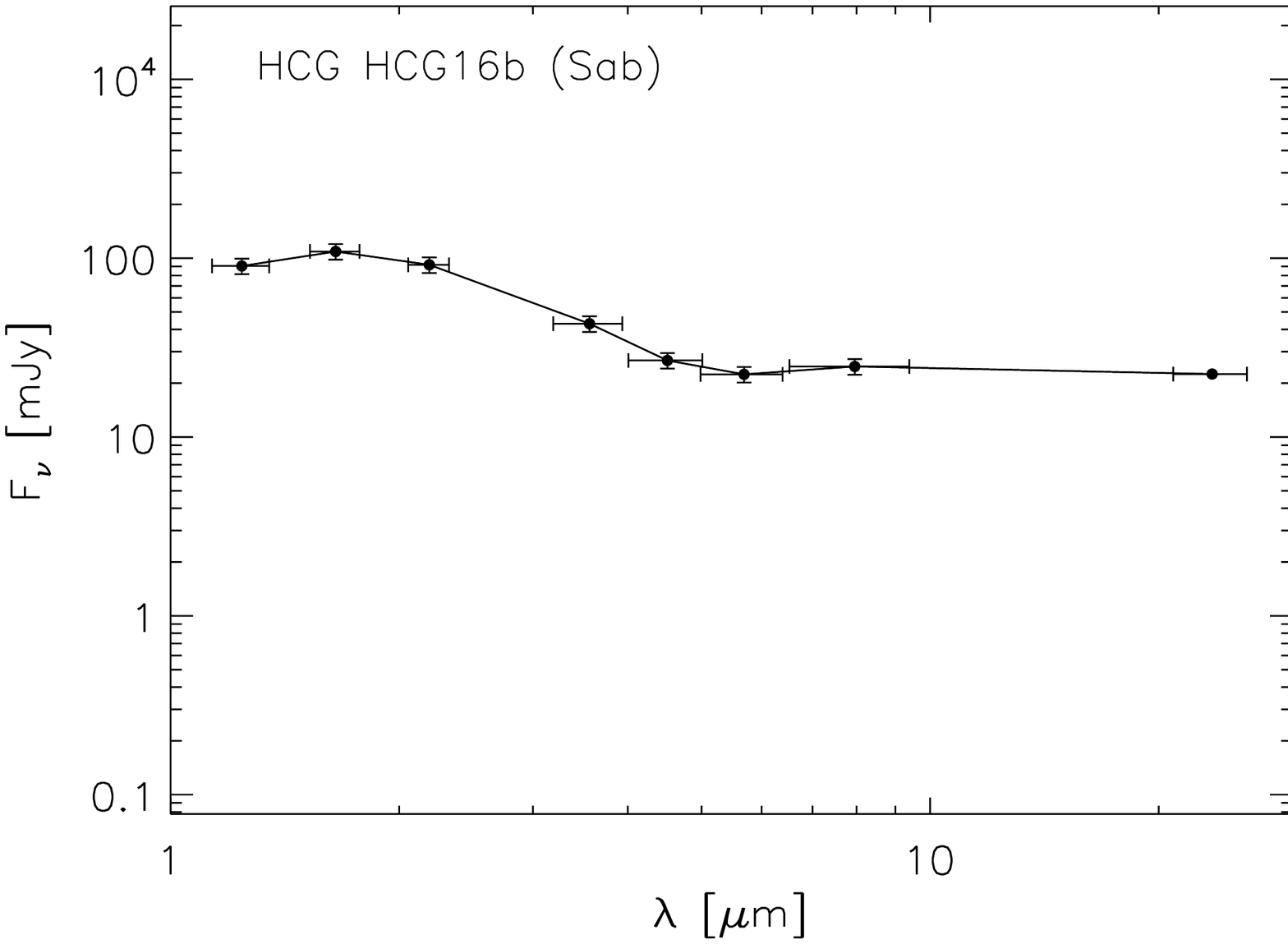}\\
\plottwo{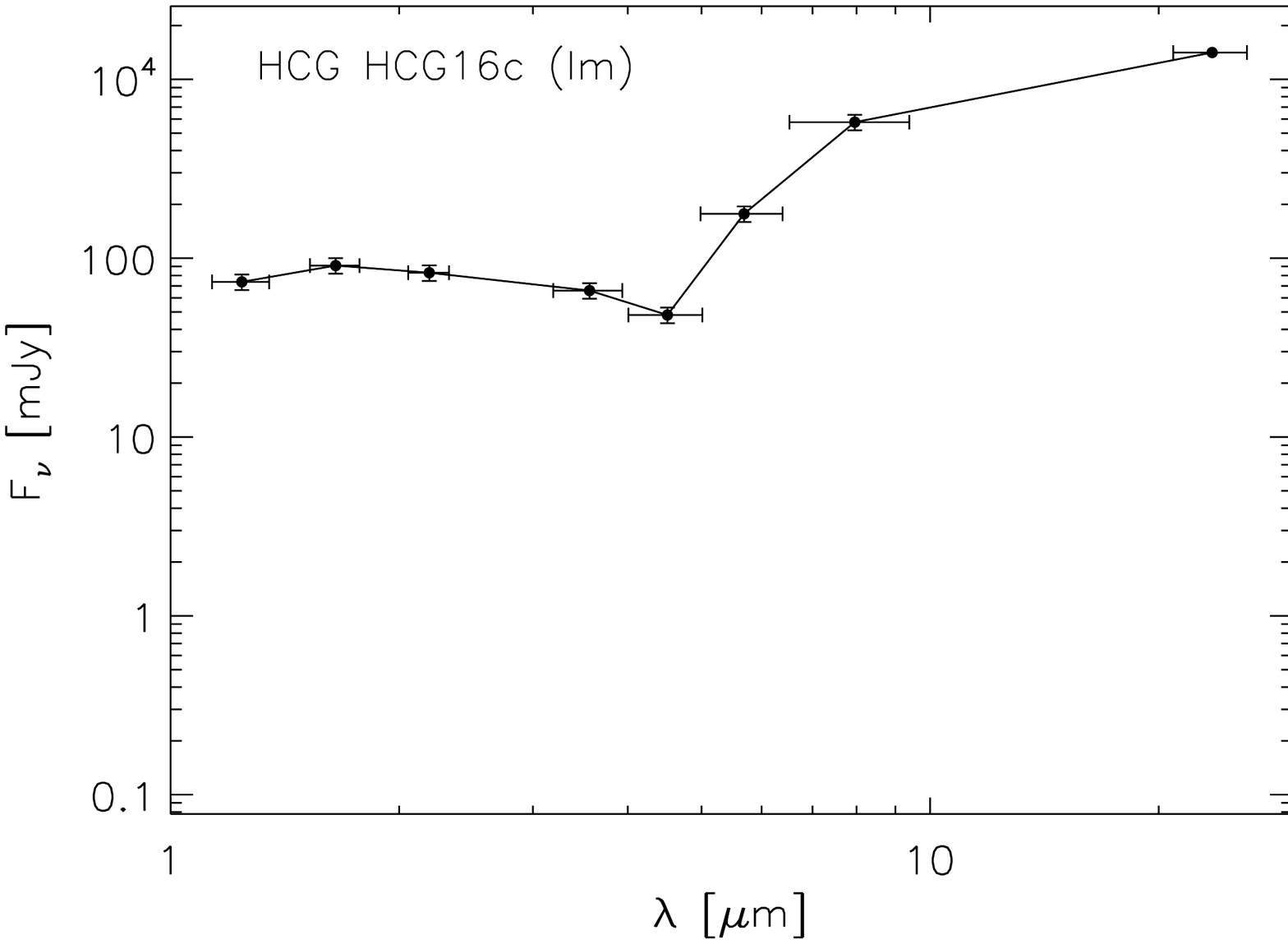}{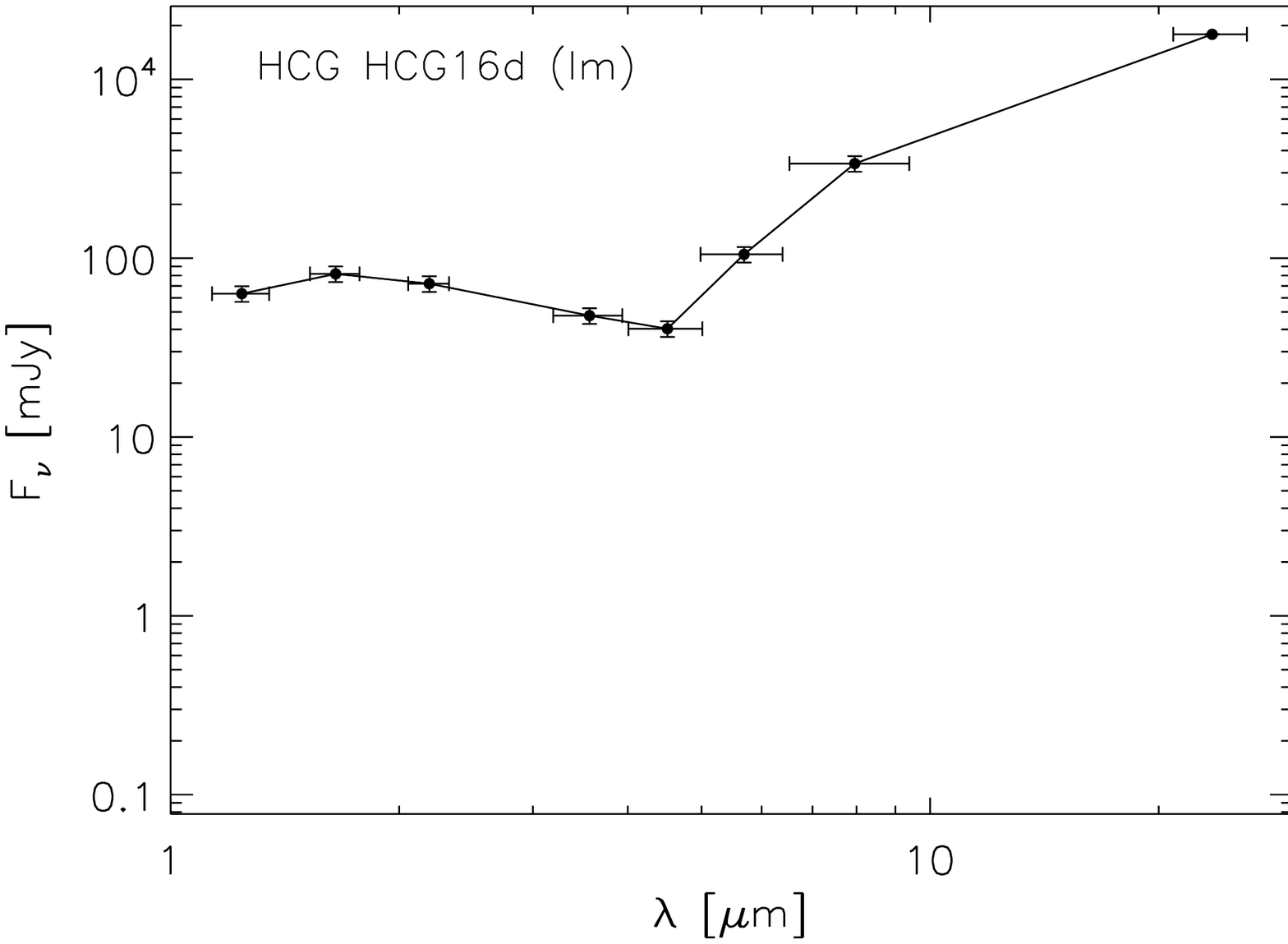}
\caption{Infrared spectral energy distributions for the galaxies
in HCG~16.  Horizontal error bars reflect the filter widths. \label{hcg16}}
\end{figure}

The spectral energy distributions for the galaxies in HCG~16 are shown
in Figure~\ref{hcg16}.  This group is classified as type~I according
to ${\rm log}(M_{{\rm HI}})/{\rm log}(M_{{\rm dyn}})$, indicating it
is relatively gas-rich.  \citet{ribeiro98} observe this group to be a 
core$+$halo system, suggesting the compact group is part of a larger 
collapsing structure.  Galaxies~A (NGC~835, MRK~1021) and B
(NGC~833) have extended tidal features visible in the 3.6$\mu$m and
4.5$\mu$m images without associated longer wavelength emission,
suggesting very little star formation has been triggered in the tidal
tails.  Galaxy~A does have a moderate level of star formation
activity, but galaxy~B is dominated by stellar light.  Galaxies~C
(NGC~838, MRK~1022) and D (NGC~839) are quite luminous in the
infrared, causing saturation and artifacts in the long exposure IRAC
images; both have significant levels of 24$\mu$m emission consistent
with originating from nuclear point sources.  Galaxies~A, B, and D
have previously been identified as active galaxies in this group
\citep{ribeiro96,coziol98,turner01}. \citet{mendes98} conclude that
both galaxies~C and D have recently undergone major merger events based on
their kinematics, double nuclei, and high infrared luminosities, which
is consistent with these {\it Spitzer} results.

\subsection{HCG~19}
\begin{figure}
\epsscale{1.0}
\plottwo{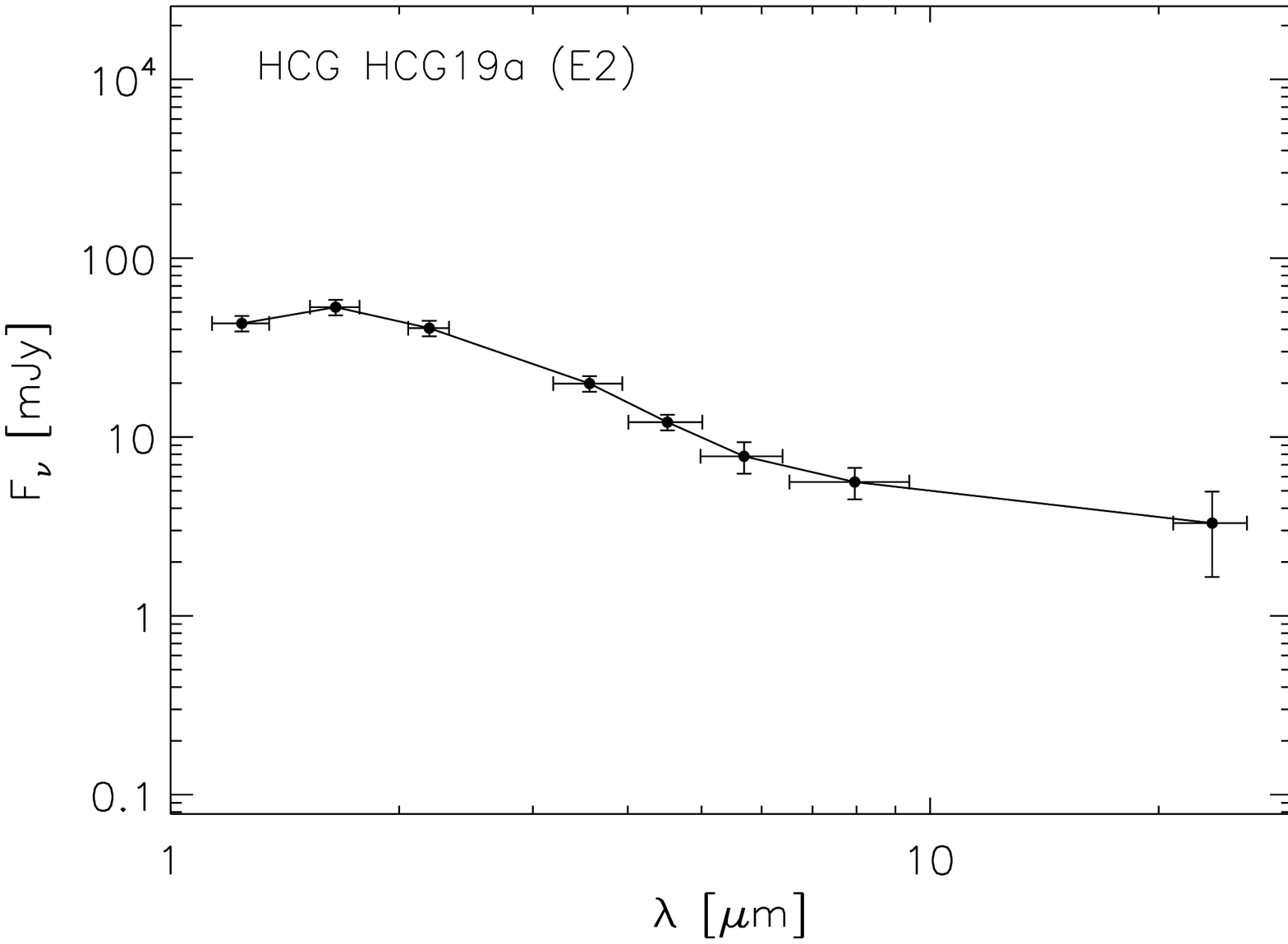}{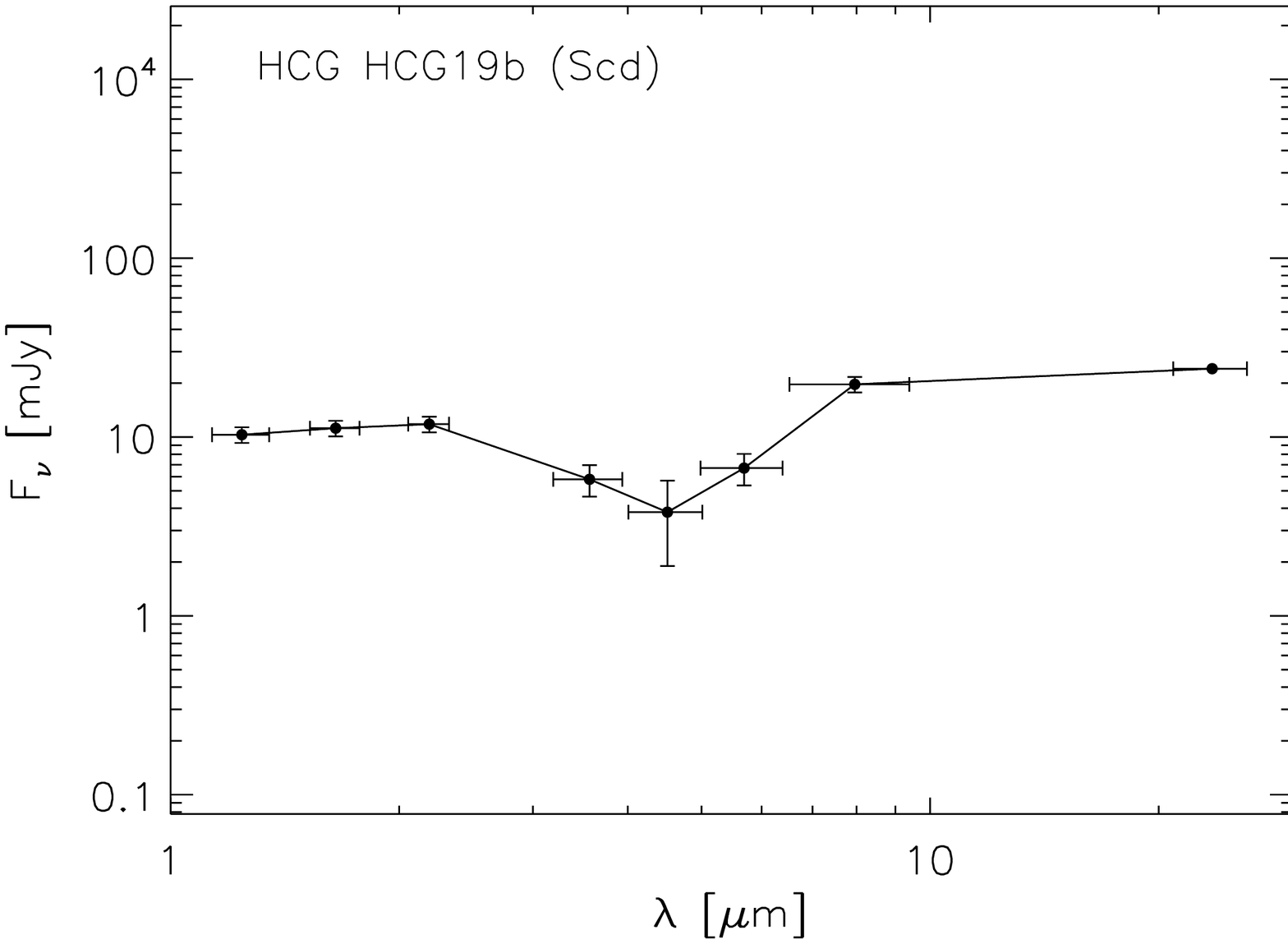}\\
\epsscale{0.46}
\plotone{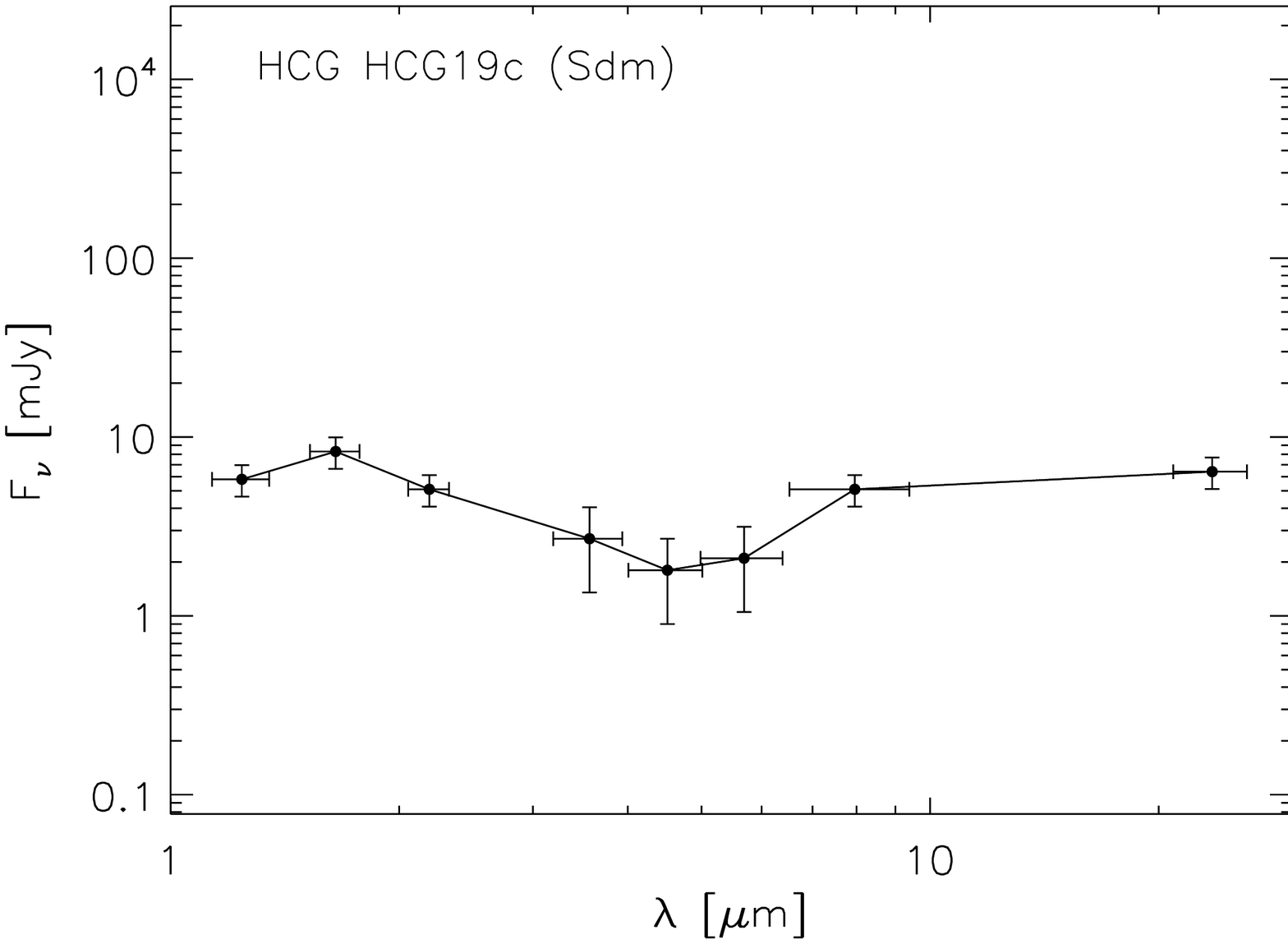}
\caption{Infrared spectral energy distributions for the galaxies
in HCG~19.  Horizontal error bars reflect the filter widths. \label{hcg19}}
\end{figure}

The spectral energy distributions for the galaxies in HCG~19 are shown
in Figure~\ref{hcg19}.  This group is classified as type~II according
to ${\rm log}(M_{{\rm HI}})/{\rm log}(M_{{\rm dyn}})$, indicating a
moderate amount of gas present in the group.  Galaxy~A is dominated by
stellar light.  Galaxy~B has moderate star formation activity
throughout, and hosts a number of compact star forming knots in the
distorted spiral arms.  Galaxy~C is quite diffuse, with a hint of
elevated star formation on its south-west edge (facing the group).
Galaxy~D is a background object.

\subsection{HCG~22}
\begin{figure}
\epsscale{1.0}
\plottwo{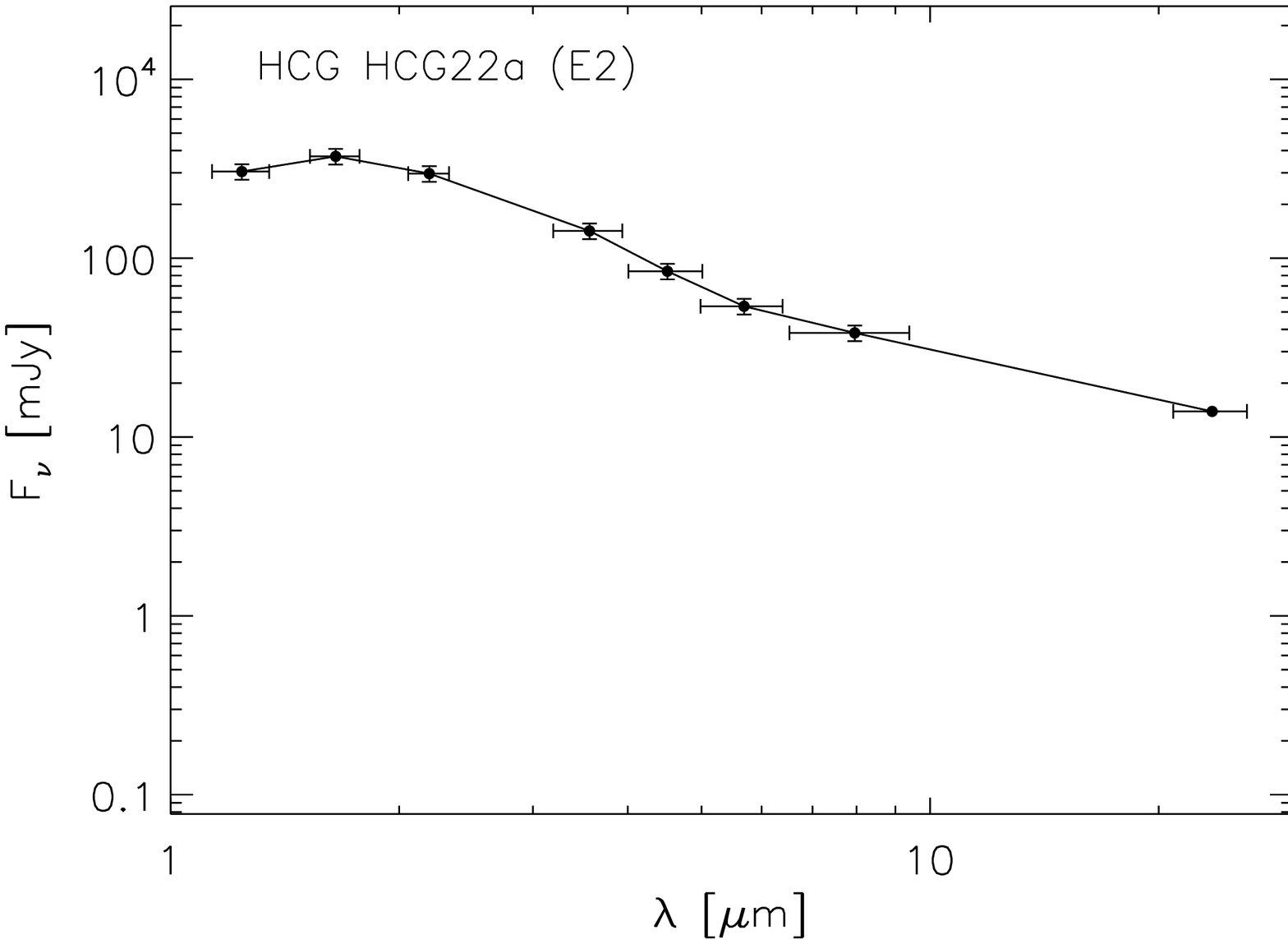}{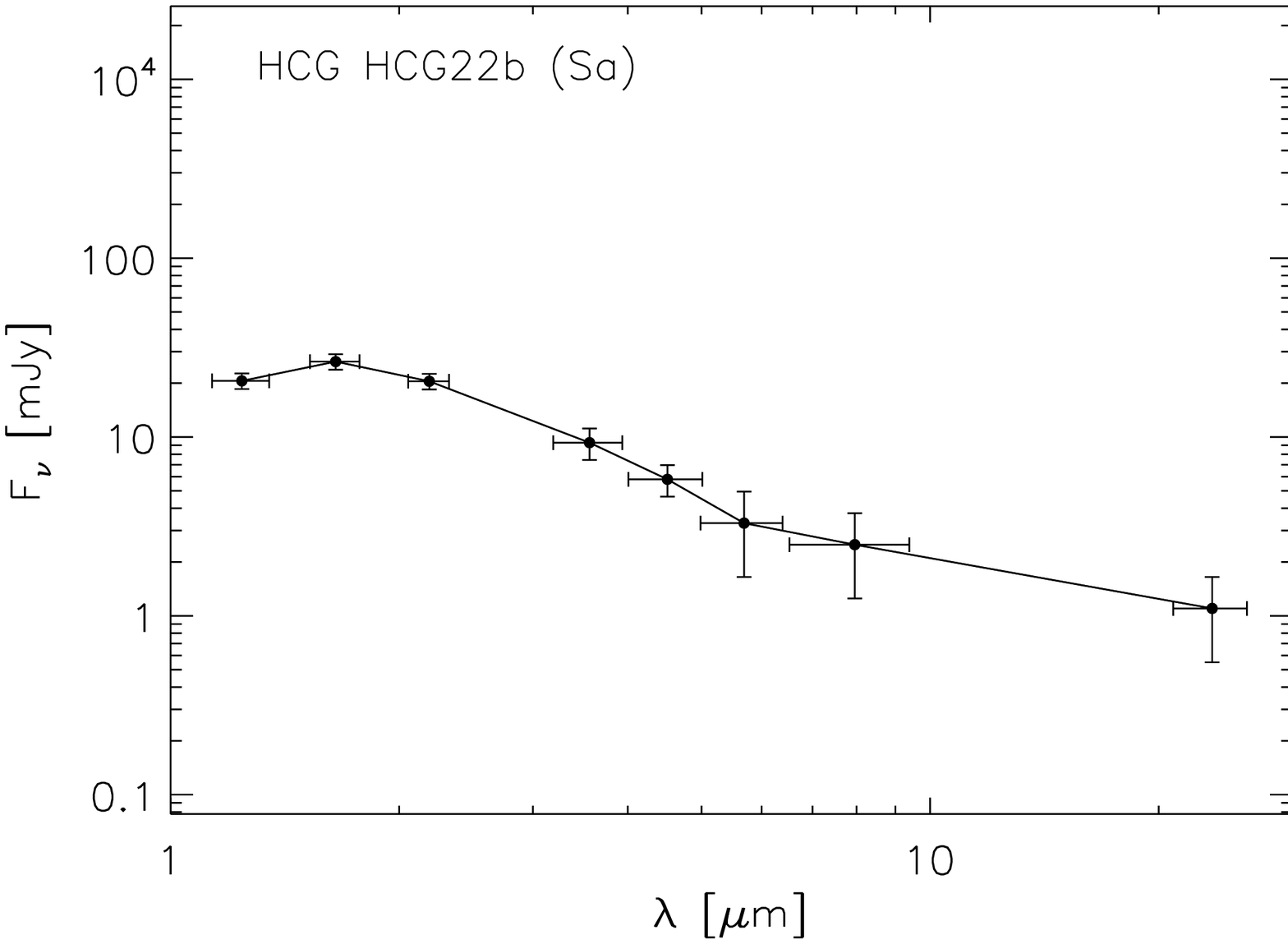}\\
\epsscale{0.46}
\plotone{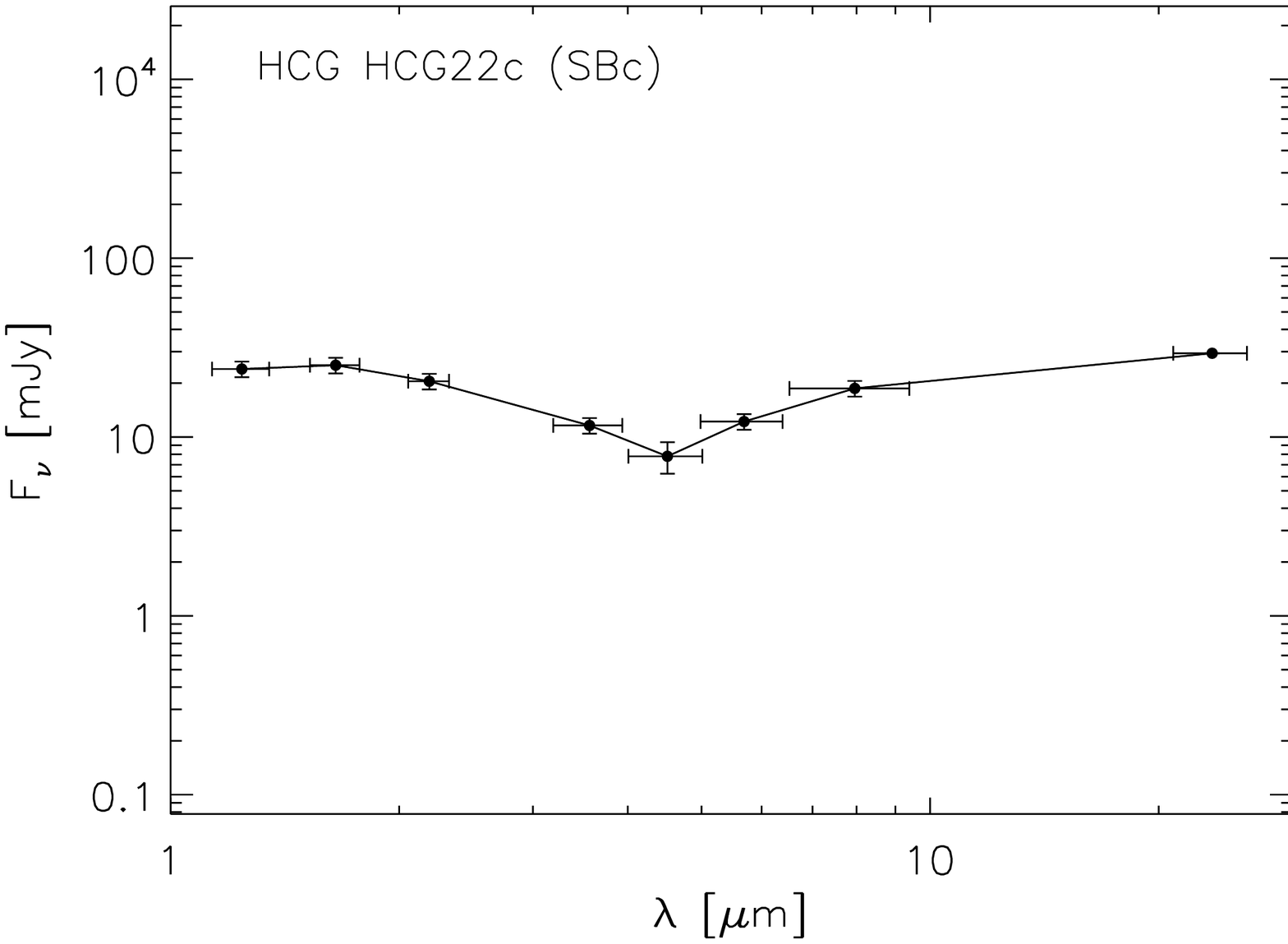}
\caption{Infrared spectral energy distributions for the galaxies
in HCG~22.  Horizontal error bars reflect the filter widths. \label{hcg22}}
\end{figure}

The spectral energy distributions for the galaxies in HCG~22 are shown
in Figure~\ref{hcg22}.  This group is classified as type~II according
to ${\rm log}(M_{{\rm HI}})/{\rm log}(M_{{\rm dyn}})$, indicating a
moderate amount of gas present in the group.  HCG~22 may be associated
with a larger structure \citep{rood94}, although \citet{ribeiro98}
identify it as a ``real'' compact group.  Galaxies~A (NGC~1199) and B
(NGC~1190) are dominated by stellar light with little long wavelength
emission.  Galaxy~C (NGC~1189) has extended clumpy star formation
throughout its spiral arms with remarkably little associated stellar
light, which is striking in the color images.  Galaxies~D (NGC~1191)
and E (NGC~1192) are background objects.

\subsection{HCG~31}
\begin{figure}
\epsscale{1.0}
\plottwo{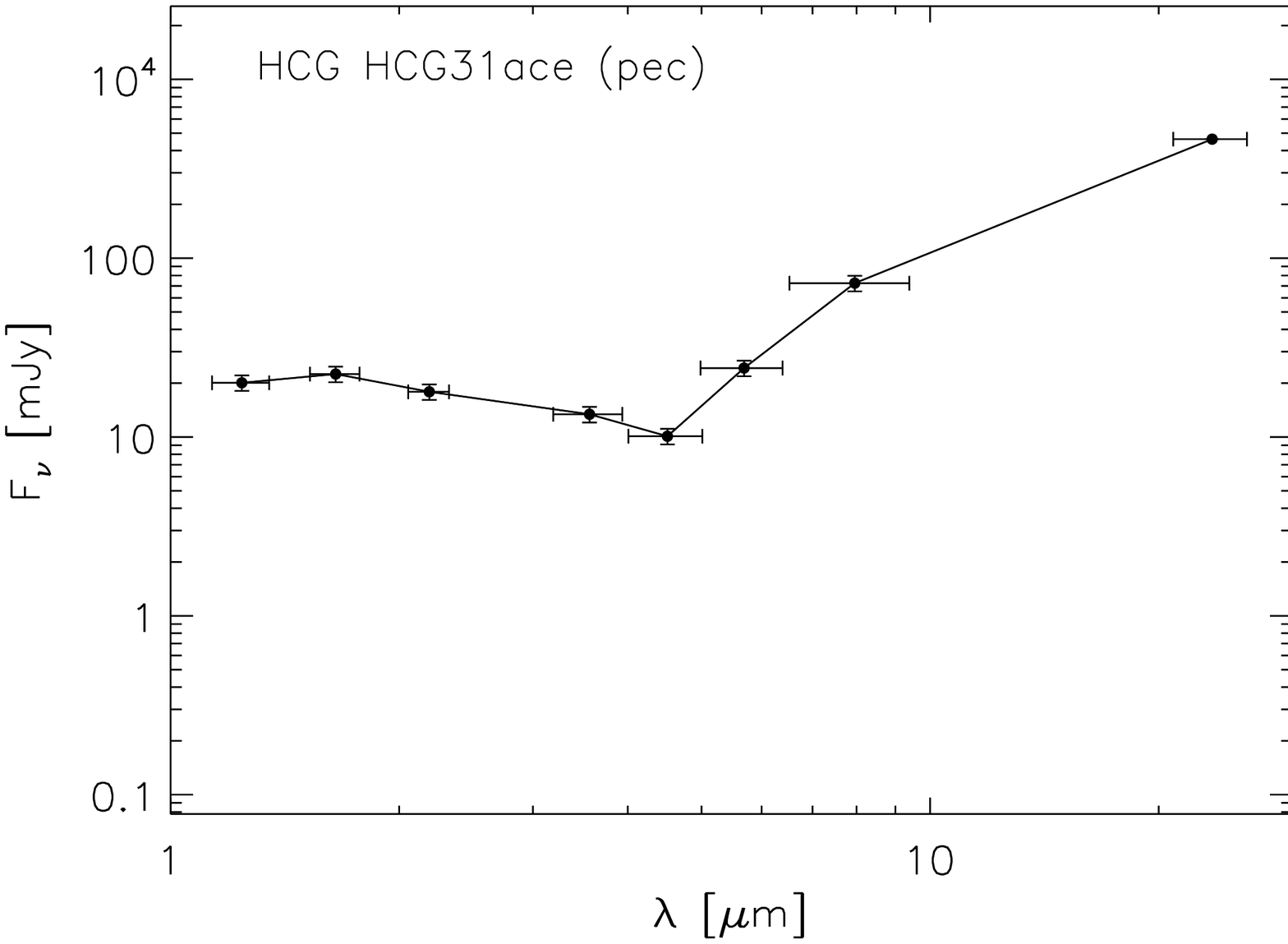}{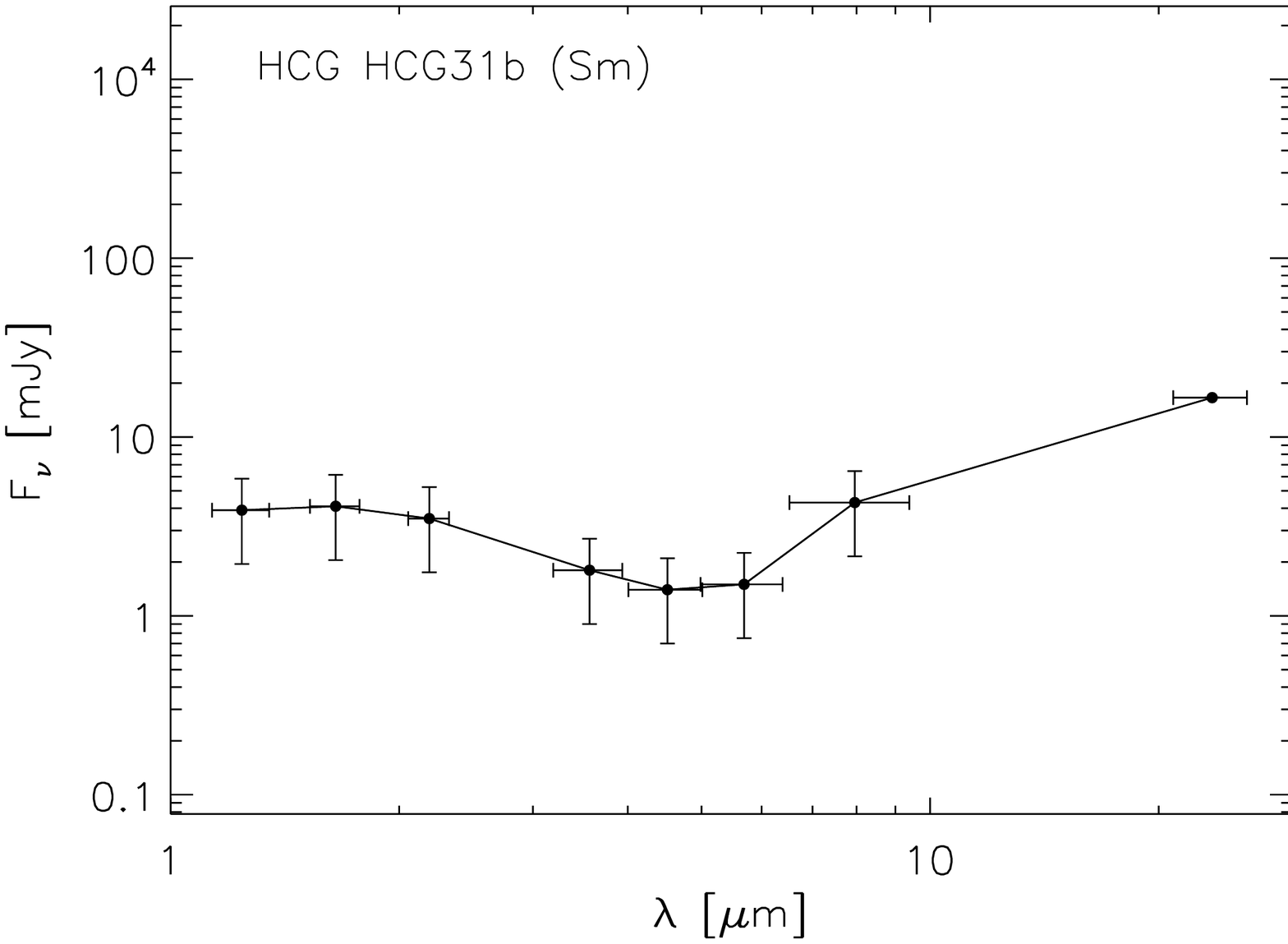}\\
\plottwo{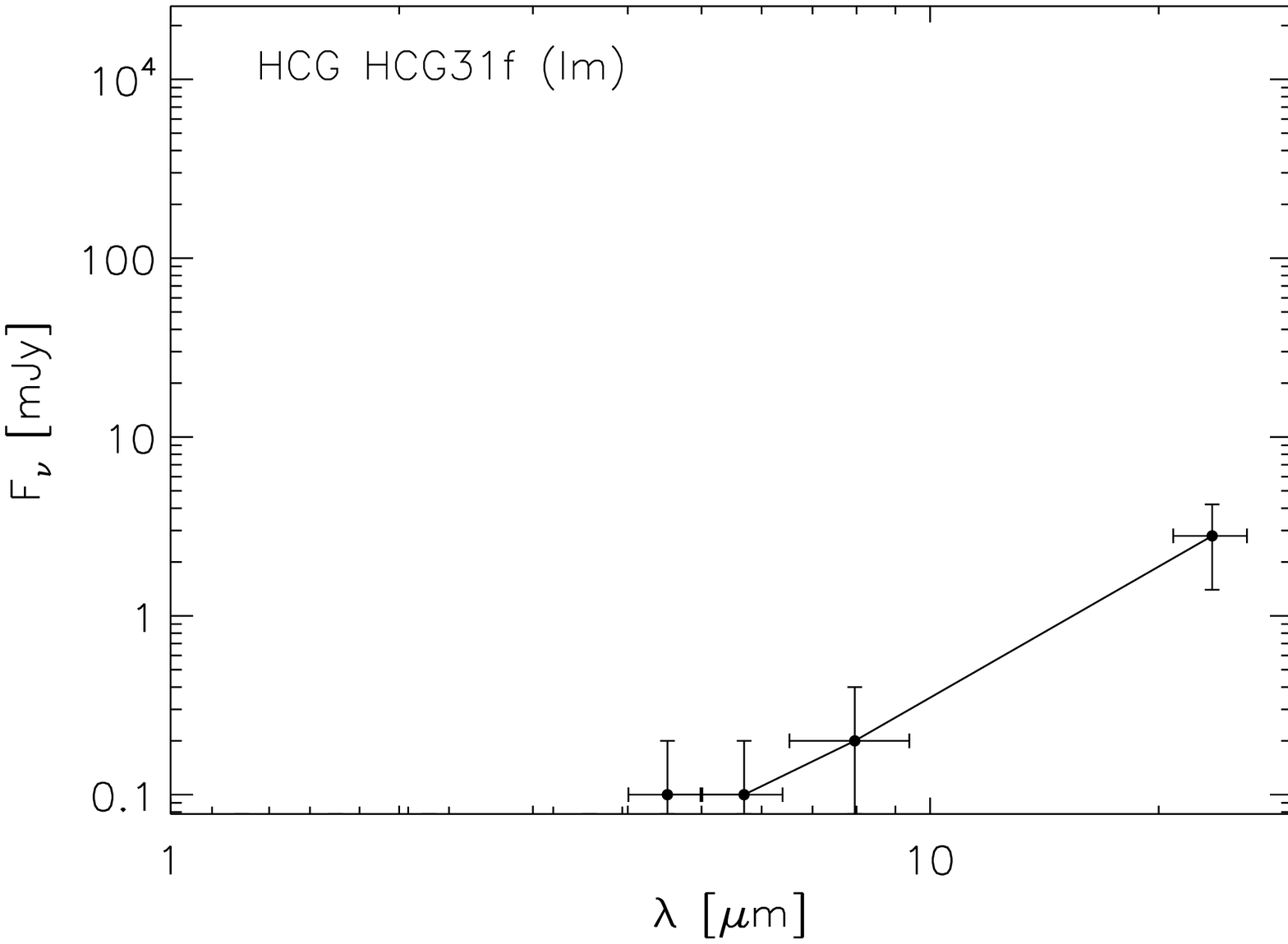}{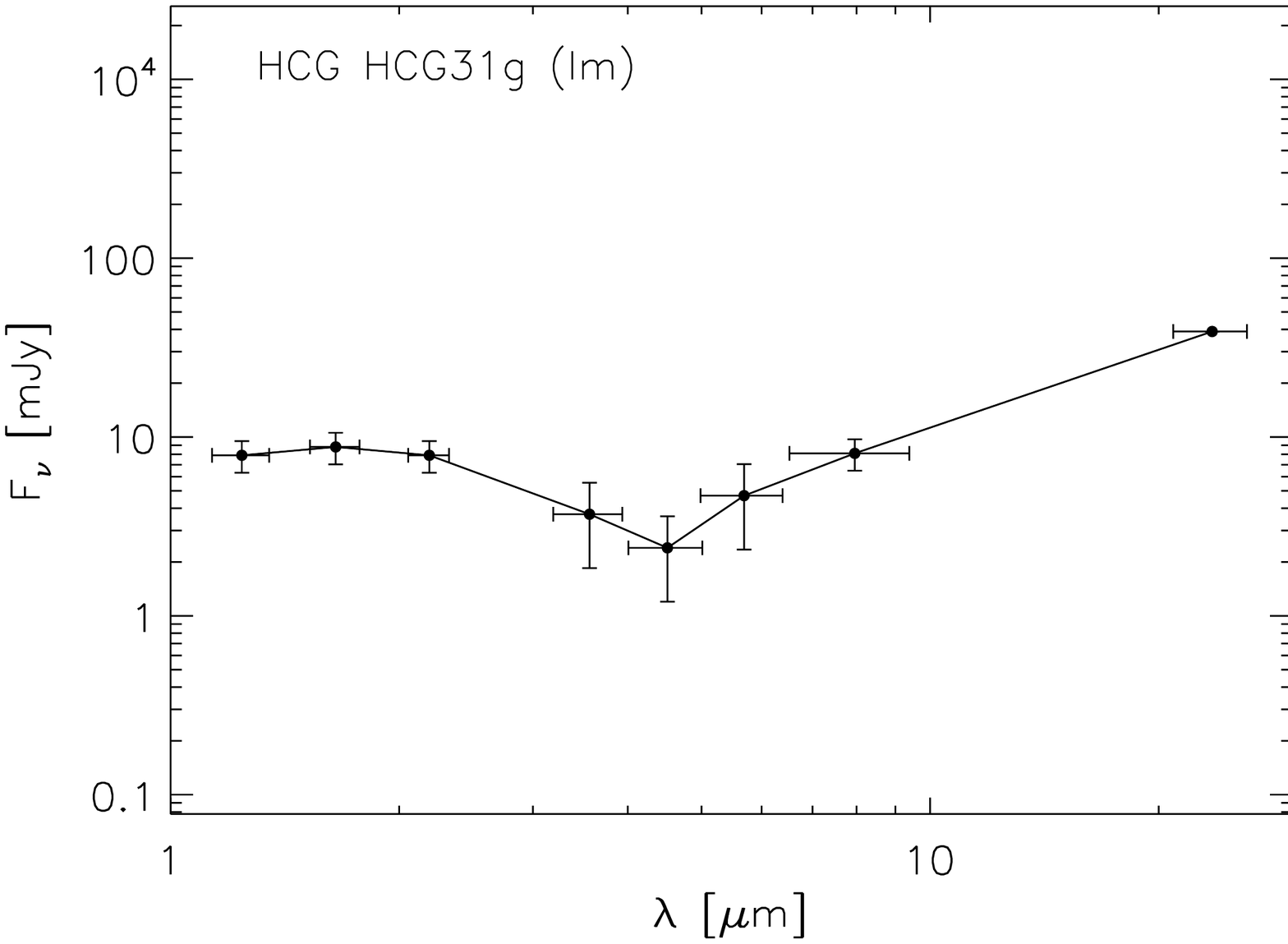}\\
\epsscale{0.46}
\plotone{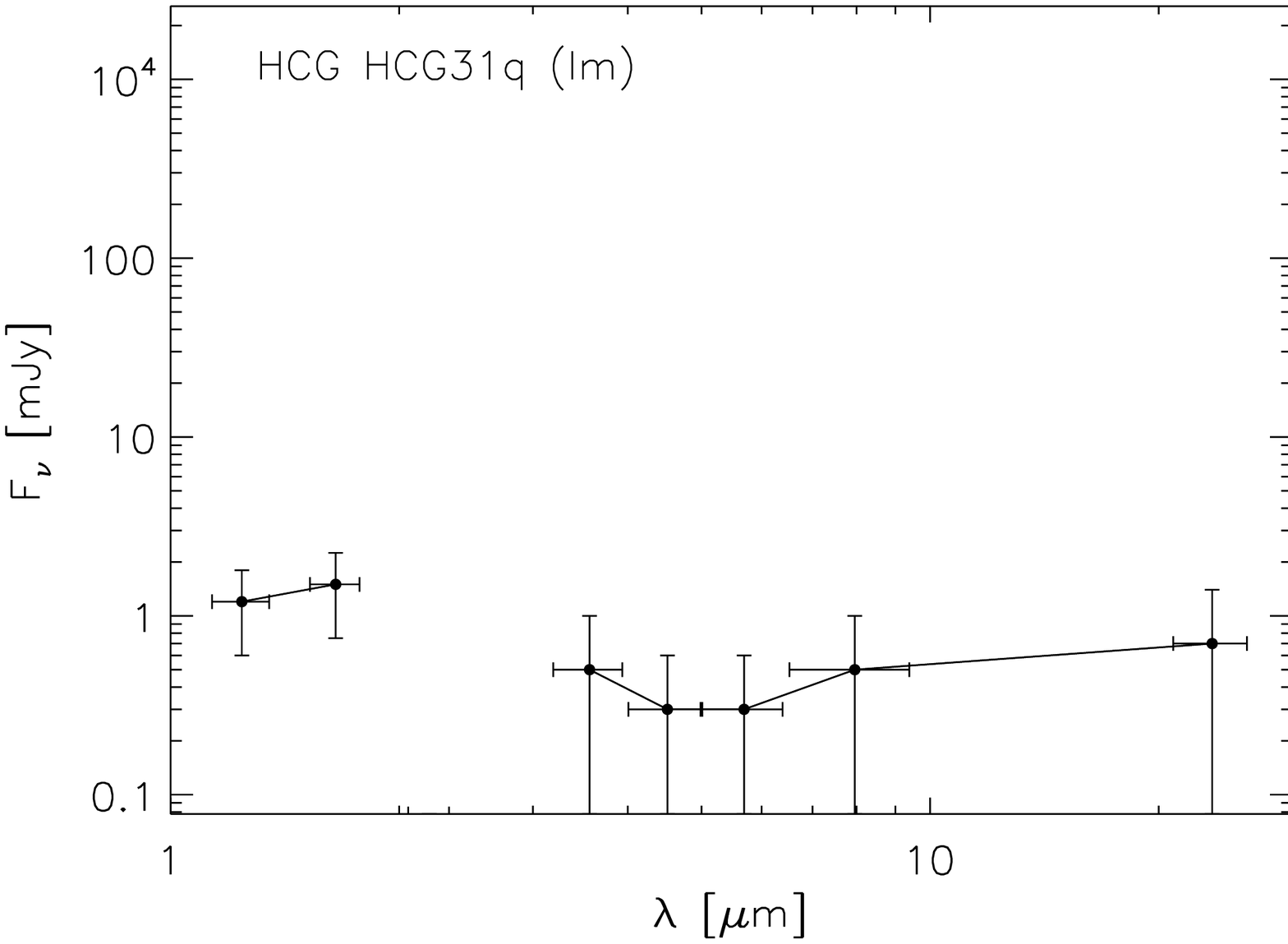}
\caption{Infrared spectral energy distributions for the galaxies
in HCG~31.  Horizontal error bars reflect the filter widths. \label{hcg31}}
\end{figure}

The spectral energy distributions for the galaxies in HCG~31 are shown
in Figure~\ref{hcg31}.  This group is classified as type~I according
to ${\rm log}(M_{{\rm HI}})/{\rm log}(M_{{\rm dyn}})$, indicating it
is relatively gas-rich.  HCG~31 is a highly disturbed with elevated
levels of infrared emission throughout most the group.  Due to the
close interaction between galaxies~A, C, and E, they were not
separable at low contour levels and their photometry was combined into
a single data point. Galaxy~F was not detected in the 2MASS images
at J,H, and K-bands, and galaxy~Q was not detected at K-band.
Galaxies~A and C (collectively NGC~1741, MRK~1089) are undergoing the
strongest interaction, and the location of their collision is a strong
24$\mu$m source.  Galaxy~B has moderate infrared emission clumped in
on the edges of its disk and nuclear region.  Galaxy~D is a background
source.  Galaxies~E and F are possible tidal dwarfs, and have moderate
levels of star formation.  Galaxy~F, in particular, is relatively
luminous at 24$\mu$m given its small size and faint near-IR and
optical emission.  Galaxy~G (MRK~1090) has several star forming knots
with moderate infrared emission on its north-west edge (facing the
group).  These results are generally consistent with conclusions based
on optical observations of this group \citep[e.g.][]{johnson99,
johnson00}.  Galaxy~Q has been recently classified as a group member
\citep{verdes-montenegro05}, but appears to be dominated by stellar
light and is not particularly disturbed in these infrared images.

\subsection{HCG~42}
\begin{figure}
\epsscale{1.0}
\plottwo{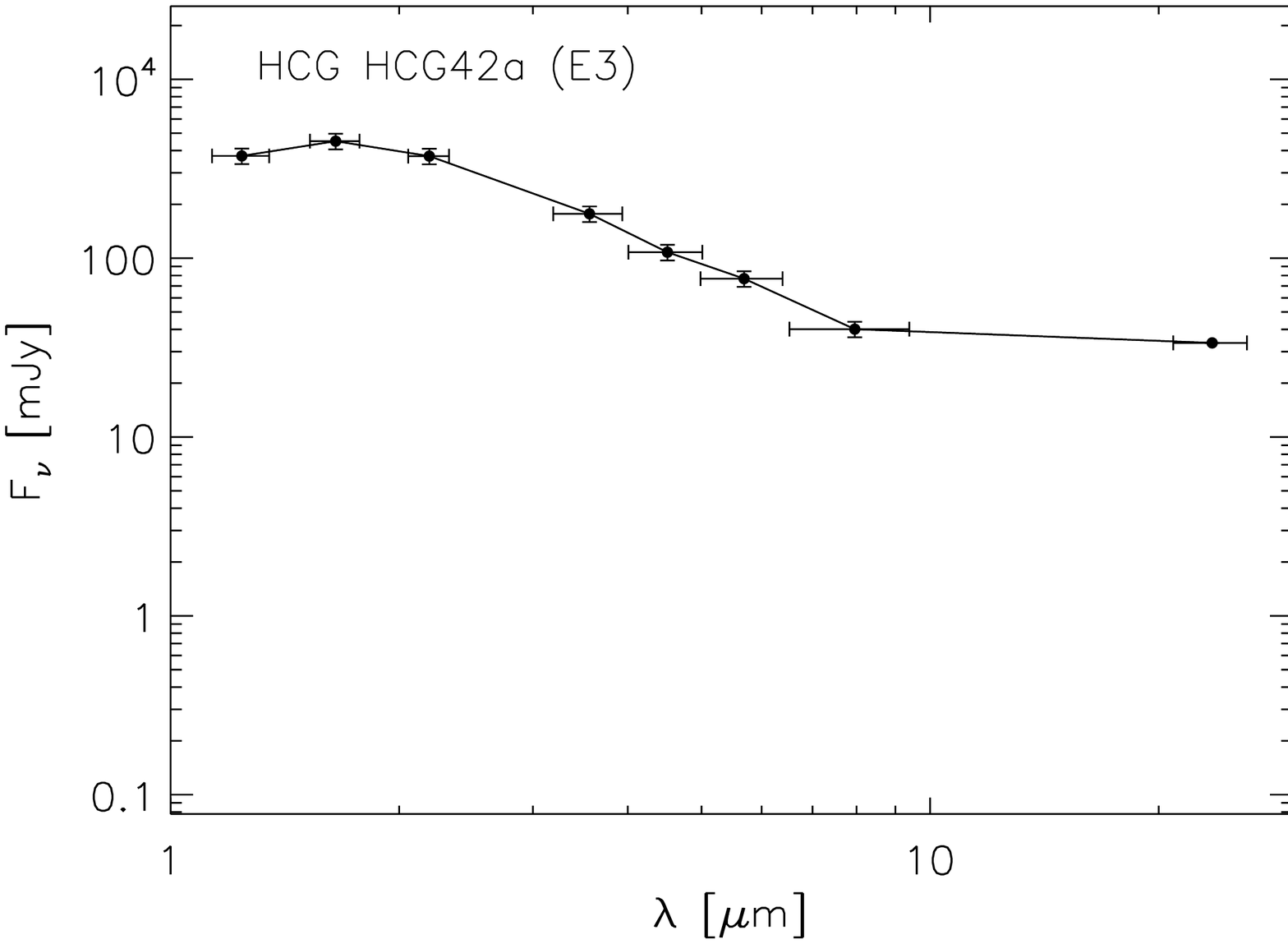}{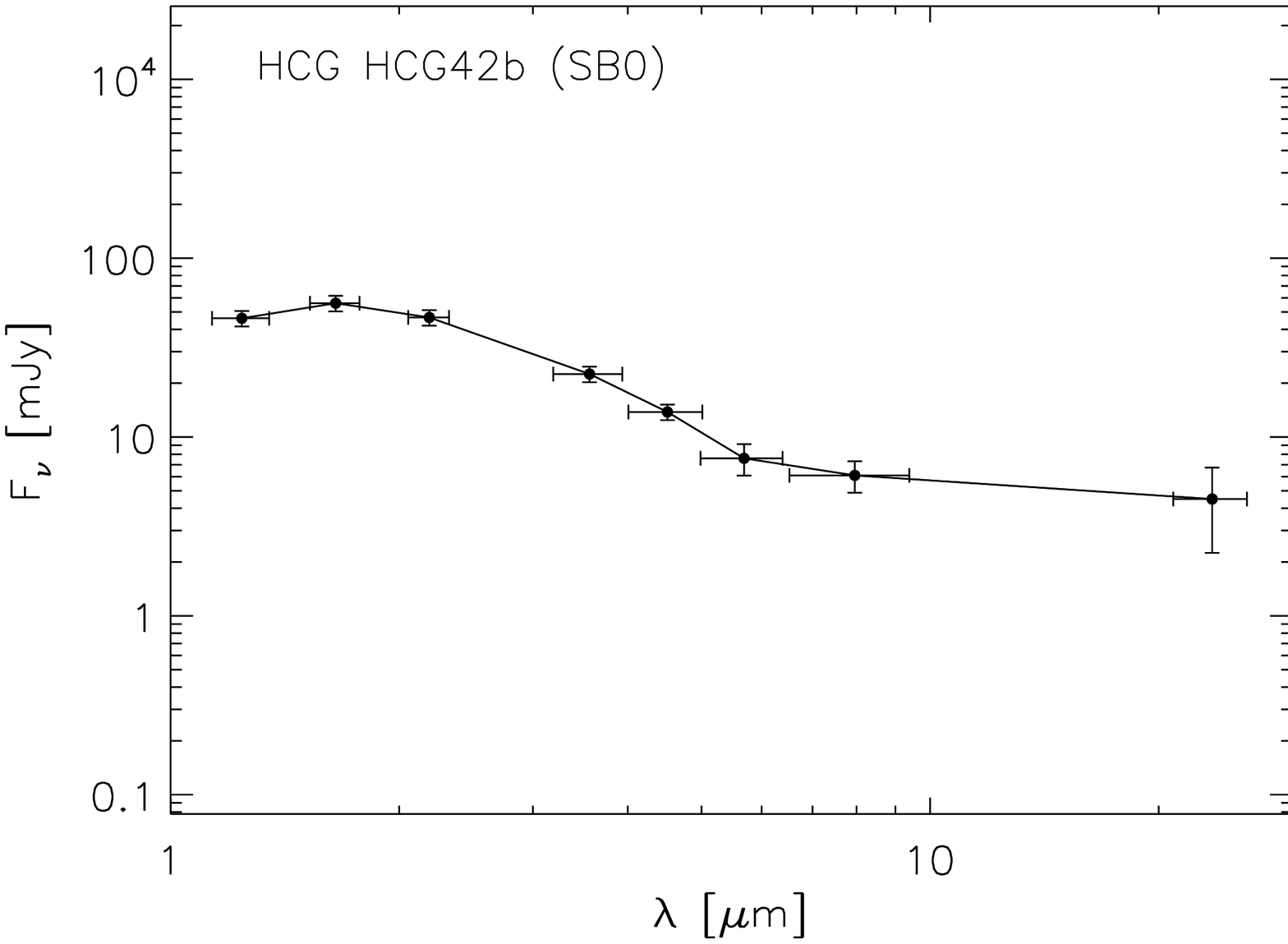}\\
\plottwo{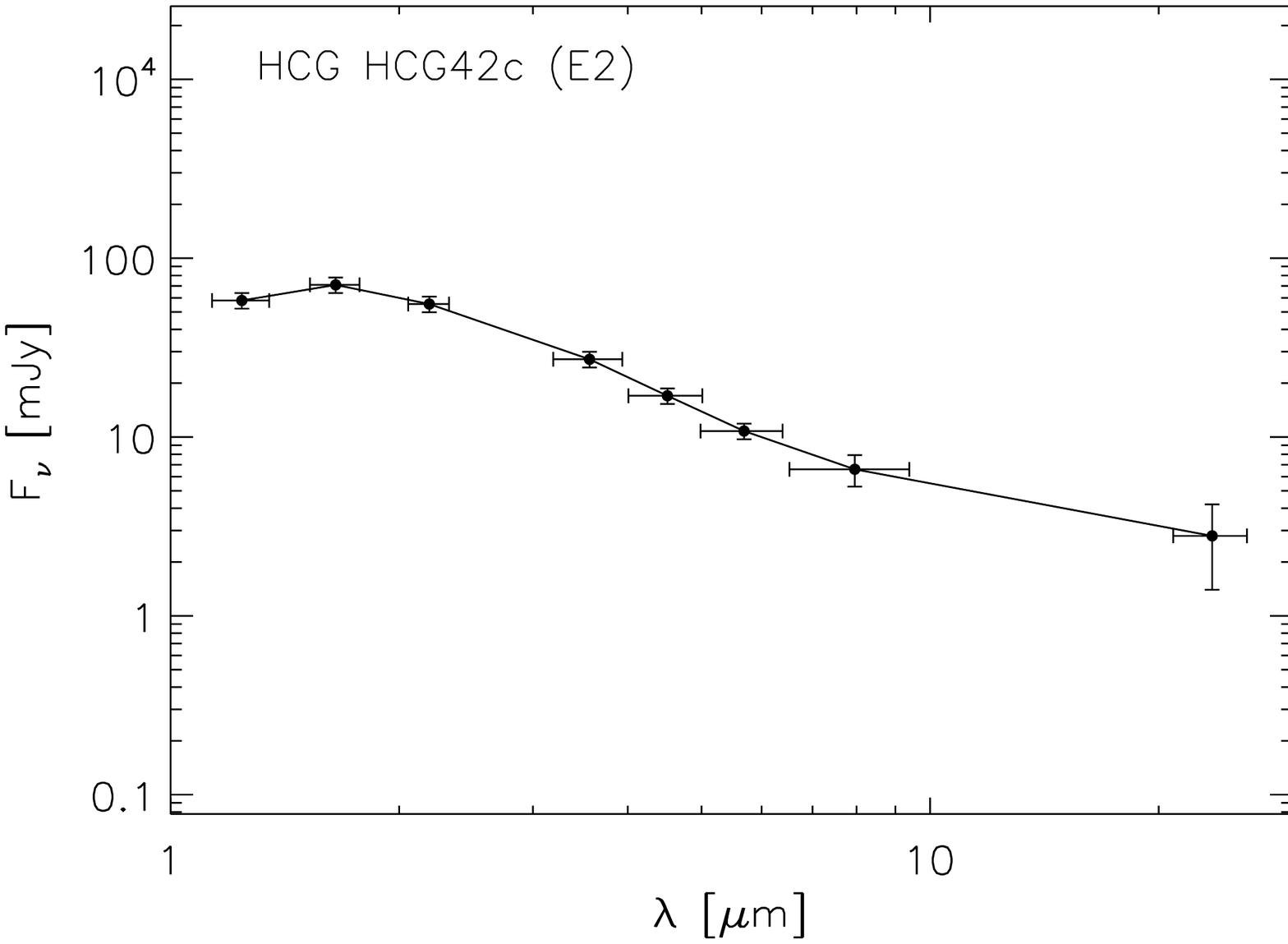}{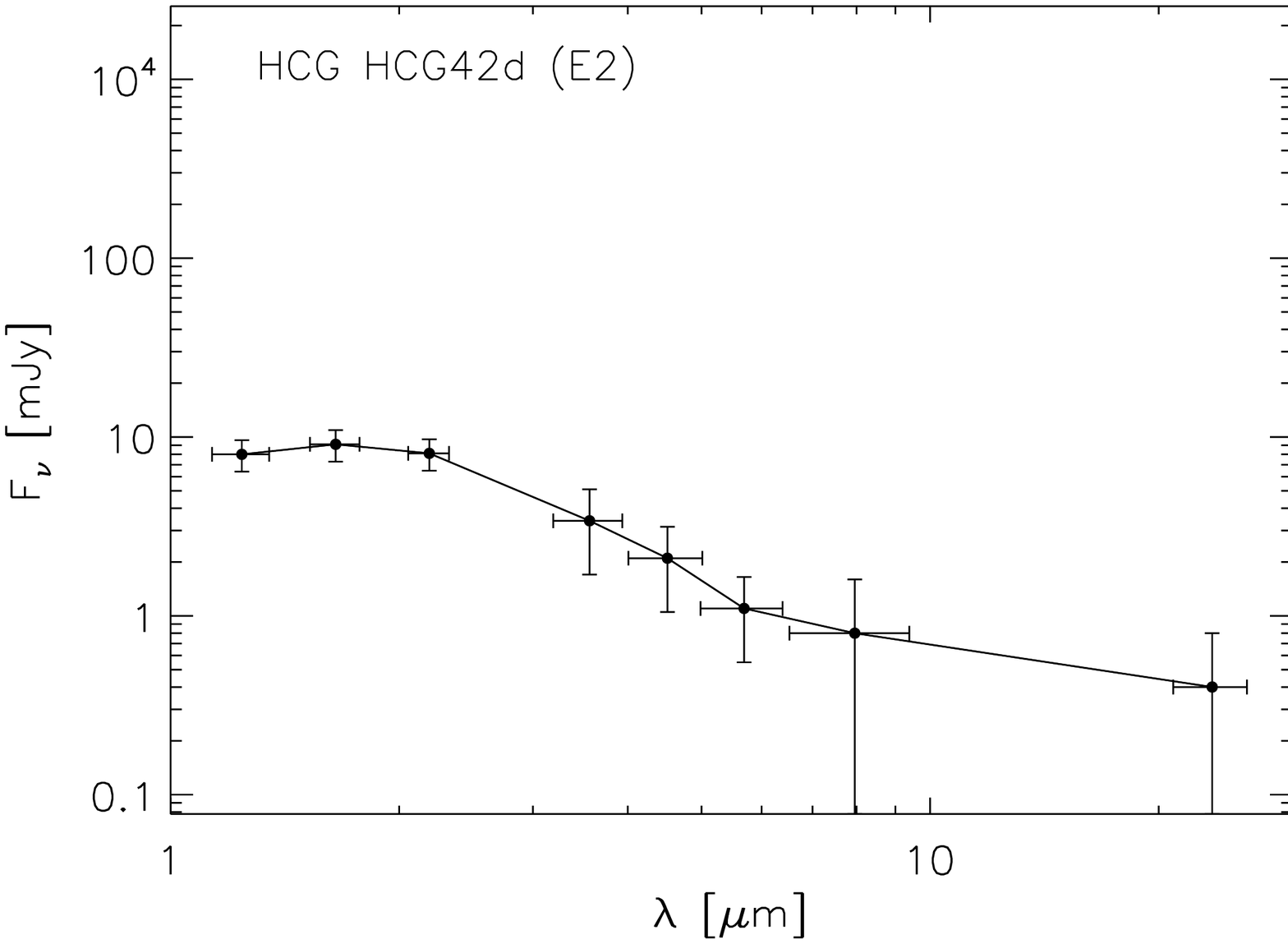}
\caption{Infrared spectral energy distributions for the galaxies
in HCG~42.  Horizontal error bars reflect the filter widths. \label{hcg42}}
\end{figure}

The spectral energy distributions for the galaxies in HCG~42 are shown
in Figure~\ref{hcg42}.  This group is classified as type~III according
to ${\rm log}(M_{{\rm HI}})/{\rm log}(M_{{\rm dyn}})$, indicating that
the group is relatively gas-poor.  \citet{ribeiro98} identify HCG~42
as being part of a loose group.  Galaxy~A is also known as NGC~3091,
and galaxy~B is also known as NGC~3096.  The galaxies in HCG~42 are
all dominated by stellar light with little evidence for recent star
formation activity. The lack of 24$\mu$m emission from any of the
group members suggests little AGN activity as well.  A number of
luminous infrared sources are apparent in the background of this
group.

\subsection{HCG~48}
\begin{figure}
\epsscale{1.0}
\plottwo{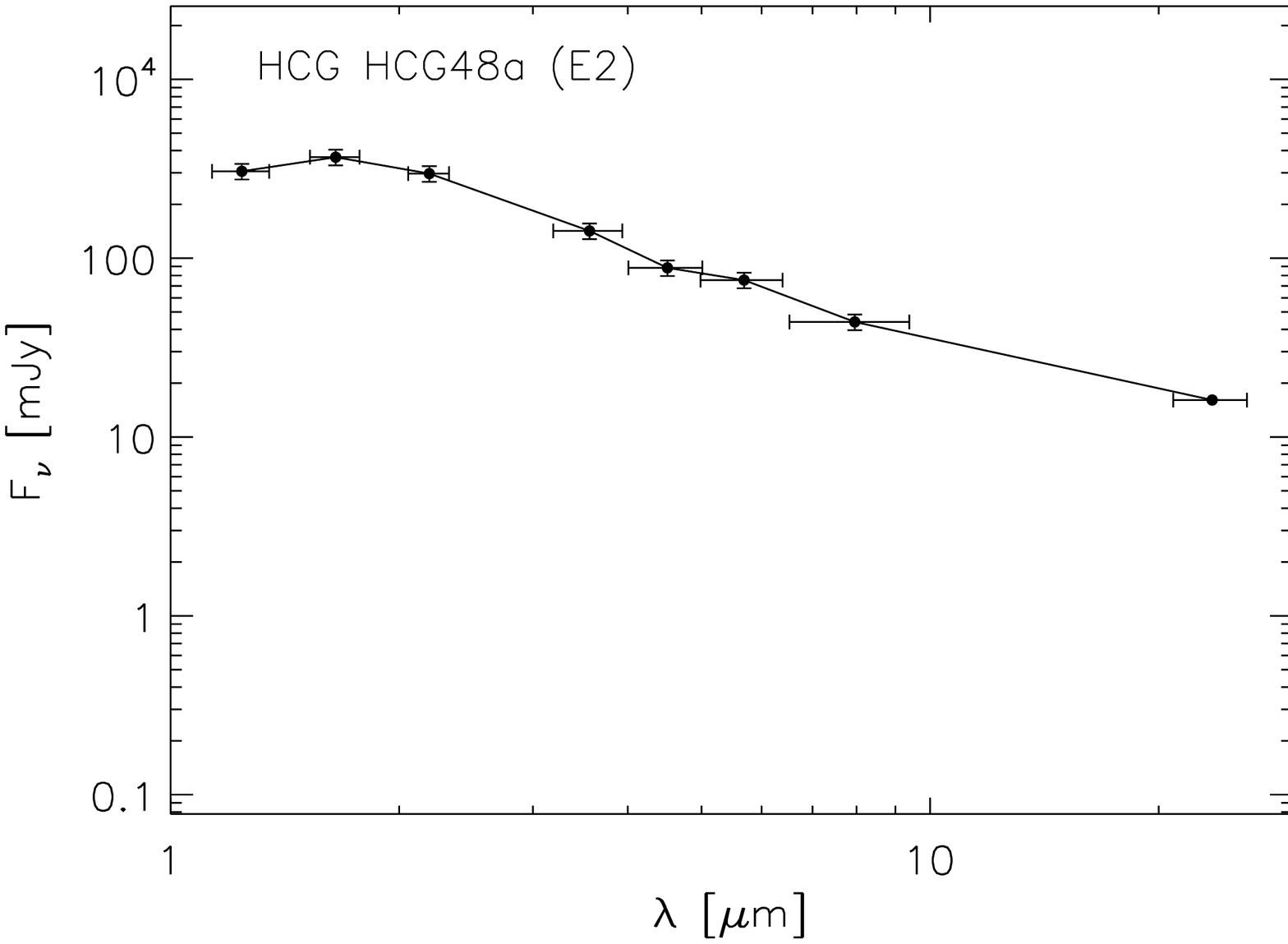}{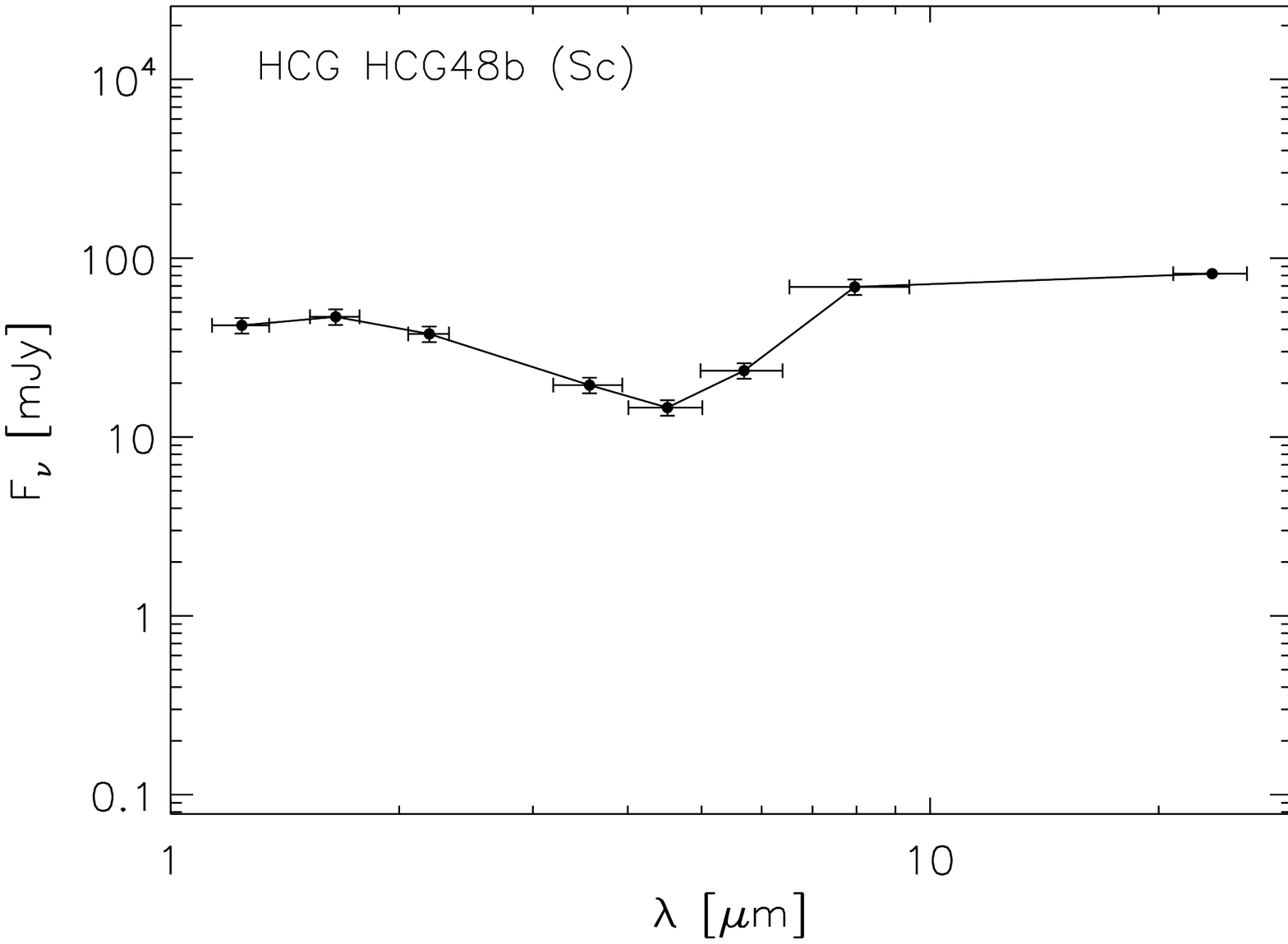}\\
\plottwo{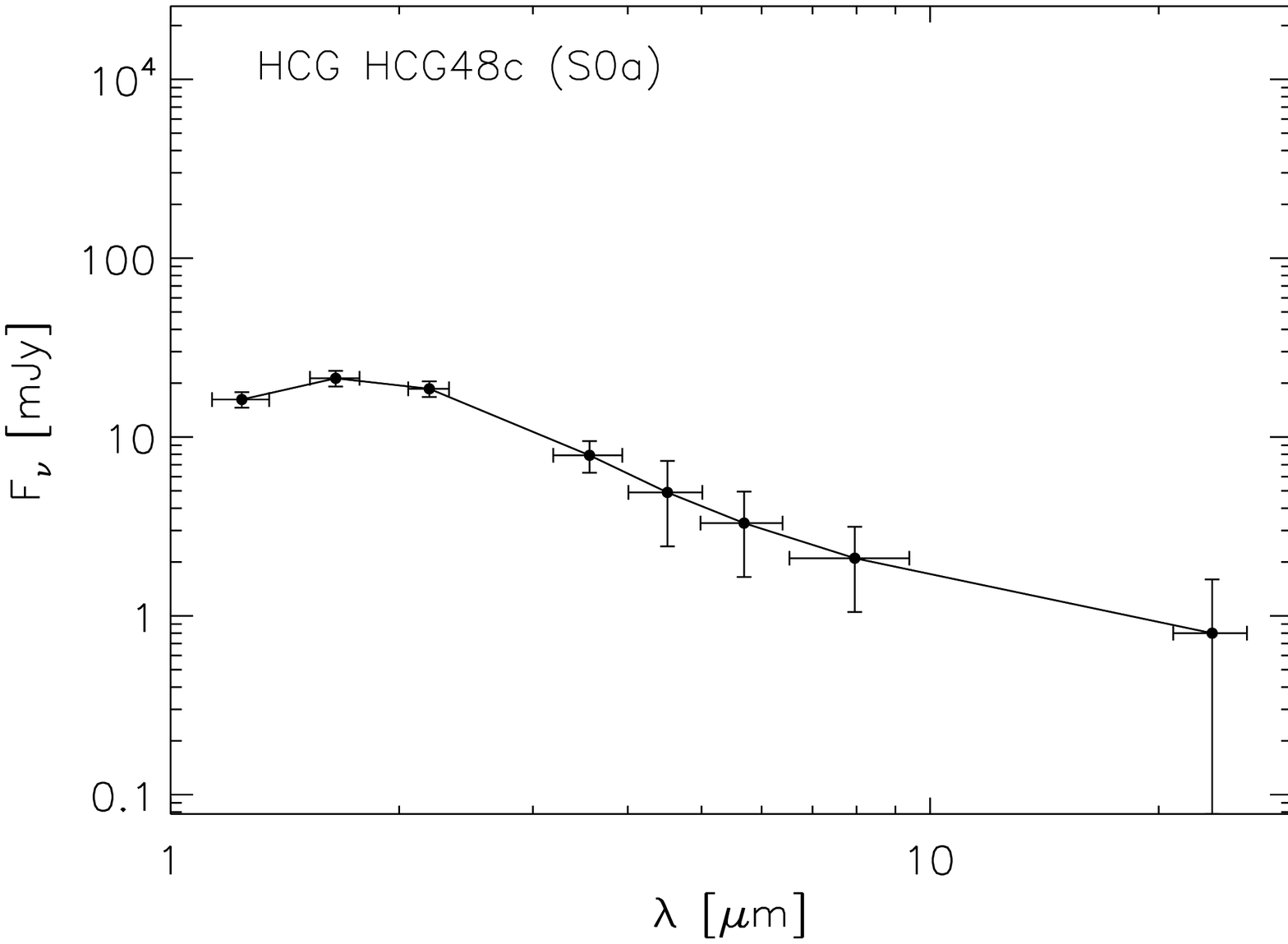}{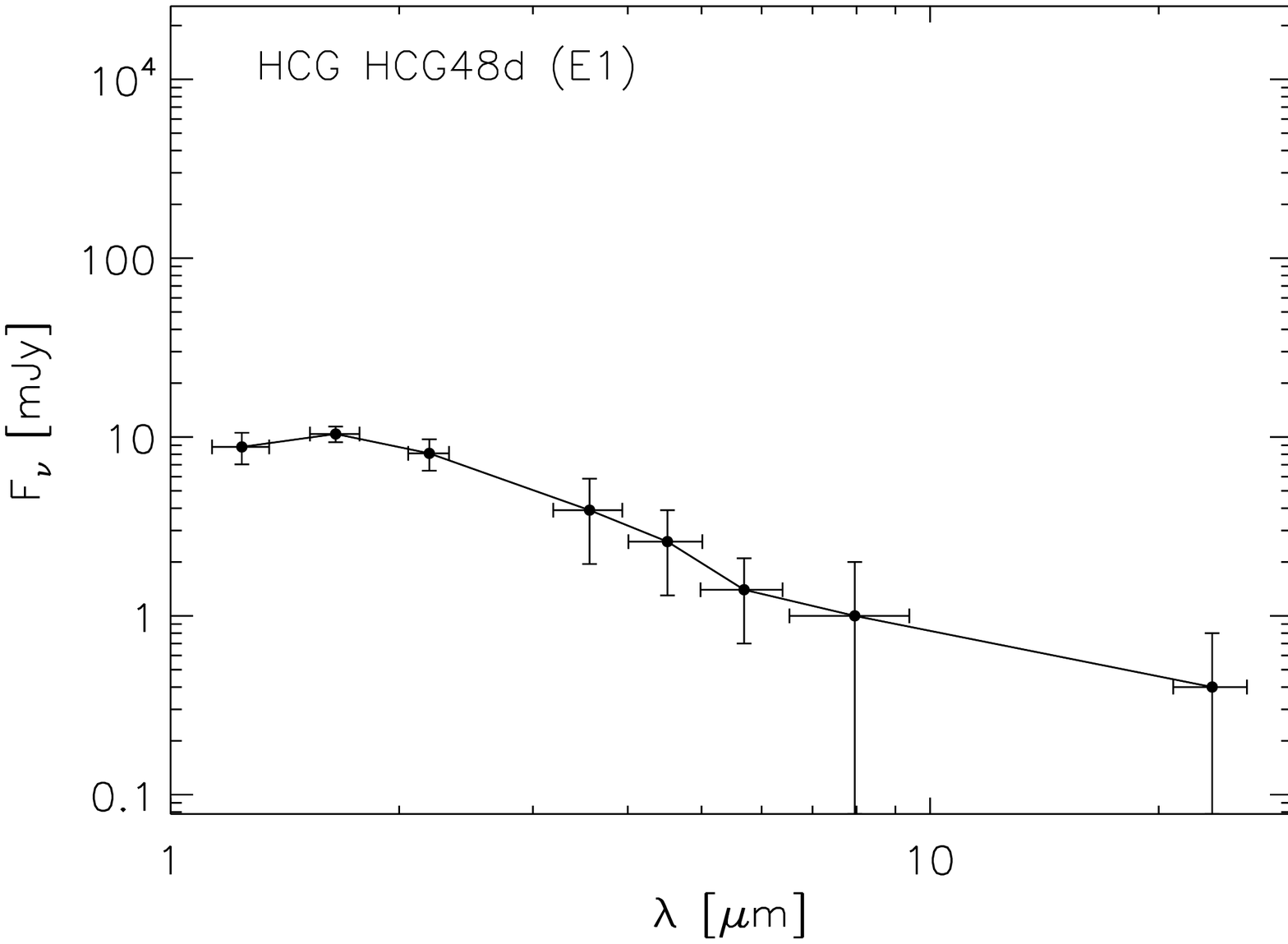}
\caption{Infrared spectral energy distributions for the galaxies in
HCG~48.  Note that galaxies~B and C may not be proper group members.
Horizontal error bars reflect the filter widths. \label{hcg48}}
\end{figure}

The spectral energy distributions for the galaxies in HCG~48 are shown
in Figure~\ref{hcg48}.  This group is classified as type~III according
to ${\rm log}(M_{{\rm HI}})/{\rm log}(M_{{\rm dyn}})$, indicating that
the group is relatively gas-poor.  HCG~48 is embedded in the larger
galaxy group Abell~1060, and the kinematical analysis of
\citet{ribeiro98} indicates that this group is divided into three sets
of objects at slightly different velocities; galaxies~A and D are
associated, but galaxies~B and C appear to be in different subgroups,
in which case the reality of HCG~48 as a compact group is in
question. We include these photometric measurements of galaxies~A, B,
C, and D in Table~\ref{fluxes} for completeness, but exclude the data
points from galaxies~B and C in Figures~\ref{plotIRAC1}-\ref{plotJHK2}.
Galaxies~A, C, and D are dominated by stellar light with little sign
of recent star formation or AGN activity.  Galaxy~B has infrared
emission from throughout the galaxy and moderate 24$\mu$m emission
associated with the nuclear region.

\subsection{HCG~59}
\begin{figure}
\epsscale{1.0}
\plottwo{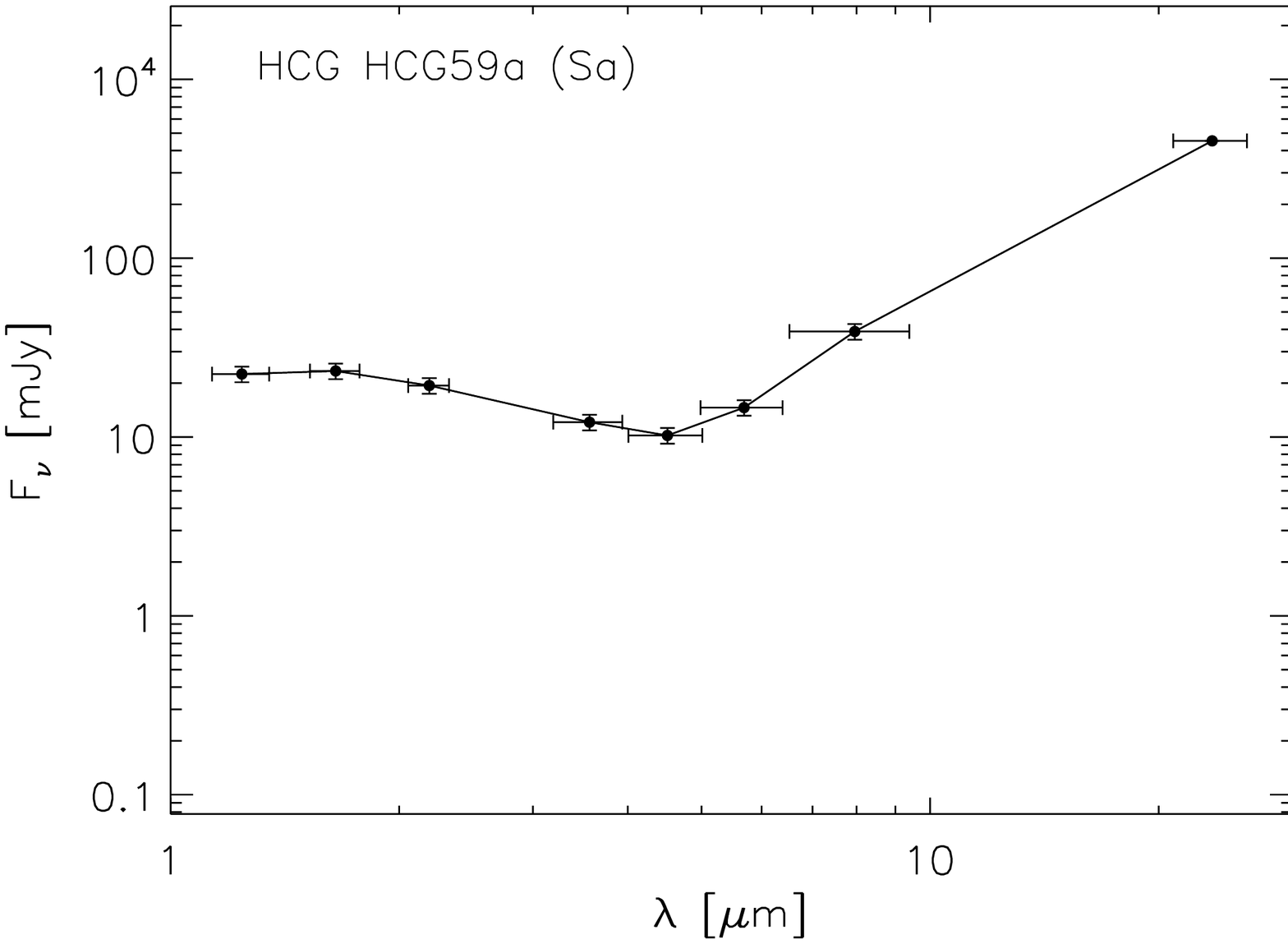}{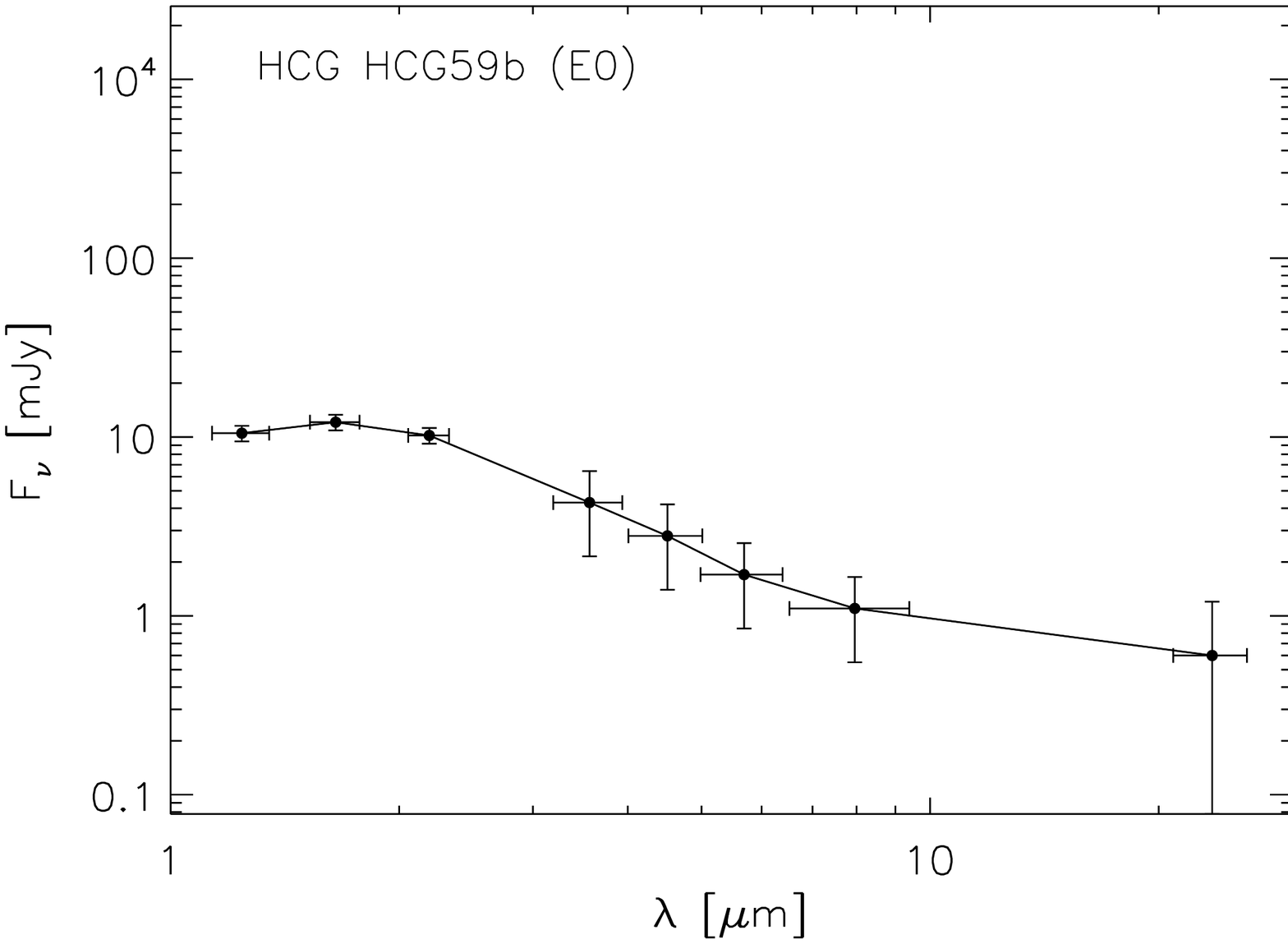}\\
\plottwo{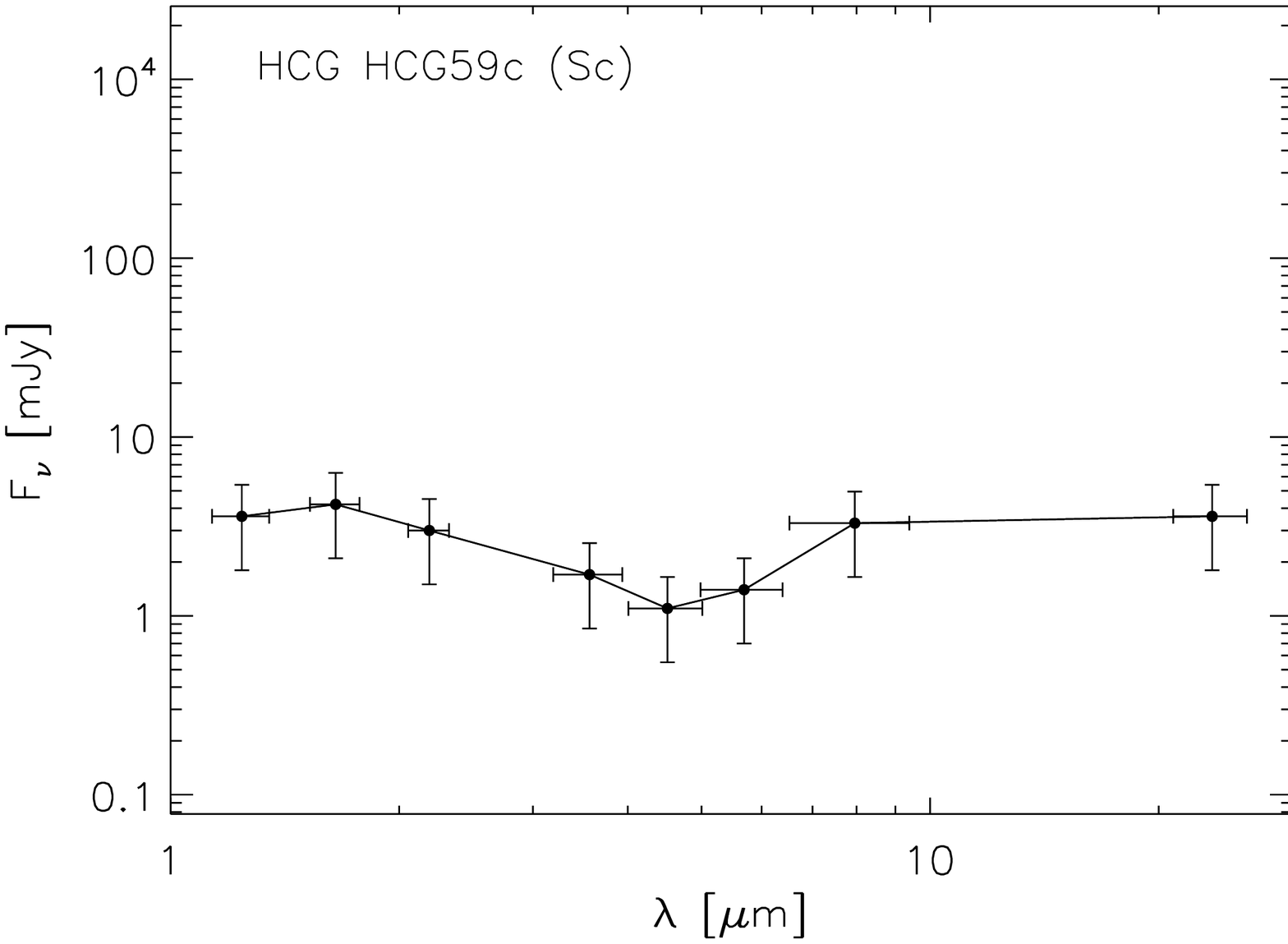}{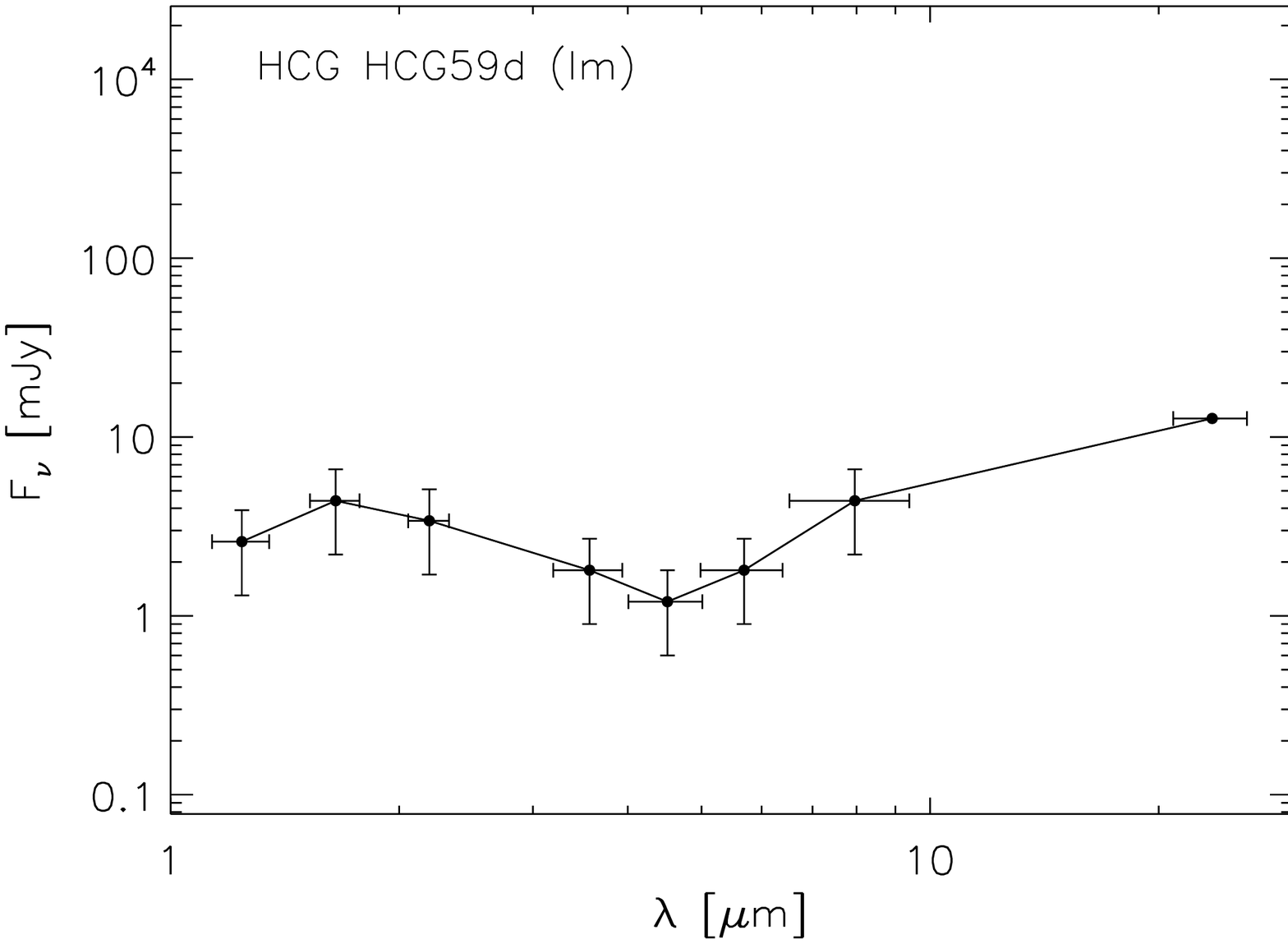}
\caption{Infrared spectral energy distributions for the galaxies
in HCG~59.  Horizontal error bars reflect the filter widths. \label{hcg59}}
\end{figure}

The spectral energy distributions for the galaxies in HCG~59 are shown
in Figure~\ref{hcg59}.  This group is classified as type~II according
to ${\rm log}(M_{{\rm HI}})/{\rm log}(M_{{\rm dyn}})$, indicating a
moderate amount of gas present in the group.  Galaxy~A (IC~737) is
point-like and quite luminous in the thermal infrared.
Galaxy~B (IC~736) is dominated by stellar light.  Galaxy~C has
moderate infrared emission and appears to be slightly disturbed.
Galaxy~D is diffuse and highly disturbed, with clumps of infrared
emission on its southern and northern edges.  Galaxy~E is a background
object.

\subsection{HCG~61}
\begin{figure}
\epsscale{1.0}
\plottwo{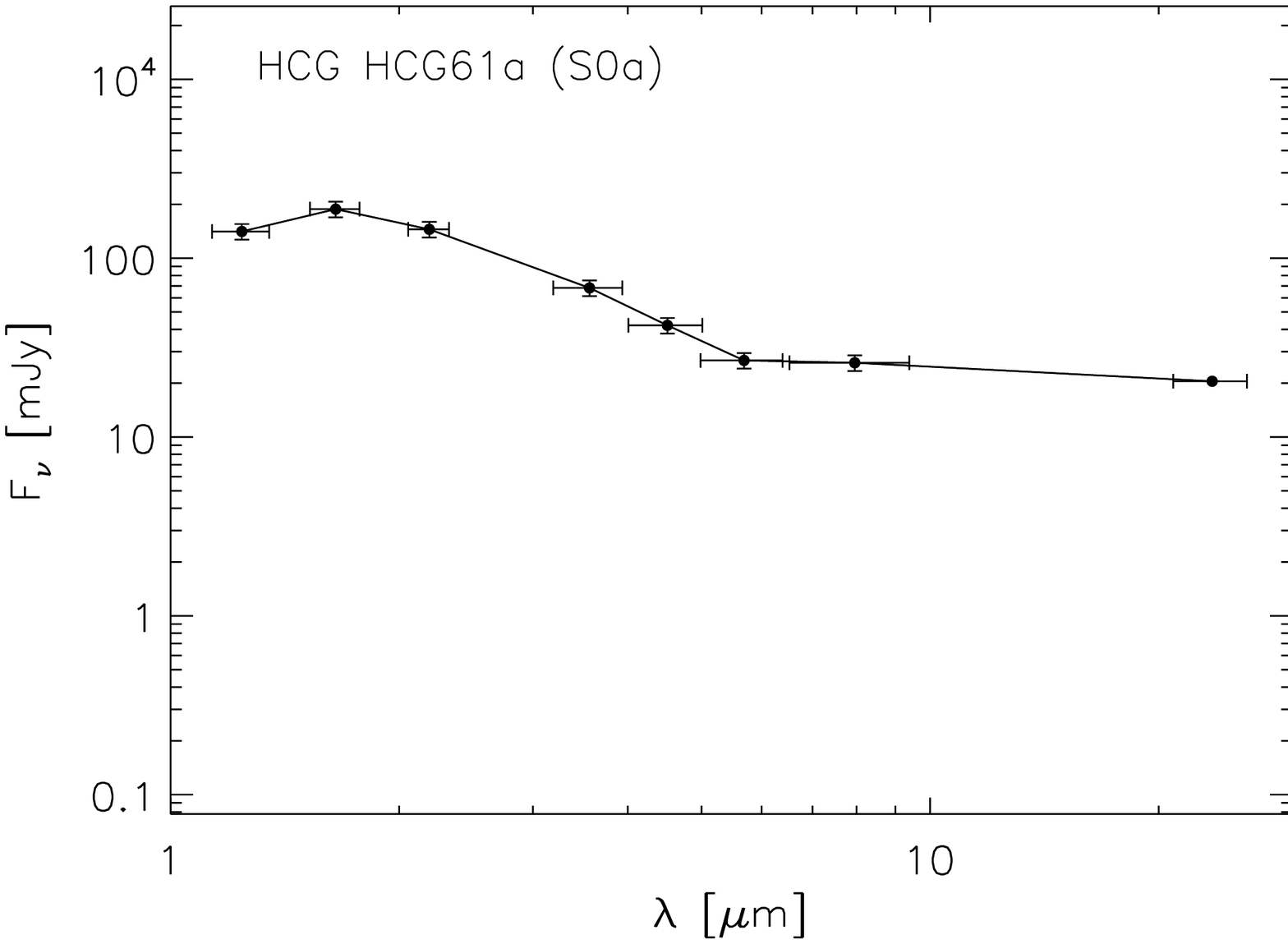}{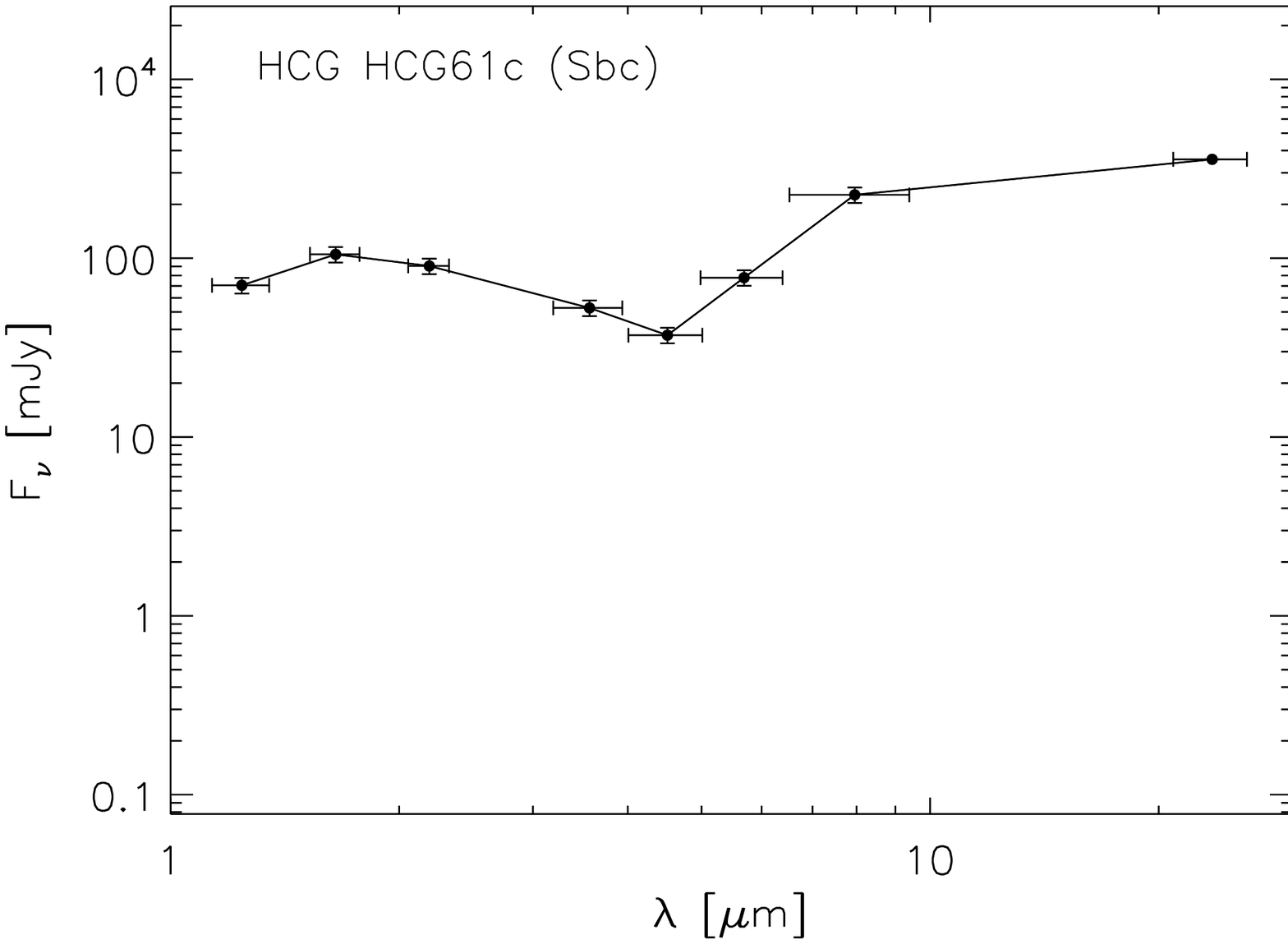}\\
\epsscale{0.46}
\plotone{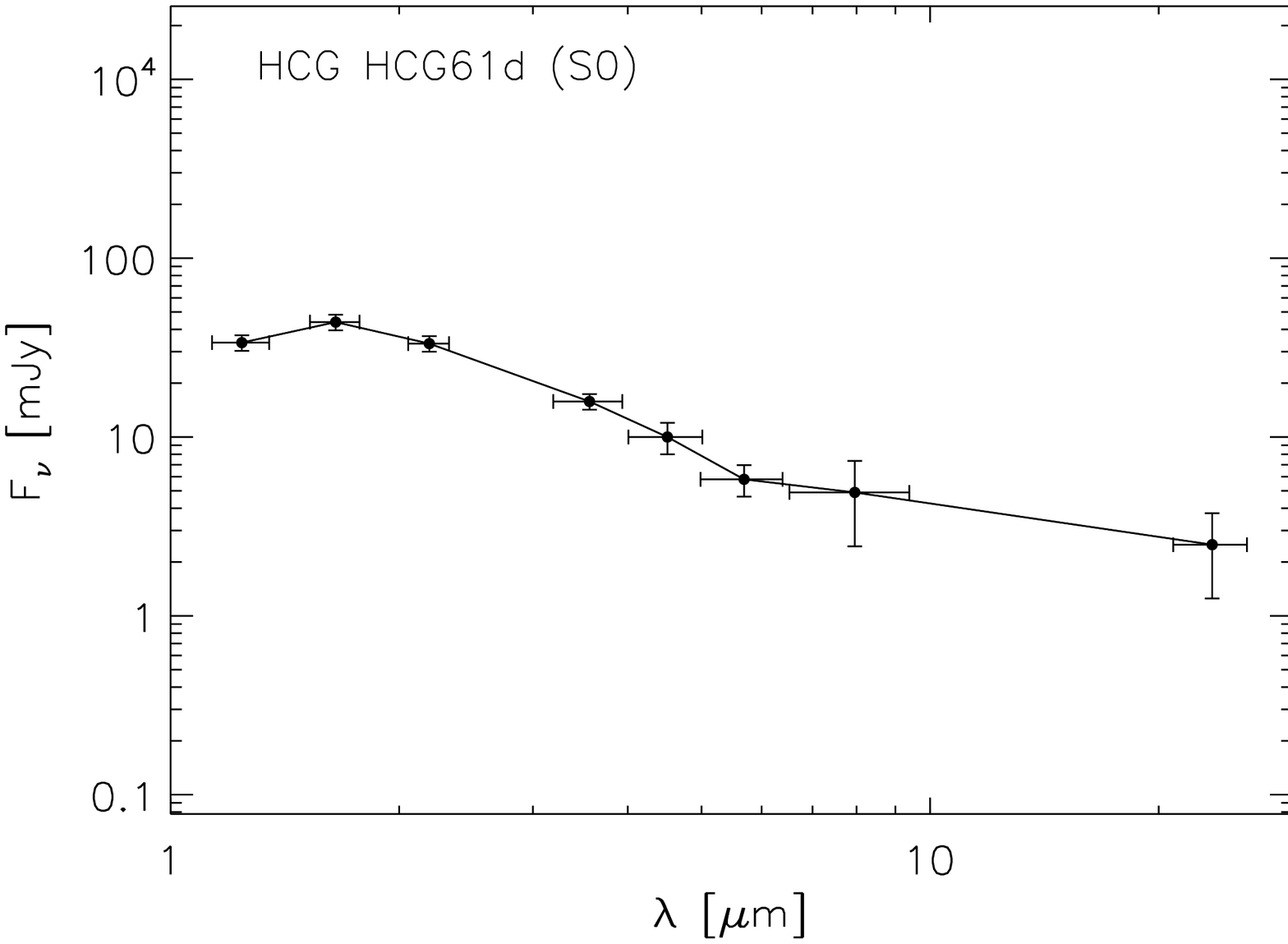}
\caption{Infrared spectral energy distributions for the galaxies
in HCG~61.  Horizontal error bars reflect the filter widths. \label{hcg61}}
\end{figure}

The spectral energy distributions for the galaxies in HCG~61 are shown
in Figure~\ref{hcg61}.  This group is classified as type~I according
to ${\rm log}(M_{{\rm HI}})/{\rm log}(M_{{\rm dyn}})$, indicating it
is relatively gas-rich.  Galaxies~A (NGC~4169) and D (NGC~4174) are
dominated by stellar light. Galaxy~B (NGC~4173) is a foreground
object.  Galaxy~C (NGC~4175) has a disturbed morphology near the
nuclear region and has moderately strong nuclear 24$\mu$m emission.
Galaxies~A and C have been identified as active galaxies \citep{white97}.

\subsection{HCG~62}
\begin{figure}
\epsscale{1.0}
\plottwo{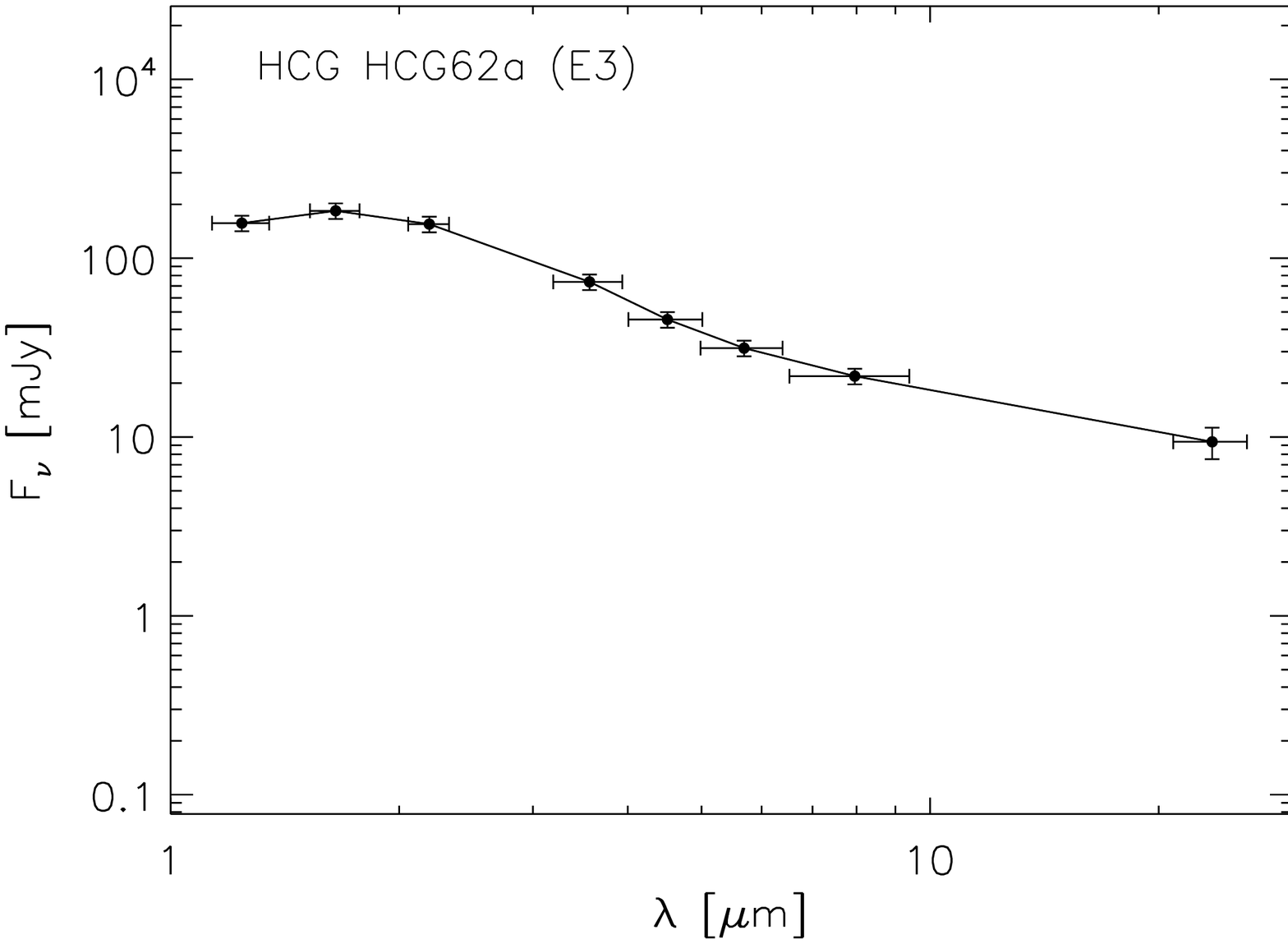}{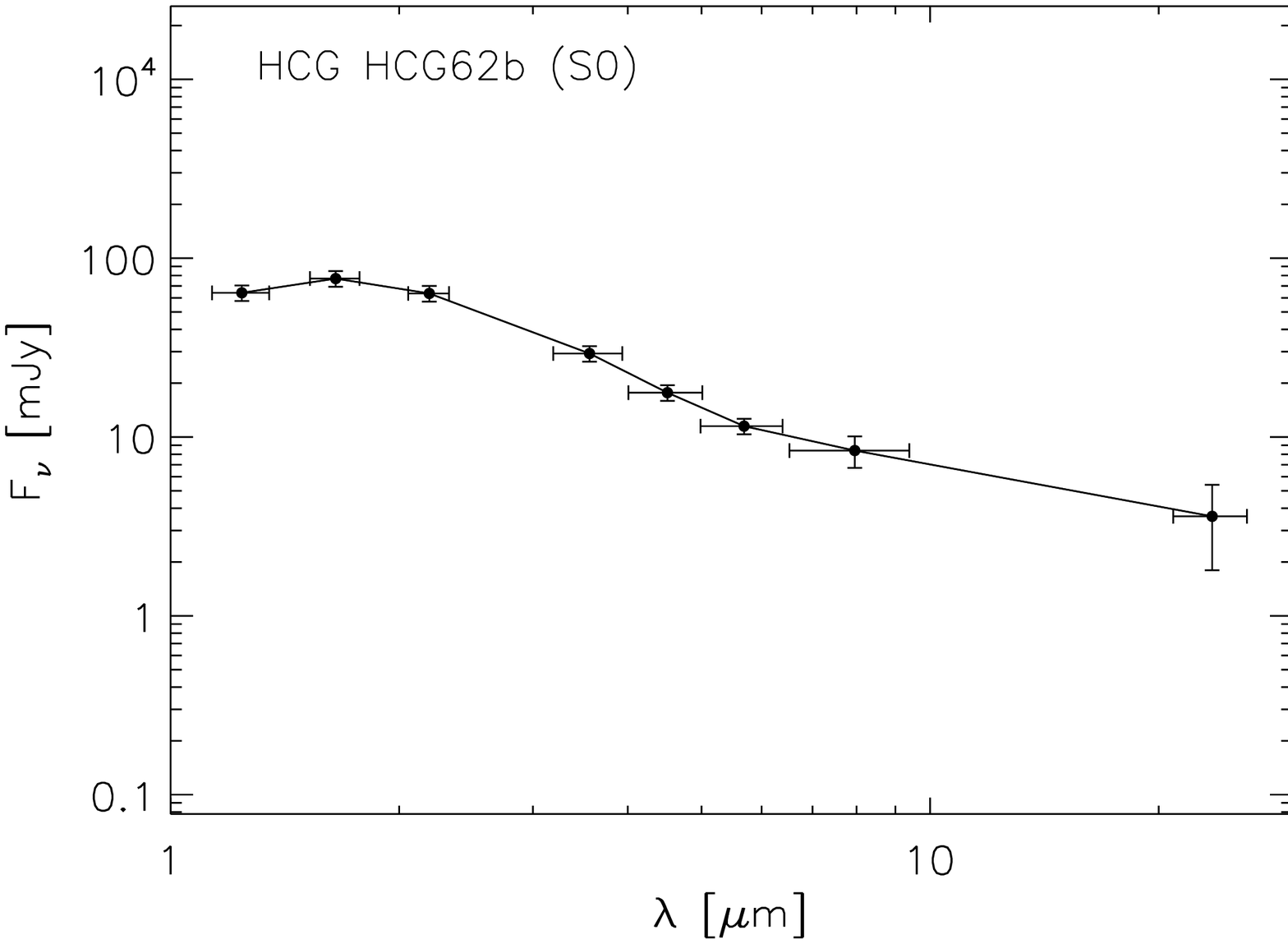}\\
\plottwo{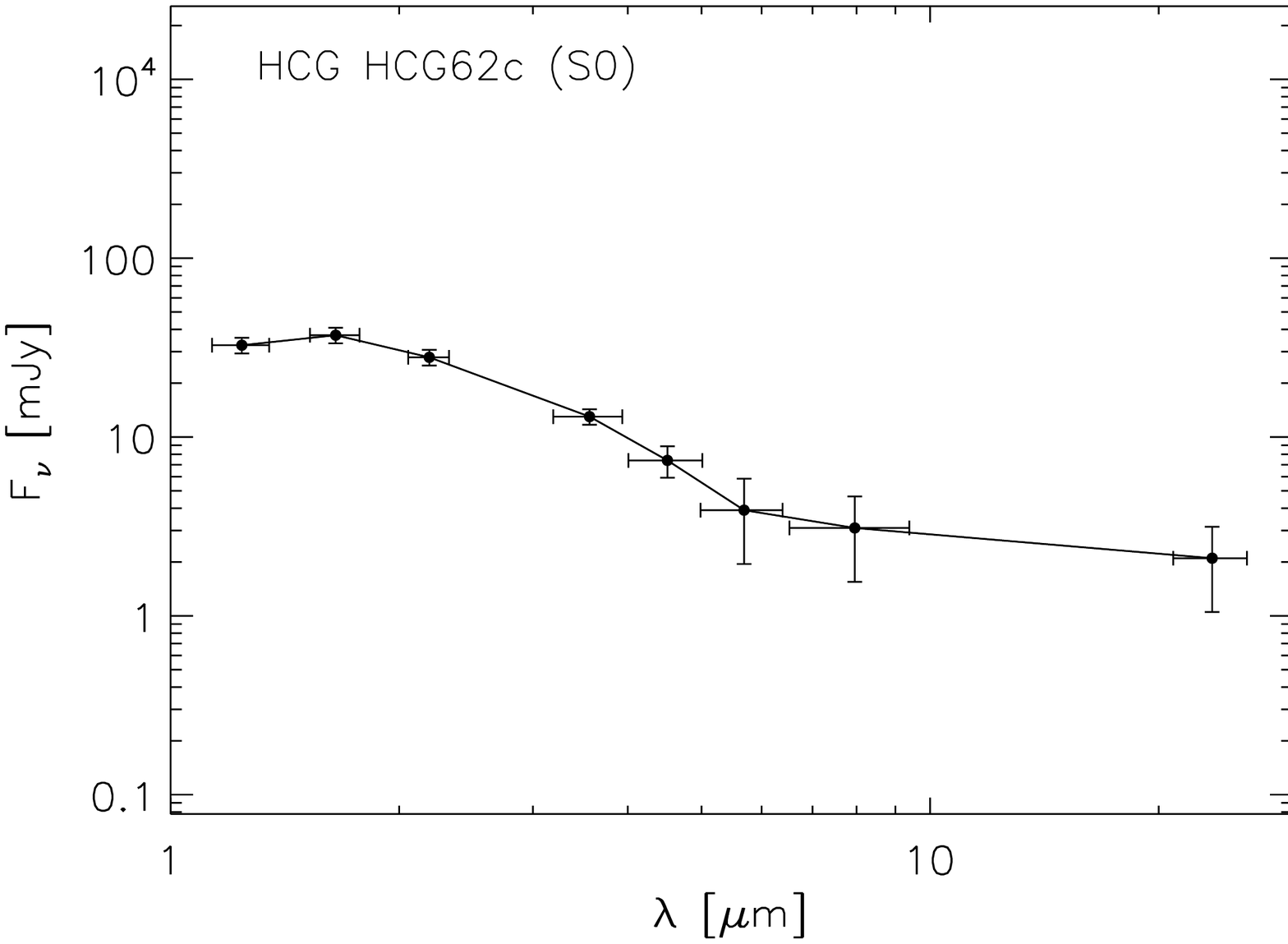}{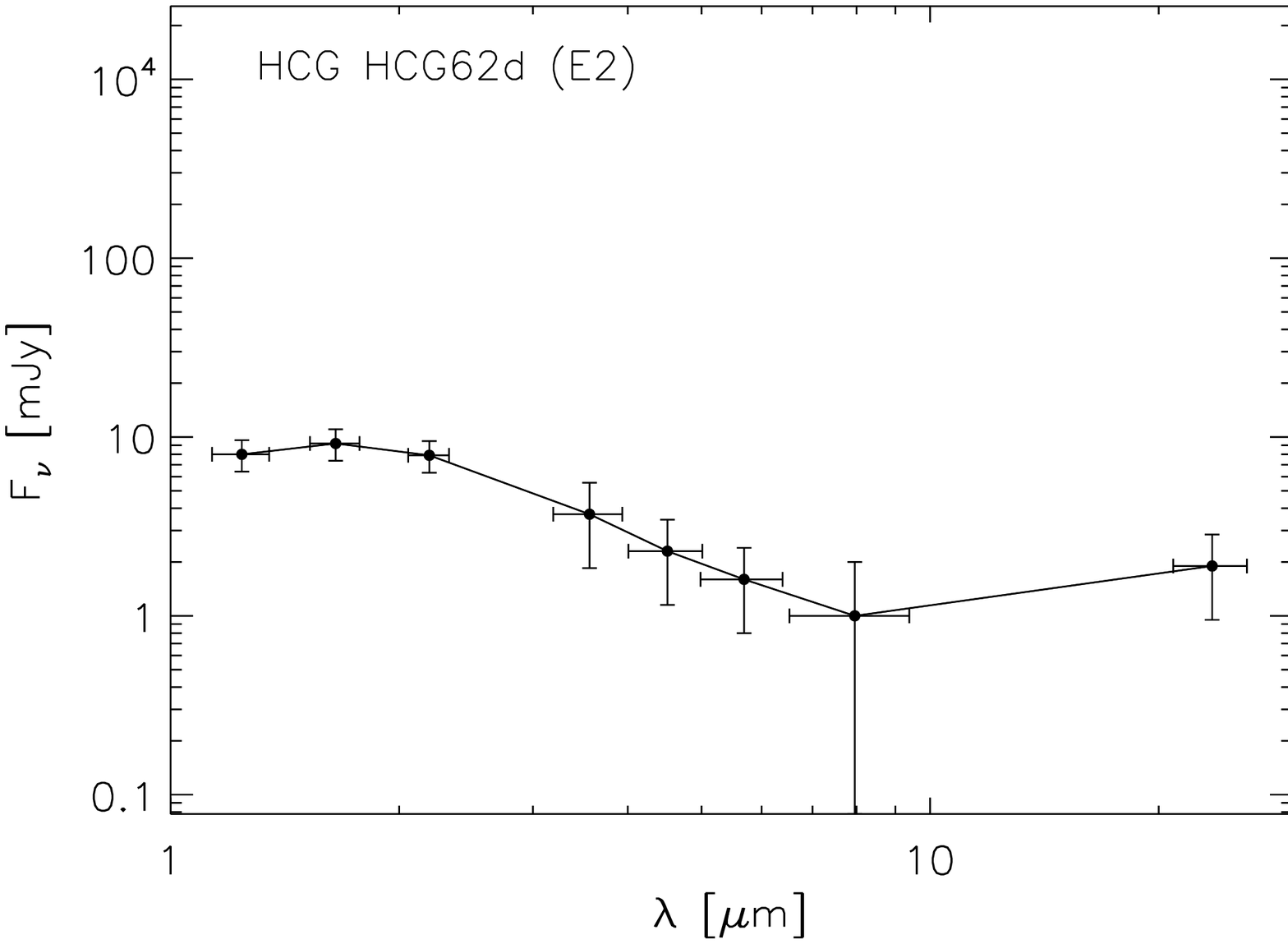}
\caption{Infrared spectral energy distributions for the galaxies
in HCG~62.  Horizontal error bars reflect the filter widths. \label{hcg62}}
\end{figure}

The spectral energy distributions for the galaxies in HCG~62 are shown
in Figure~\ref{hcg62}.  This group is classified as type~III according
to ${\rm log}(M_{{\rm HI}})/{\rm log}(M_{{\rm dyn}})$, indicating that
the group is relatively gas-poor.  \citet{ribeiro98} identify HCG~62
as being part of a loose group.  Galaxies~A (NGC~4778) and B
(NGC~4776) are strongly interacting, but are dominated by stellar
light with little recent star formation activity, although galaxy~B
has a modest amount of 24$\mu$m emission associated with its nuclear
region.  The kinematical properties of galaxy~A (NGC~4778), including
nuclear counter-rotation suggest that it may have undergone a minor
merger \citep{spavone06}, although based on the relative lack of
infrared luminosity, these {\it Spitzer} observations imply any such
merger had little effect on star formation, perhaps indicating that
the gas in these galaxies was already depleted.  Galaxies~C (NGC~4761)
and D (NGC~4764) are also dominated by stellar light, with little
evidence for recent activity.  

\subsection{HCG~90 \label{hcg90sed}}
\begin{figure}
\epsscale{1.0}
\plottwo{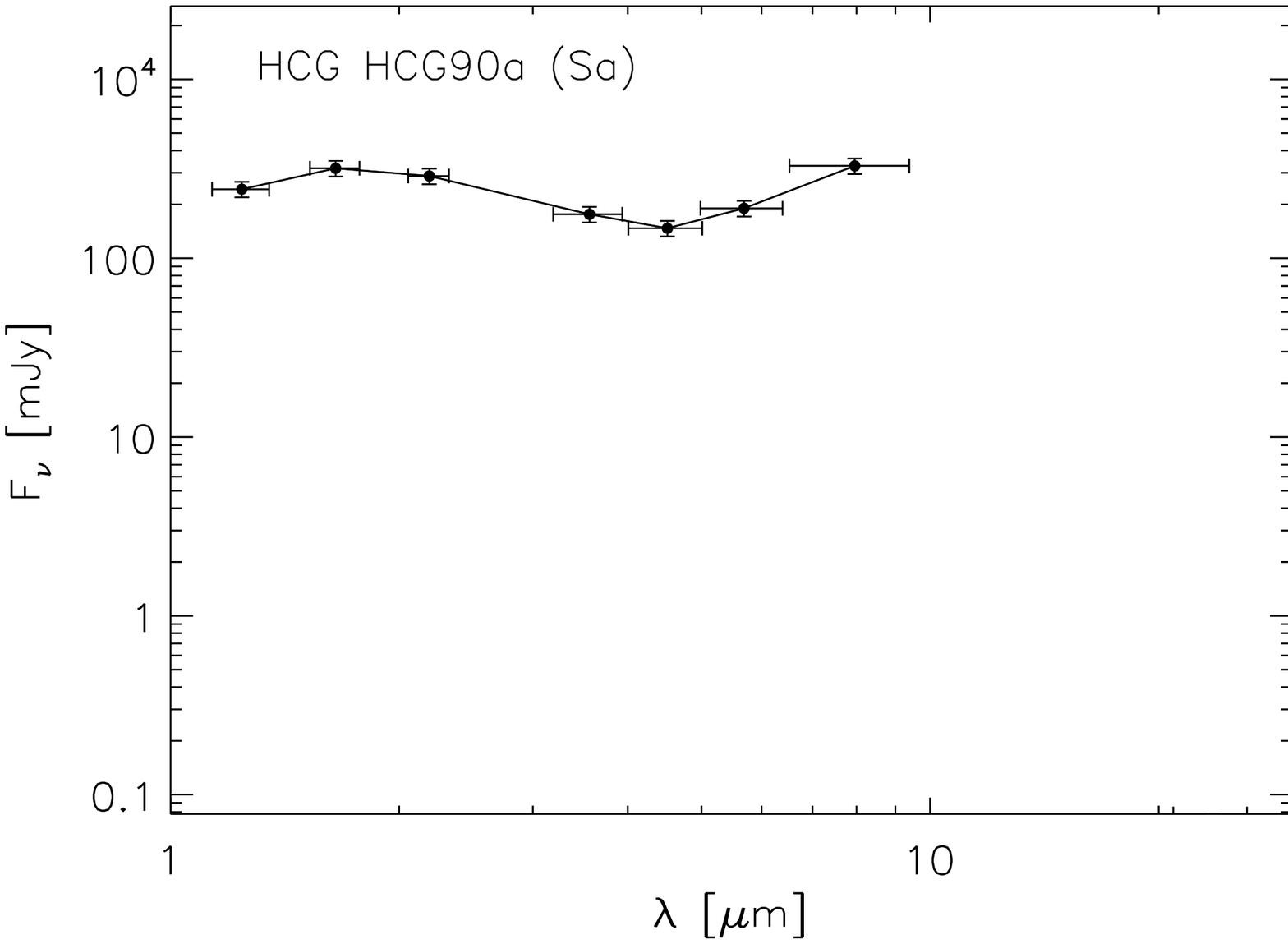}{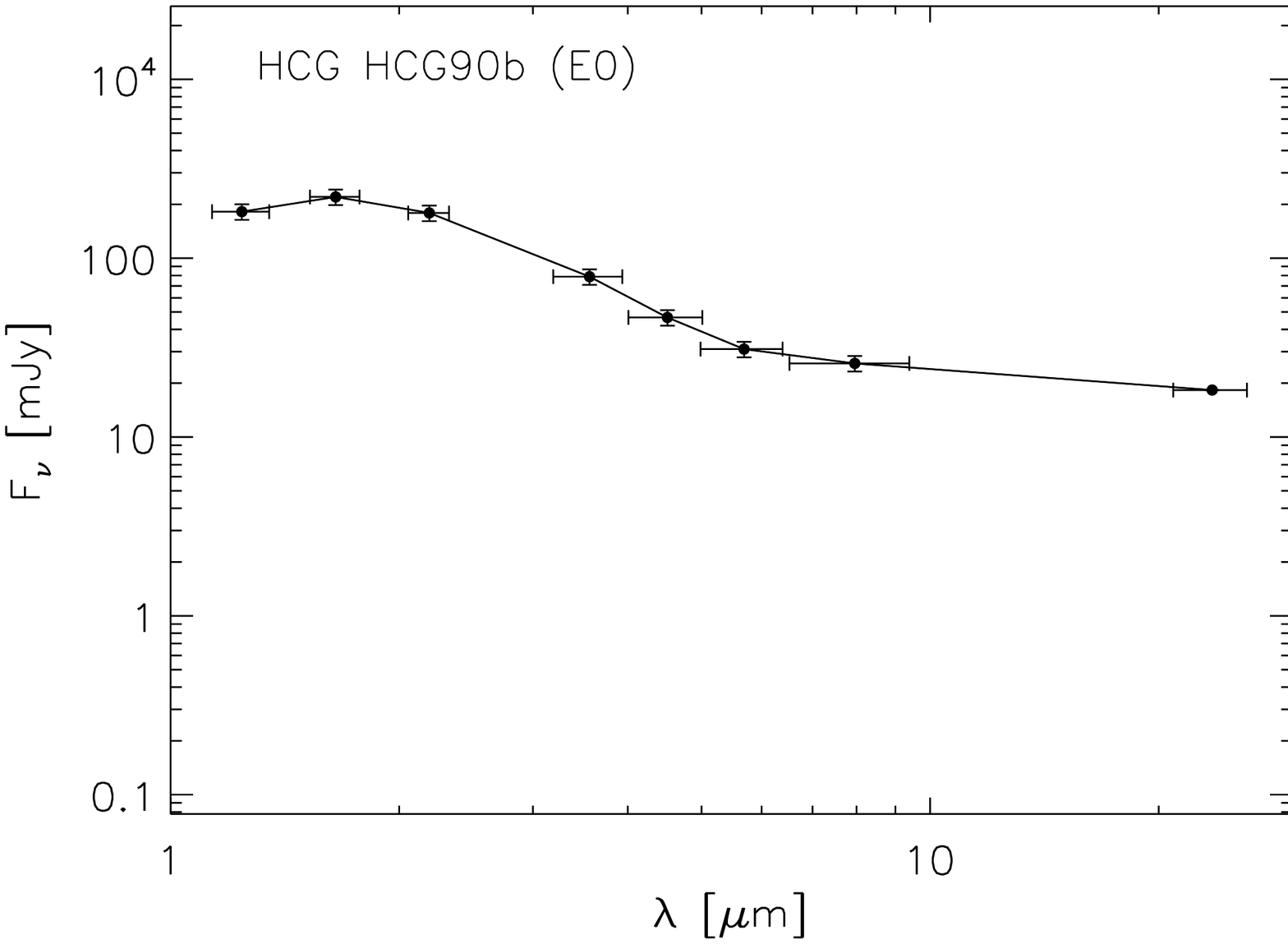}\\
\plottwo{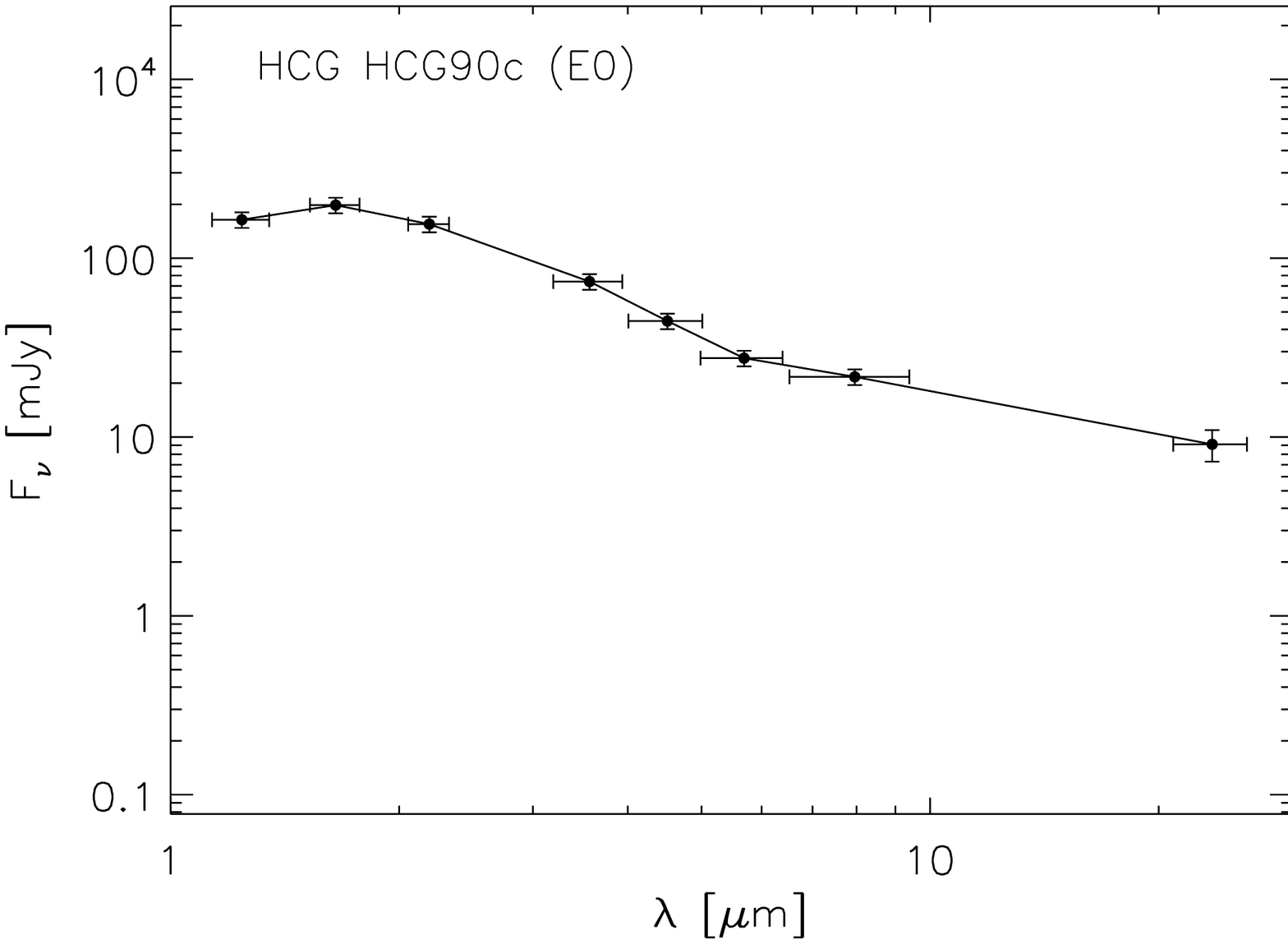}{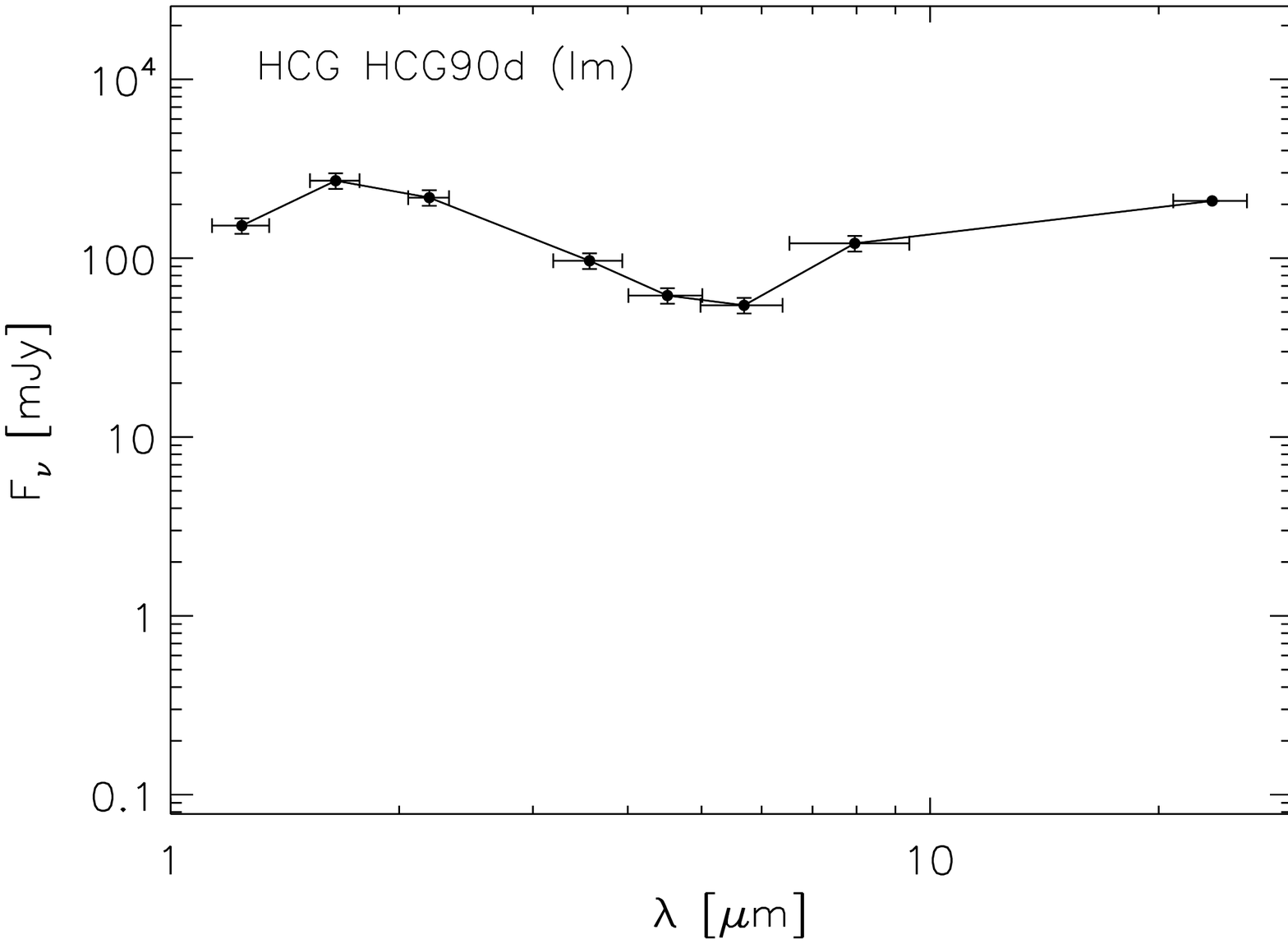}
\caption{Infrared spectral energy distributions for the galaxies
in HCG~90.  Horizontal error bars reflect the filter widths. \label{hcg90}}
\end{figure}

The spectral energy distributions for the galaxies in HCG~90 are shown
in Figure~\ref{hcg90}.  This group is classified as type~III according
to ${\rm log}(M_{{\rm dyn}})/{\rm log}(M_{{\rm HI}})$, indicating that
the group is relatively gas-poor.  \citet{ribeiro98} identify HCG~90
as being part of a loose group.  Due to the close interaction between
galaxies~B and D, they were not separable at low contour levels and
their photometry was combined into a single data point.  Galaxy~D fell
off the detector in the 24$\mu$m observations, and is therefore
lacking this data point. Galaxy~A (NGC~7172) is somewhat disturbed and
has clumpy infrared emission throughout the galaxy.  Galaxies~B
(NGC~7176) and C (NGC~7173) are dominated by stellar light, with
little evidence for recent activity.  Galaxy~D (NGC~7174) is fairly
disturbed, with a spur of infrared emission originating from its
western edge.  The nuclear region of galaxy~D is also a luminous
24$\mu$m source.  These results are consistent with the kinematic
H$\alpha$ observations of \citet{plana98}, who conclude that galaxy~D
is a warm-gas reservoir in the group.  Galaxies~A and D have been
identified as active galaxies by \citet{coziol98}, and these are the
outlying type~III galaxies in Figs.~\ref{plotIRAC2}-\ref{plotJHK2}.

\onecolumn

\twocolumn

\end{document}